\documentclass[11pt]{iopart}

\expandafter\let\csname equation*\endcsname\relax
\expandafter\let\csname endequation*\endcsname\relax

\usepackage{comment}
\usepackage{psfrag,graphicx}
\usepackage{amsfonts,amssymb,amsmath,mathtools} \usepackage{enumitem}     
\usepackage{graphicx}
\usepackage{dcolumn}
\usepackage{bm}
\usepackage{float}
\usepackage{hyperref}
\usepackage{accents}
\usepackage{bbding}
\usepackage{twemojis}
\usepackage{color,soul}
\usepackage{epstopdf}
\usepackage{changepage}
\usepackage{bm}
\usepackage{subcaption}
\usepackage{orcidlink}

\usepackage[normalem]{ulem}

\usepackage{xr}
\usepackage[capitalize,noabbrev]{cleveref}

\crefname{section}{Sec.}{Secs.}
\Crefname{section}{Sec.}{Secs.}

\crefname{subsection}{Sec.}{Secs.}
\Crefname{subsection}{Sec.}{Secs.}

\crefname{equation}{Eq.}{Eqs.}
\Crefname{section}{Equation}{Equations}

\newcommand{\erf}[1]{Eq.~(\ref{#1})}
\newcommand{\beq}{\begin{equation}}
\newcommand{\eeq}{\end{equation}}
\newcommand{\nn}{\nonumber}

\newcommand{\dg}{^\dagger}

\newcommand{\smallfrac}[2]{\mbox{$\frac{#1}{#2}$}}

\newcommand{\bra}[1]{\langle{#1}|}
\newcommand{\ket}[1]{|{#1}\rangle}

\newcommand{\dbd}[1]{\frac{\partial}{\partial {#1}}}

\newcommand{\cu}[1]{\left\{ {#1} \right\}}
\newcommand{\ro}[1]{\left( {#1} \right)}
\newcommand{\an}[1]{\left\langle{#1}\right\rangle}

\newcommand{\s}[1]{\hat{\sigma}_{#1}}

\newcommand{\dd}{\text{d}}

\newcommand{\ie}{{\em i.e.}}
\newcommand{\eg}{{\em e.g.}}

\newcommand{\ea}{{\em et.~al.}}

\newcommand{\boss}{\boldsymbol}

\newcommand{\union}{\cup}
\newcommand{\past}[1]{\overleftarrow{{#1}}} %\overleftarrow{#1}}
\newcommand{\fut}[1]{\overrightarrow{{#1}}}
\newcommand{\both}[1]{\overleftrightarrow{{#1}}}

\newcommand{\inv}{^{-1}}

\newcommand{\nullset}{{\emptyset}}

\renewcommand{\Pr}{{\rm Pr}}
\newcommand{\Su}{{\rm Su}}
\newcommand{\DAG}{G}

\definecolor{nblue}{rgb}{0.3,0.3,1.0}%229
\definecolor{ngreen}{rgb}{0,0.7,0}%272%161
\definecolor{nred}{rgb}{0.8,0.1,0.1}%711&900%1022
\definecolor{nyellow}{rgb}{0.9,0.7,0.2}%772
\definecolor{npurple}{rgb}{0.6,0.0,0.6}%606
\definecolor{nrose}{rgb}{0.5,0.3,0.3}
\definecolor{nbackground}{rgb}{1,1,1}
\definecolor{ngrey}{rgb}{0.5,0.5,0.5}
\definecolor{nbrown}{rgb}{0.5,0.4,0.0}
\definecolor{nblack}{rgb}{0,0,0}
\definecolor{ncyan}{rgb}{0.1,0.5,0.5}
\definecolor{npale}{rgb}{0.9,0.9,0.9}
\definecolor{sky}{rgb}{0,0.6,0.8}%159
\definecolor{nmauve}{rgb}{0.1,0.0,0.7}
\definecolor{norange}{rgb}{1.0,0.4,0}
\definecolor{napricot}{rgb}{1,0.6,0.5} %10,8,7
\definecolor{golden}{rgb}{0.75,0.6,0.15}
\definecolor{nbrick}{rgb}{0.65,0.25,0.15}

\definecolor{nmagenta}{rgb}{0.7,0.0,0.3}

\newcommand{\blk}{\color{nblack}}

\usepackage[autostyle, english = american]{csquotes}
\MakeOuterQuote{"}

\begin{document}

\title[Counterfactual quantum measurements: a general treatment]{Counterfactual quantum measurements: a general treatment using hybrid causal models}
%more general causal structures, more interesting summary statistics, and more parties $\twemoji{partying face}\twemoji{party popper}$}

\author{Ingita Banerjee$^{1}$, Kiarn T. Laverick\orcidlink{0000-0002-3688-1159}$^{2,3}$, Howard M. Wiseman\orcidlink{0000-0001-6815-854X}$^{1}$}
\address{$^1$Quantum and Advanced Technologies Research Institute, Griffith 
University, Yuggera Country, Brisbane, Queensland 4111, Australia}
\address{$^2$MajuLab, CNRS-UCA-SU-NUS-NTU International Joint Research Laboratory}
\address{$^3$Centre for Quantum Technologies, National University of Singapore, 117543 Singapore, Singapore}
%\address{Centre for Quantum Computation and Communication Technology 
%(Australian Research Council), \\ Quantum and Advanced Technologies Research Institute, Griffith 
%University, Yuggera Country, Brisbane, Queensland 4111, Australia}

\ead{h.wiseman@griffith.edu.au}

\date{\today}

\begin{abstract}

Counterfactual reasoning, of an informal sort, is used ubiquitously to answer questions like ``what would have happened if I had done $A'$ instead of $A$?'' In previous work [Phys. Rev. A {\bf 113}, 042206 (2026)] we introduced a rigorous calculus for answering counterfactual questions in orthodox quantum theory, where the locus of change, $\bm{A}$,  from actual to counterfactual world, is one or more measurement setting. This involved introducing the concept of {\em supposability}, for the supposed probability of outcomes in the counterfactual world conditioned on evidence in the actual world. In this paper, we extend that work and explore new applications. First, for scenarios with parties at different locations, we show how the supposabilities are evaluated when the parties communicate. Second, we show how to ground the calculus in a general hybrid (quantum--classical) causal model rather than relying on space-time relations to enforce causal independence. Third, rather than calculating simple time-local averages of a continuous measurement record, according to its supposability distribution, we calculate more interesting counterfactual summary statistics of such records. Specifically, we apply our theory to
the solution of quantum trajectory equations, which describe the real-time knowledge
of different parties about the quantum system under investigation: a resonantly driven atom. Its fluorescence is subject to photon detection by two parties, one of whom asks what would have happened if she had used homodyne detection instead.

\end{abstract}

\maketitle

\section{Introduction} 
Counterfactual reasoning---the consideration of alternate possibilities to events that have already happened---is both a ubiquitous tendency in human thought and a central concept in modern causal inference and statistical analysis.
It has been used in fields as diverse as policy making~\cite{tetlock1996counterfactual}, game theory~\cite{epstude2008functional}, weather and climate~\cite{otto2017attribution}, artificial intelligence~\cite{celar2023people}, philosophy~\cite{lewis1979counterfactual,butterfield1992david,pearl2018book} and causal discovery~\cite{pearl2009causal,Imbens_Rubin_2015,repec:ris:nobelp:2021_002}. In quantum theory, however, the status of counterfactuals has long been unclear. The Einstein-Podolsky-Rosen (EPR)~\cite{PhysRev.47.777} and Bohr~\cite{PhysRev.48.696}, which centered around the counterfactual result of a measurement that could have been performed but was not. Since then, considerations of counterfactuals in quantum mechanics have mostly concerned hidden variables ~\cite{stapp1997nonlocal,PhysRevA.59.126,mermin1998nonlocal,stapp1998meaning,AHARONOV2002130,vaidman2009counterfactuals,vaidman2007counterfactualsquantummechanics}, 
%Counterfactual statements have been prominent in the foundations of quantum theory literature ever since the 1935 debate between Einstein-Podolsky-Rosen (EPR)~\cite{PhysRev.47.777} and Bohr~\cite{PhysRev.48.696}, which centered around the counterfactual result of a measurement that could have been performed instead of the one actually done. This was 
and the controversy over whether `counterfactual definiteness' (that counterfactual propositions have definite truth values) was an implicit assumption in the derivation of Bell's theorem~\cite{skyrms1982counterfactual,PhysRevD.3.1303,lambare2021note,Zukowski2014}, and this has also inspired other theorems~\cite{wang2025quantumexperimentjointexogeneity}. Very recently, however, some works~\cite{suresh2023semantics,banerjee2025counterfactualquantummeasurements} have shown that counterfactual reasoning can be formalized entirely within orthodox quantum theory (OQT), without recourse to hidden-variable models.

In our previous work~\cite{banerjee2025counterfactualquantummeasurements}, we developed a calculus for quantum counterfactuals, extending David Lewis' analysis~\cite{lewis1979counterfactual}---originally formulated for deterministic laws of nature---to the indeterministic setting of quantum theory. In our formalism, the counterfactual antecedent is a choice of measurement setting, and the deterministic truth-values of classical counterfactual propositions are replaced by counterfactual probabilities, which we termed \emph{supposabilities}. We first illustrated the calculus with a simple counterfactual proposition in a Bell--CHSH~\cite{PhysRevLett.23.880} scenario, showing that counterfactual questions admit non-trivial answers even in this minimal setting. We then formalized the framework in precise mathematical language, identifying the essential ingredients that constitute a quantum counterfactual---in particular, the \emph{fixtures}: the classical events that remain fixed between the actual and the counterfactual world. Finally, we demonstrated the power of the formalism in a much less trivial, and also physically interesting, scenario involving continuous monitoring of atomic fluorescence~\cite{Carmichael1993,PhysRevA.47.1652,PhysRevLett.81.3635}. There, the framework allowed us to meaningfully answer questions like: ``Given that a photon-detector, observing an atom's fluorescence, clicked at a certain time, what would a field-quadrature detector have measured, if it had been used instead?''

Although our previous paper showed the power of our formalism to answer interesting questions in physically relevant measurement scenarios, our examples were not as general as the formalism allowed. First, they were restricted to cases in which there was no communication between the observers involved. That is, the parties were always assumed to remain causally separated from each other during the course of the experiment. Second, we limited our discussion to situations where the causal structure was taken to be determined by the space-time arrangement of events. That is, the reader may have got the impression that our formalism relied upon such space-time arrangements. Third, we restricted our calculation to what we call \emph{suspectation values}---the counterfactual analogues of expectation values---of measurement outcomes. For the CHSH example, because Alice's measurement is dichotomic, knowing the suspectation value of Alice's counterfactual result is equivalent to knowing its full \emph{suppposability distribution}---the counterfactual analogue of a probability distribution. But for the atomic fluorescence measurement, the suspectation value of Alice's counterfactual homodyne current by no means exhausts the information in the supposability distribution for that record.

 In the current work, we address all three of the above limitations. First, we introduce more general scenarios, involving more than two parties, or involving communication between the parties (which is allowed if we change the space-time arrangement of events). This  breaks the simple structure of our previous examples where the evidence, antecedent and consequent referred to one party (Alice) and the fixtures referred to the other (Bob). Second, we show how our calculus can be grounded by any causal structure, described by a hybrid (quantum--classical) causal model without specifying a space-time arrangement. We provide explicit causal diagrams for the cases studied previously, as well as for the cases introduced here. 
 %where we explore a more general class of scenarios which we discussed in the conclusion of our previous work. In particular, we show that our calculus naturally accommodates situations 
  Finally, we further illustrate the generality of the framework by computing suspectation values of more informative summary statistics derived from the full measurement record, rather than merely of the instantaneous record at each time. Specifically, we apply our counterfactual reasoning to the solution of quantum trajectory equations~\cite{wiseman2009quantum}, which describe the real-time {\em knowledge} of certain parties about the quantum system under investigation. This fits within our paradigm for counterfactuals in quantum mechanics because states of knowledge are themselves classical objects.

\subsection{Outline} 

The paper is organized as follows.
In \cref{sec:2}, we review the notion of counterfactuals, beginning with the classical frameworks of Lewis and Pearl, and then tracing their development in quantum theory. We discuss early quantum counterfactual reasoning arising from the Einstein–Bohr debate, Bell’s theorem, and subsequent extensions, before summarizing a recent ``semantic'' approach to quantum counterfactuals~\cite{suresh2023semantics}.

In \cref{sec:CalculusandDag}, we first present the intuition, based on Lewis, behind the counterfactual calculus of our previous work~\cite{banerjee2025counterfactualquantummeasurements}. Then we introduce Hybrid Structural Causal Models (HSMs), which embed quantum and classical nodes within a unified causal framework\footnote{After this work was complete, we became aware of the diagrammatic {\bf QPhys} model introduced by Selby \ea~\cite{Selby2026timesymmetryin}, which works in essentially the same way.}. Using the HSM, we extend our earlier work to general causal structures and  introduce the labeled directed acyclic graphs (L-DAGs) that serve as the central representational tool throughout the paper. 

% In \cref{sec:CalculusandDag}, we introduce our counterfactual quantum measurement calculus in the language of quantum causal directed acyclic graphs (Q-DAGs). 
Having presented the general formalism, we illustrate it with two examples: a Bell--CHSH scenario and a driven atom, whose resonance fluorescence in different directions is monitored by different observers. Both of these were introduced in Ref.~\cite{banerjee2025counterfactualquantummeasurements}, but here we give an L-DAG treatment, and also introduce (for notatational convenience) new names for the two observers: Tara and Vlad.    

In \cref{sec:overnew}, we overview the new scenarios appearing later in this paper, all more complex versions of the resonance fluorescence system. First, we discuss how the calculus also allows one to break the standard two-party correspondence, in which one party's results are unknown fixtures. We then consider suspectation values for more informative summary statistics of the observer's record than simply the instantaneous current, namely the filtration of that current that yields the conditioned state. Finally, we show how a multi-observer framework (involving a third party, Godot) accommodates counterfactual questions that naturally use certain party's assigned conditional states to break the standard party correspondence.

The treatment of the driven atom in \cref{sec:CalculusandDag}, and the discussion in \cref{sec:overnew}, was qualitative only. In \cref{sec:QMT}, we review the techniques from continuous measurement and estimation required for a quantitative treatment. This includes open quantum systems, quantum state filtering, effect retro-filtering, and quantum state smoothing. We end the section by commenting on how the smoothed state can be interpreted as a supposable state within a counterfactual framework, slightly extending the work of Ref.~\cite{liu2026unifyingquantumsmoothingtheories}.

In \cref{sec:BreakParty} we return to the breaking of party correspondence using a  %mathematically evaluating a suspectation value for a , and discussing how it differs from prior results. We show how the 
modified causal structure from Ref.~\cite{banerjee2025counterfactualquantummeasurements} which changes the membership of the sets used to define the counterfactual. Here the counterfactual can naturally be posed by one party (Tara) in the experiment, and we evaluate the suspectation of her counterfactual homodyne current. Comparing the result with the strict spacelike-separated case in~\cite{banerjee2025counterfactualquantummeasurements}, we find that conditioning on the other party's (Vlad's) causally accessible record acts as an additional pre-selection on the compatible ensemble, producing features diagnostic of weak-values---a transient excursion of the suspected current beyond the eigenvalue range of the measured operator.

In \cref{sec:issi} we extend the calculus from estimation of counterfactual records to estimation of counterfactual states. We give a unified expression for the suspected state estimates an observer can assign via the counterfactual setting calculus, and identify which such estimates qualify as legitimate consequents of the counterfactual. We then compute the simplest case: the suspectation value of Tara's counterfactual conditional (filtered) state, which is purely a function of variables in the consequent. 
We compare this to Tara's actual filtered state, highlighting the similarities and differences.  
%For conceptual convenience and readability, we only stick to the computation of counterfactual estimate of the case that preserves the strict space-like separation in this section. We discuss the rest in a separate section. 
%In both the sections the resulting suspectation values are then compared with the various conditioned states  that the same observer infers in the actual scenario, illuminating the relationship between actual and counterfactual quantum inferences. \blk

%\how{This paragraph and the following paragraph seem to mix up a description of Secs. 7 and 8. Please rewrite, one paragraph each. I removed references to subsections because we don't do that in any other paragraph.}
%In \cref{sec:QTestimates} we show explicitly how the counterfactual calculus is applied to evaluate the suspected estimates. We identify the evidence, antecedent, and fixtures for this question, and 
% Finally, we
% compute the resulting estimate as a nested average over the fixed records and the counterfactual records. 
% \blk
% In \cref{sec:3party} we exploit the full classical-processing freedom afforded by the HSM within the counterfactual calculus. By classically processing local measurement readouts, observers can compute conditioned states that have interesting informative summary statistics. We define the suspected estimates accessible to the primary observer, extend them to the other parties in the scenario, and present their explicit evaluation. 

 In \cref{sec:3party} we return to the two less simple counterfactual state estimates from the preceding section. These estimates involve  Godot, as introduced in \cref{sec:overnew}, and thus break the party correspondence. In one of these cases, when the antecedent corresponds to Tara's performing of no measurement, the suspected state reduces to a type of {\em smoothed} estimate of Vlad's filtered state in the actual world. In the other case, where the antecedent corresponds to Tara's performing a homodyne measurement (as usual), it corresponds to a smoothed estimate of the {\em wrong smoothed} state of Vlad. The most useful comparison for these cases is Tara's actual world smoothed state. 
 
 We end in \cref{sec:conc} with our final remarks and scope for future work. 
\blk

\section{Counterfactuals: classical to quantum \label{sec:2}}
\subsection{Lewis \label{sec:2a}}

David Lewis's analysis of counterfactuals~\cite{lewis1979counterfactual} has been one of the most influential approaches for counterfactual reasoning in philosophy. Lewis aimed to give a rigorous, principled account of the meaning and truth conditions of statements about what would have happened in situations that did not actually occur, using comparative analysis between possible worlds. The counterfactuals that he used were of the form: ``If it were that $A$, then it would be that $C$''. Here $A$ is the antecedent--- the alternative hypothetical scenario --- and $C$ is the consequent --- the events that follow as a result of the antecedent. The antecedent in Lewis' scenario is a plausible alternative within the laws of the theory that requires some `minimal changes' from the actual world. According to his analysis, the truth value of this counterfactual depends on what happens in the closest possible world to the actual world where the antecedent $A$ is true. This was achieved by comparing a set of possible alternative worlds to the actual world and single out the one that is closest to the actual world. To identify this closest world Lewis laid down the following hierarchy of desiderata:
\begin{enumerate}
    \item Avoid big, widespread, diverse violations of law. 
    \item It is of second importance to maximize the spatio-temporal region throughout which perfect match of particular fact prevails.
    \item It is of third importance to avoid even small, localized, simple violations of law.
    \item It is of little or no importance to secure approximate similarity of particular fact, even in matters that concern us greatly. 
\end{enumerate}
However, the lack of a proper metric to measure the closeness  drew a lot of criticisms to his analysis~\cite{Percival1999-PERANO-2,Elga2001-ELGCDA,horwich1987asymmetries,suresh2023semantics}, particularly that it is not always clear as to what should remain ``matched'' when proposing the antecedent. 

There are two major hurdles to extending Lewis's methods to quantum theory~\cite{butterfield1992david}. First, it assumes a classical ontology, where there is a matter of fact about the physical state of the world at any point in space-time. Second, his analysis is built on deterministic laws of nature, whereby Lewis concludes that  counterfactual statements are always either true or false. In OQT,  however, most physical quantities do not have determinate properties and the theory is inherently probabilistic. We will return to this  
issue in Sec.~\ref{sec:ReturnToLewis}, after we have reviewed other work.

\subsection{Pearl \label{sec:1b}} 
Judea Pearl's analysis of counterfactuals~\cite{pearl2000models} is another influential approach, which aims to be empirically useful as well as rigorous. In Pearl’s framework, counterfactuals are defined with respect to a {\em Classical  Structural causal Model (CSM)}, that represents the causal dependencies in the situation under study. Structural causal models sit at the top of Pearl’s “ladder of causation,”~\cite{pearl2018book} a hierarchy in which progressively more sophisticated models are needed to answer progressively more sophisticated causal questions. Such models are formulated on causal directed acyclic graphs (DAGs). A DAG $G$ is a graph whose vertices represent events or variables and whose directed edges reflect causal relations; crucially, these directed connections never form closed loops (hence ``acyclic''). (See Sec.~\ref{sec:Bell} below for an example of a DAG.) In this framework, counterfactuals still constitute an antecedent and a consequent, but their evaluation is fully grounded in the causal model. The antecedent is implemented by performing an {\em intervention}, thereby severing the causal connections into the intervened variable, and the consequent is then obtained by propagating this change through the model. This is very different from Lewis’s ``minimal change'' criterion, since Pearl specifies the hypothetical scenario by explicitly modifying the causal mechanism rather than seeking a closest possible world. 

Pearl's calculus of counterfactuals is formulated within a {\em probabilistic structural causal model (PSM)} defined by the pair $\langle M,{\bf p}\rangle$, %P(\mathbf{u})
%\how{Does Pearl allow $P(\mathbf{u})$ to describe correlated $u$s? Or is it, like we assume, really ${\bf p} = (\dots,p_l(u_l)\dots)$, \ie, factorized?}\ing{I think correlated u's because there is no factorization of the probability anywhere.}\how{Eric said factorized. So I have changed.} 
where $M$ is a triple $\langle \boss{U}, \boss{V},\mathbf{F}\rangle$. (Here, following our earlier paper~\cite{banerjee2025counterfactualquantummeasurements}, we are using italic bold for a set of variables that can be associated with space-time events, though this is not something Pearl discusses.) 
First, $\boss{V}=\{V_1,\ldots,V_n\}$ is a set of {\em endogenous} variables, which are in one-to-one correspondence with the nodes of the graph $G$. By contrast, $\boss{U}=\{U_1,\ldots,U_n\}$ is a set of {\em exogenous} variables, again one for each node, whose causes (if any) lie outside the model. They may be thought of as unknown background factors, or, in physical terms, local noise variables. The edges of the graph determine the causal relations between the $V_j$s, which is described in full detail by their set of functions, $\mathbf{F}=\{f_1,\ldots,f_n\}$. That is, each $f_j$ gives $V_j$ in terms of $U_j$ and the other $V_k$ which are connected to $V_j$ by an in-coming edge. Finally, 
${\bf p}=(p_1,\ldots,p_n)$ (our notation from Ref.~\cite{banerjee2025counterfactualquantummeasurements}) 
is a set probability distributions, one for each exogenous variable $U_j$.

Now we can explain how Pearl treats counterfactuals.  
A counterfactual starts with some evidence $\bm{E}=\bm{e}$, where 
 $\bm{E} \subseteq \bm{V}$ is the set of observable 
(endogenous) variables in the model. For what comes later, it is useful 
to split the evidence variables into two disjoint sets, $\bm{X}$ and $\bm{Y}$. 
Then, if $\bm{x}$ and $\bm{y}$ are the particular realizations of those variables known to pertain, the evidence is expressed as $\bm{X}=\bm{x}$ and $\bm{Y}=\bm{y}$. 
Pearl considers counterfactuals of the form ``$\bm{Y}$ {\em would have been} $\bm{y}'$ 
had $\bm{X}$ {\em been} $\bm{x}'$, given the observed evidence $\bm{X}=\bm{x}$ and $\bm{Y}=\bm{y}$''.  He denotes the probability for this counterfactual to be true 
by $P(\bm{Y}_{\bm{x}'}=\bm{y}'|\bm{E}=\bm{e})$, which can be evaluated systematically by following a three-step process:
\begin{itemize}
    \item {\em \textbf{Abduction}}: The observed data $\bm{E}=\bm{e}$ is used for Bayesian inference to update the distribution $P(\bm{u})=\prod_j p_j(u_j)$ in the {\em PSM}, to the conditioned probability $P(\bm{u}|\bm{e})$, which need not factorize any more.
    \item {\em \textbf{Action}}: A {\em do-intervention} is performed to set the antecedent variables do$(\bm{X}=\bm{x})$, severing all the incoming causal arrows. This corresponds to replacing the structural equation for $\bm{X}$ (namely $f_{\bm{X}}(\boss{V},\boss{U})$) by a constant assignment $\bm{x}'$. This results in modifying the underlying model to a submodel $M_{\bm{x}'}$.
    \item {\em \textbf{Prediction}}: Finally, in the modified submodel $\langle M_{\bm{x}'},P(\bm{u}|\bm{e})\rangle$, compute the probability of the consequent $\bm{Y}$ as
    \begin{equation}
        P(\bm{Y}_{\bm{x}'}=\bm{y}'|\bm{X}=\bm{x},\bm{Y}=\bm{y})=\sum_{\bm{u}|\bm{Y}_{\bm{x}'}(\bm{u})=\bm{y}'} P(\bm{u}|\bm{X}=\bm{x},\bm{Y}=\bm{y}).
    \end{equation}
    %\how{Or is it a counterfactual probability, as Eric would say, I think? What does Pearl say?} \ing{So, I wrote down this part entirely from Eric's paper itself, where it has been specified as `probability of the counterfactual', I think counterfactual probability is the concept that Eric introduces to extend it to the quantum case. Pearl never talked about `counterfactual probabaility'.}
\end{itemize}
That is, the probability of the counterfactual to be true is the probability (updated by $\bf{e}$) that ${\bf u}$ is such that, in the sub-model $M_{\bm{x}'}$, it would have given rise to $\bm{Y}=\bm{y}$.

\subsection{History of counterfactuals in quantum mechanics} 
    \subsubsection{Einstein--Bohr debate} 

    The Einstein--Bohr debates were a series of public disagreements between Einstein and Bohr about quantum mechanics. A central point that it revolved around was the status of the wave function. % that was, at that time, taken to completely characterize the state of a quantum system. 
    Einstein, Podolsky and Rosen (EPR)~\cite{PhysRev.47.777} famously argued that the wave function does {\em not} provide a complete description of physical reality, which according to their criterion is ``a sufficient condition for the reality of a physical quantity is the possibility of predicting it with certainty, without disturbing the system.'' They considered an entangled wave function for two different systems (although the term entanglement had not yet been introduced). They say, ``as a consequence of two different measurements [$A$ or $B$] on the first system, the second system can be left in states described by two different [types of] wave functions.'' Moreover, as they show, these wave functions can correspond to the eigenfunctions of two non-commuting operators, $\hat Q$ and $\hat P$. They go on,  
    ``Thus, measuring either $A$ or $B$ we are in a position to predict with certainty, and without in any way disturbing the second system, either the value of the quantity $P$ (that is $p_k$) or the value of the quantity $Q$ (that is $q_r$).'' Hence, the values of both of those physical quantities must be elements of reality, which clearly contradicts the idea that the wavefunction gives a complete description of reality. 
    %Crucially, the reasoning used in the EPR paper relies on an implicit assumption of  counterfactual definiteness, namely assigning reality to the outcome of a  measurement that has not been performed but {\em could} have been, since for them, the outcomes of measurements were predefined properties of a physical system that are merely revealed by the act of measurement.  This is how they argued about the incompleteness of quantum mechanics stating that it is an aggregate behaviour of a collection of physical systems.

Bohr’s reply~\cite{PhysRev.48.696} to EPR is infamous for its resistance to apprehension. In the view of one of us~\cite{WISEMAN2013361}, the technical content of his counter-argument was his rejection of EPR's assumption that a measurement of the first system does not disturb the second. 
%focused precisely on attacking the counterfactual assumption underlying the EPR conclusion. Because his argument was expressed in the highly compressed style characteristic of his writings, its interpretation was later elucidated by 
In the view of Bohr's collaborator Rosenfeld~\cite{Rozental1967-ROZNBH,Rozental1967}, however, 
his real objection was to the counterfactuality inherent in considering both measurements $A$ and $B$. As Rosenfeld wrote, %who explained: 
``In the case of the two particles, it is true that the measurement carried out on the first particle does not cause any direct physical disturbance of the second; but the measurement decisively affects the nature of verifiable predictions we will be able to make about this second particle. ...
%[A]s long as we do not carry out any measurement ... we have no control at all over this correlation [between the two particles]. If we really want the system to be subject to study and communication, we must carry out some measurement. 
If we now observe the position of the first particle, the correlation between the positions of the particles can be used to give us information about where the second particle is, but we have no way of making use of the correlation between the pulses of the particles .... If we observe the momentum of the first particle, it is just the opposite. We retain control over the momentum correlation, but lose it over the position correlation. The two different measurements define two complementary phenomena that can never be reconciled into a single description of the given two-particle system.'' 

%This can be understood as an explicit illustration of Bohr’s principle of complementarity. For Bohr, the quantum state is a contextual and epistemic, that is, its meaning cannot be separated from the actual measurement arrangement. In this view what changes under different measurements is not the physical disturbance of the distant system but the conditions under which predictions can meaningfully be made. Hence, 
That is, in Rosenfeld's interpretation, EPR’s argument rests on counterfactual assignments of outcomes to incompatible measurements, %which is precisely the structure on which , are 
and these are illegitimate.
%because they break the necessary link between the preparation, the observable, and the meaning of the quantum state.

    \subsubsection{Bell} \label{sec:Bell} 
    
\begin{figure}[H]
    \centering
    \includegraphics[width=0.5\textwidth]{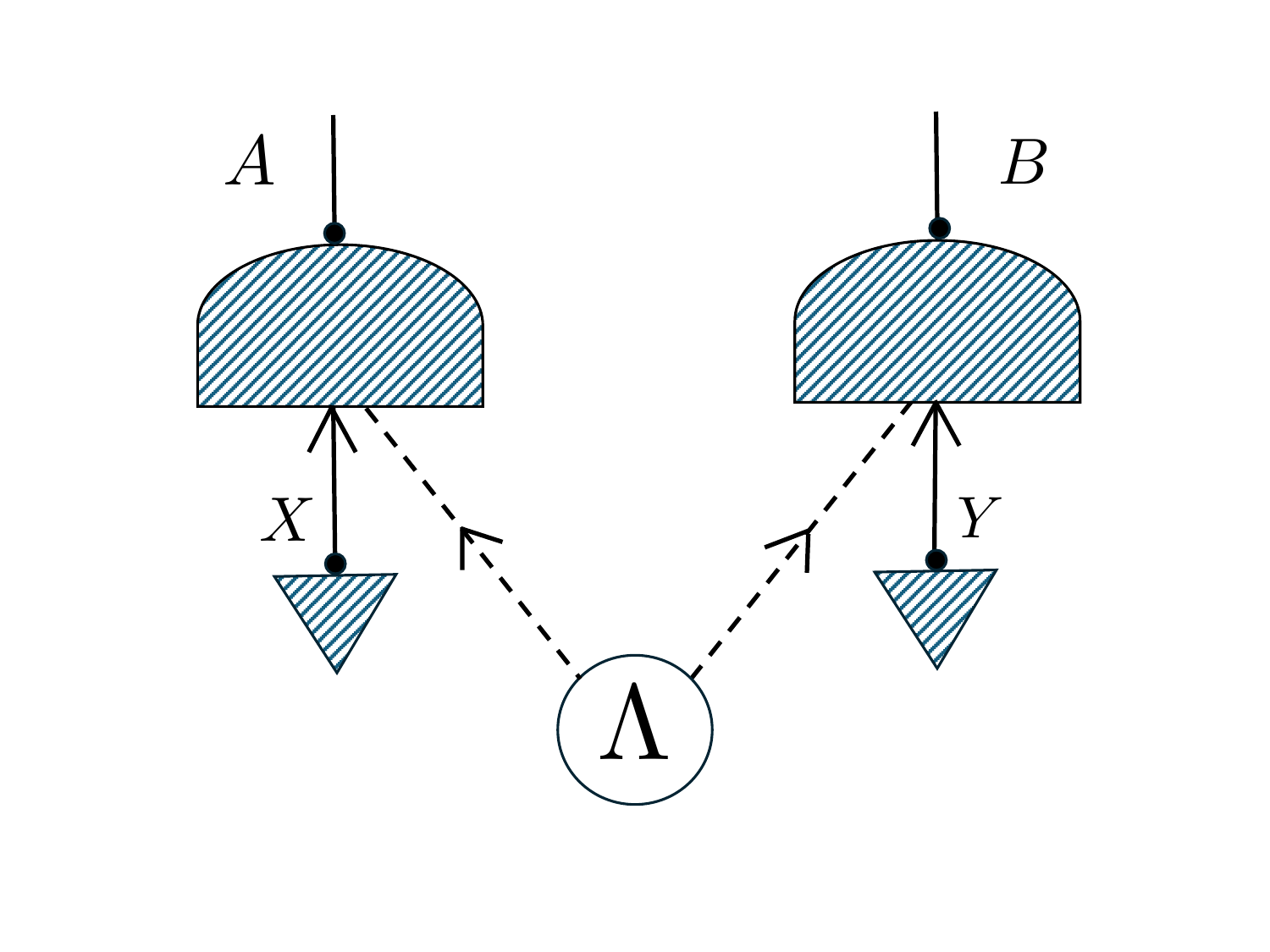}
    \caption{L-DAG representation of the causal structure of a Bell scenario that is inconsistent with QM. Details of the graphical notation used here are provided in Fig.~\ref{Terms}.
    \label{Bell DAG}}
\end{figure}

    John Bell's famous 1964 paper~\cite{PhysicsPhysiqueFizika.1.195} was inspired by the EPR paradox (as it is sometimes called), but he shifted the focus from quantum states to {\em operational} quantities: the measurement settings and the observed outcome statistics.  Experimental data ultimately consists of conditional probabilities $ \wp (a,b|x,y)$, of the outcomes given the settings. Since EPR had concluded that quantum theory was incomplete, Bell asked whether a local hidden variable theory could reproduce these empirically observable correlations. In modern approaches~\cite{wiseman2016causarum,PhysRevX.8.021018}, such theories have been expressed in terms of classical causal models. For a Bell experiment that is performed by two observers, Alice and Bob, a hidden variable model can be represented by a DAG as illustrated in Fig.~\ref{Bell DAG} (using graphical conventions we will introduce in Sec.~\ref{Sec:HSM}). This has nodes for the measurement settings $\{X,Y\}$ and the measurement outcomes $\{A,B\}$ and a common cause $\Lambda$ \cite{PhysRevX.8.021018}. $X$ and $Y$ do not have any relevant causes, making them the exogenous variables in the DAG. This encodes the assumption that the measurement choices by Alice and Bob are ``free choices''. 
    
%Bell then used Einstein's proposed assumptions (realism, locality and freedom) 
%\how{this was awkward, since you just introduced a causal model description, but now you are going back to other language.} 
Bell's theorem is, in this language, that there are quantum correlations, from measurements on an entangled state, that are 
incompatible with this classical causal model~\cite{wiseman2016causarum,PhysRevX.8.021018,PRXQuantum.2.040323,brukner2014quantum}.
%and demonstrated that together with the principles of quantum mechanics, they lead to a contradiction and therefore one of the three assumptions must be given up. 
Just as for EPR's argument, the question was raised as to whether Bell was using an illegitimate form of counterfactual reasoning~\cite{skyrms1982counterfactual,PhysRevD.3.1303,lambare2021note,Zukowski2014}. 
Specifically, it has often been claimed that he used an extra, implicit,  assumption, of counterfactual definiteness, first so called  in~\cite{skyrms1982counterfactual}.
%Following this, there were subsequent debates whether Bell assumed counterfactual definiteness in his derivation of the inequalities.
For example, Richard Gill~\cite{Gill_2014} argues that ``Instead of assuming quantum mechanics and deriving counterfactual definiteness, Bell turned the EPR argument on its head. He assumes three principles which Einstein would have endorsed anyway, and uses them to get a contradiction with quantum mechanics; and the first is counterfactual definiteness. We must first agree that though, say, only $A$ and $B$ are actually measured in one particular run, still, in a mathematical sense, $A'$ and $B'$ also exist (or at least may be constructed) alongside of the other two.'' In Gill's interpretation, counterfactual definiteness is equivalent to ``realism'', while the other two assumptions are locality (encapsulated by the DAG) and freedom of choice, 
%; and moreover, they may be thought to be located in space and time just where one would imagine. Only after that does it make sense to discuss locality: the assumption that which variable is being observed at Alice’s location does not influence the values taken by the other two at Bob’s location.'' Now, a local hidden-variable (LHV) model that Bell used corresponds to imposing two main conditions on this graph: (i) {\em locality}, that is, no direct causal influence from $X$ to $B$ or from $Y$ to $A$, and (ii) {\em realism}, the existence of a hidden variable (local) which can be deterministic or stochastic. A third assumption is a form of {\em measurement independence} or “freedom of choice”, 
namely that 
$\Lambda$ is statistically independent of the settings $X$ and $Y$. This assumption is interpreted by Gill as the standard causal requirement from randomized experiments: ``the statistical independence between the randomization in a randomized experiment and the set of outcomes of each experimental unit under all possible treatments.'' Under this interpretation, Gill concludes that the assumption most naturally relinquished is counterfactual definiteness, that is, the idea that all potential outcomes have definite values, whether or not they are actually measured. 
%emphasizing that this is a claim about a model of reality containing definite counterfactual outcomes, rather than about reality itself.

\subsubsection{Extension to Bell by Wang and Zhang} 

    % Zhang and Wang~\cite{wang2025quantumexperimentjointexogeneity}, in their work, they start from the (quantum) instrumental scenario and show that the classical formulation of counterfactuals, using the joint exogeneity assumption, can lead to incorrect estimations of the average causal effect (ACE) in the quantum regime. 

Recently, Wang and Zhang~\cite{wang2025quantumexperimentjointexogeneity} have proposed a novel extension of the discussion about quantum counterfactuals from a Bell- or EPR-scenario to the so-called (quantum) instrumental scenario~\cite{chaves2018quantum}. They show that the classical formulation of counterfactuals, in terms of a joint exogeneity assumption, can lead to  contradiction with experimental results in the instrumental scenario. Specifically, they say such experiments would rule out the conjunction of the following two assumptions:
\begin{itemize}
    \item ``potential outcomes exist'' (POE), and 
    \item ``realism'', which they use to mean, more or less, freedom-of-choice.
\end{itemize}

In the instrumental scenario, there is a binary setting $Z$, leading to a binary outcome $X$, which itself serves as a setting for a second measurement yielding a binary outcome $Y$. In the quantum version, the two measurements are perfomed on different parts of an entangled state~\cite{chaves2018quantum}.
In this context, Ref.~\cite{wang2025quantumexperimentjointexogeneity} take POE to mean the existence of latent variables $Y(x,z)$ for $x, z \in \{0, 1\}$ satisfying these two conditions: 
%only two minimal conditions,
\begin{enumerate}[label=(\alph*)]
\item %consistency: 
The observed outcome $Y$ equals the latent variable appropriate to the actual settings:
    $
        Y = X Z Y(1, 1) + X (1-Z) Y(1, 0) + (1 - X) Z Y(0, 1) + (1 - X) (1 - Z) Y(0, 0).
    $
\item %interventional interpretation: 
For every fixed $x,z \in \{0, 1\}$, there is a joint distribution $Q_{xz}$ of random variables $Y(x, z), X$, and $Z$ that has the following property:
% distribution is exactly the outcome distribution produced by the ideal
% intervention $\mathrm{do}(X = x', Z = z')$, as encoded by $Q_{xz}$
\begin{equation}
    Q_{xz}(Y = y \mid X = x', Z = z') = \mathbb{P}(Y(x,z) = y \mid X = x', Z = z'),
\end{equation}
when the probability distribution $\mathbb{P}$ is obtained from the interventional experiment $\mathrm{do}(X = x', Z = z')$.
% Hence there is no joint
% distribution or independence across different $(x,z)$ is assumed.
\end{enumerate}
%These two properties are the minimal requirements for latent variables to qualify as potential outcomes. 

A natural question concerns the ontological status of the latent variables $Y(x,z)$. 
Two broad interpretations may be considered, both of which Wang and Zhang consider compatible with the above constraints~\cite{Wang2025email}. First, they may be regarded as elements of physical reality, a view commonly associated with the potential-outcomes framework developed in Ref.~\cite{Rubin01032005}. Second, they may be interpreted as purely methodological constructs to model the causal reasoning process, which is a more Humean perspective~\cite{hume1739treatise,hume1748enquiry}.  But the additional assumption required to obtain a contradiction is easier to understand under the first interpretation. It is the statistical independence of pre-existing elements of physical reality from human free choices. 
 This is the assumption Wang
and Zhang call ``realism''; in the current case it is that the joint distribution for the latent variables is independent of the actual intervention $Z$:
\begin{equation}
P(Y(1,1),Y(1,0),Y(0,1),Y(0,0)\mid Z)
=
P(Y(1,1),Y(1,0),Y(0,1),Y(0,0)).
\end{equation}

% Consequently, joint exogeneity arises only when either of the two conditions are considered: (i) potential outcomes are interpreted as elements of the natural world, and (ii) realism is assumed. 

An explanatory test of the strength of these assumptions is provided by Bohmian mechanics~\cite{Bohm1952a,Bohm1952b}, taken here as a representative deterministic hidden-variable theory. In the Bell scenario, such a theory does exhibit a joint distribution over all potential outcomes: writing
$\lambda$ for the Bohmian hidden variable at some initial time, the outcomes on each wing are deterministic functions $A(x,y) = f(x,y,\lambda)$, $B(x,y) = g(x,y,\lambda)$ of the joint settings, and a joint distribution over all counterfactual outcomes follows immediately by marginalising over $P(\lambda)$.
 
In the instrumental scenario, by contrast, the second setting $X$ is itself determined by the hidden variable and $Z$, the actual free choice or exogenous variable. That is, $X = f(\lambda, Z)$, with $Y = g(\lambda, X, Z)$. An assignment of the form $Y(x,z) = g(\lambda, x, z)$ then cannot be coherently defined: for a given $\lambda$, the counterfactual value $X = x$ may simply not be realized, and there is no principled way to fill in the corresponding argument of $g$. Bohmian mechanics therefore fails to satisfy the POE assumption of Wang and Zhang, and is consequently not a counterexample to their no-go result. This observation, however, cuts both ways: even a theory that fully determines what would have occurred under every intervention---as Bohmian mechanics does---fails to satisfy POE as defined in Ref.~\cite{wang2025quantumexperimentjointexogeneity}, showing how strong an assumption it is. 
%\red Moreover the ``realism'' (freedom) assumption is not even decidable because the random variables $Y(x,z)$ do not exist. 
\blk

\subsection{The semantics for quantum counterfactuals of Suresh \ea}\label{sec:Suresh}

Despite the extensive discussions and debates on counterfactuals in quantum theory, no one attempted to create a mathematical theory of quantum counterfactuals, comparable to Lewis's or Pearl's theory of classical counterfactuals, until very recently~\cite{suresh2023semantics,banerjee2025counterfactualquantummeasurements}. The first of these formalizations, by 
Suresh \ea~\cite{suresh2023semantics}, extends Pearl's semantics for counterfactuals from PSMs to Quantum Structural causal Models (QSMs). 
A QSM, $M_Q$, is a way of describing orthodox quantum theory in causal terms. Similar to Pearl's PSM, a QSM is still associated with a DAG, $G$. But now its vertices represent quantum nodes $\boldsymbol{A}$, rather than random variables, and its edges represent quantum channels transmitting quantum information between operations, rather than statistical dependencies or causal relations. 

Physically, a quantum node $A_j$ can be thought of as a general quantum instrument~\cite{Ozawa1984,DaviesLewis1970,HeinosaariZiman2012}, with a setting $z_j$ yielding a classical outcome $a_j$. The instrument is performed on a system ``living in'' an input Hilbert space, which is transformed into the system ``living in'' an output Hilbert space, which is conditioned on $a_j$, the measurement result. Note that the size of the input and output Hilbert spaces need not be the same, and the size of the setting and outcomes spaces can be trivial. Thus, for example, one can have a quantum node which simply produces a fixed quantum state, where the setting space, the measurement space and the input Hilbert space are trivial (dimension one).
In addition to its quantum inputs and a setting input, a quantum node $A_j$ also has, in general, input from exogenous classical variables $\boss{U}$, which 
play a role similar to the exogenous variables in Pearl's PSM, and have a factorized probability distribution\footnote{Ref.~\cite{suresh2023semantics} introduced a new symbol, $\boss{\Lambda}$ for these variables, and we followed suit for our own model in Ref.~\cite{banerjee2025counterfactualquantummeasurements}. In the present paper, we have chosen to stick with Pearl's notation, $\boss{U}$, because, in the context of quantum causal models, the symbol $\Lambda$ suggests a source of local hidden variables. That is, as in the classical DAG of Fig.~\ref{Bell DAG}, it represents a shared classical cause in a foil theory which Bell correlations rule out. That is {\em not} the role that these exogenous variables play in the QSM of Ref.~\cite{banerjee2025counterfactualquantummeasurements} (as will now be made apparent), or in our HSM of Sec.~\ref{Sec:HSM}. }. 

The above is not a complete description of a QSM as defined in Ref.~\cite{suresh2023semantics}. But it is sufficient for us to explain how Suresh \ea\ treat counterfactuals. First, they introduced the notion of {\em counterfactual probability} or the {\em expected value} of a counterfactual query. This is distinct from the {\em probability of a counterfactual} as in Pearl's original formalism. Pearl, like Lewis, implicitly assumes `counterfactual definiteness', but Suresh \ea\ reject this for QSMs because OQT is an inherently probabilistic theory.
Nevertheless, their evaluation of counterfactual probabilities proceeds in the same spirit as Pearl's evaluation of the probability of a counterfactual.
In their framework a standard quantum counterfactual query can be stated as follows:
{\em Given the evidence that a set of instruments $\bm{z}=(z_1,\ldots,z_n) $ has been implemented and a set of outcomes $\bm{a}=(a_1,\ldots,a_n) $ has been obtained, what is the expected probability that outcomes $\bm{c}'$ would have been obtained at a subset of nodes $\bm{C}$ had instruments $\bm{z}'=(z'_1,\ldots,z'_n) $ been implemented and outcomes $\mathbf{b}$ obtained at a set of nodes
$\bm{B}$ (disjoint from $\bm{C}$)?}

\begin{figure}[H]
    \centering
    \includegraphics[width=0.35\textwidth]{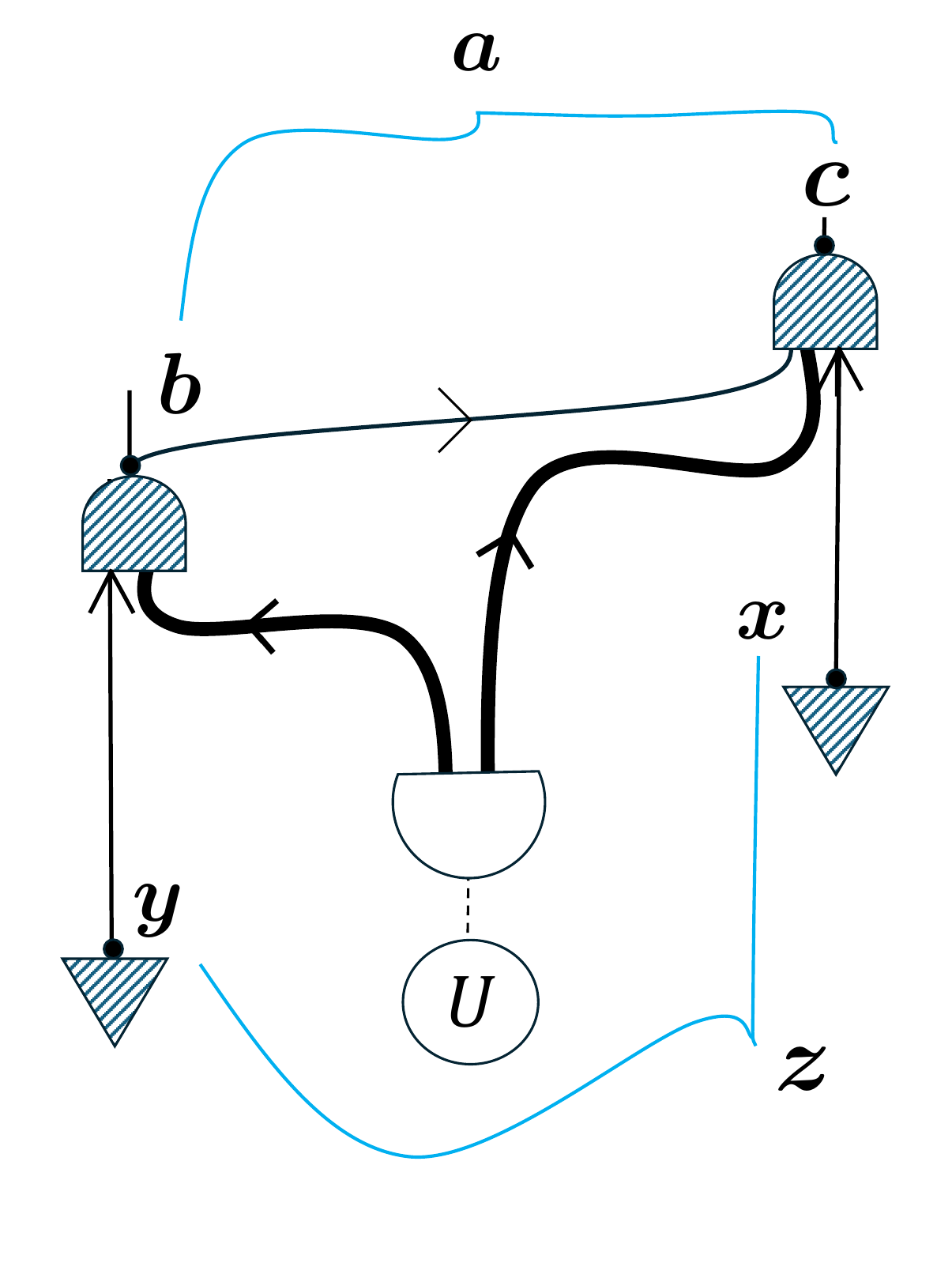}
    \caption{An example of L-DAG representation of the causal structure for the counterfactuals in quantum causal models. Schematic representation of a bipartite measurement scenario. The quantum source generates quantum outputs which is measured by the two parties and each performs an instrument chosen from the set $z=\{x,y\}$, producing the corresponding outcomes constituting the joint outcome vector $a=\{b,c\}$. The unknown exogenous variables are denoted by the node $U$. The classical output of one party can be used to adapt the measurement process of the second party depicted by the line from $\boldsymbol{b}$. (This is not mentioned in the original paper but we add it here as an obvious extension.) Details of the graphical notation used here are provided in Fig.~\ref{Terms}.} \label{Q-DAG}
\end{figure}

The evaluation of this query then proceeds according to Pearl's three-step procedure: 
\begin{itemize}
    \item \textbf{{\em Abduction}}: Updating the information about the values $u_i$ of the exogenous variables $U_i$ (see Fig.~\ref{Q-DAG} for example), given the outcomes $a_i$ that have been observed by performing the instruments $z_i$ at nodes $A_i$,
    \begin{equation}
        P_{\mathbf{z}}(\boldsymbol{u}|\boldsymbol{a})=\frac{P_{\mathbf{z}}(\boldsymbol{a}|\boldsymbol{u})P(\boldsymbol{u})}{P_{\mathbf{z}}(\boldsymbol{a})}
    \end{equation}
    \item \textbf{{\em Action}}: Modifying the instruments at the endogenous nodes (nodes that have incoming causal arrows) to what is required by the antecedent of the counterfactual query, $z_i'$.
    \item \textbf{{\em Prediction}}: Calculating the expected value of the counterfactual probability,
    \begin{equation}
        P_{\boldsymbol{b}'|\mathbf{z}'}^{\boldsymbol{a}|\mathbf{z}}(\boldsymbol{c}')=\sum_{\boldsymbol{u}\in\boldsymbol{U}} P_{\mathbf{z}}(\boldsymbol{u}|\boldsymbol{a})P_{\mathbf{z}'}^{\boldsymbol{u}}(\boldsymbol{c}'|\boldsymbol{b}').
    \end{equation}
\end{itemize}

\section{Our counterfactual quantum measurement calculus \label{sec:CalculusandDag}}

%\hmw{See how every section needs a short intro to centre the reader, to let them know where they are in the story.} 
In this section,  we introduce the counterfactual calculus of our earlier work~\cite{banerjee2025counterfactualquantummeasurements}, but with more motivation, formality, and generality. We then review the two example scenarios of Ref.~\cite{banerjee2025counterfactualquantummeasurements}. 
We begin with a return to the analysis of counterfactuals developed by David Lewis, revisiting both the challenges that this formalism faces in quantum theory and the manner in which we address them.

\subsection{Returning to Lewis \label{sec:ReturnToLewis}}
 The first challenge concerns the classical ontology presupposed by Lewis’s analysis, meaning that there exists a definite matter of fact about events occurring in space-time. Since quantum theory is fundamentally indeterministic, our application of Lewis’s analysis is necessarily restricted to the classical elements of the theory: measurement settings, measurement outcomes, and uncontrolled classical variables (noise). 

The second challenge concerns the truth-valued nature of Lewis’s counterfactuals where he assigns truth values to counterfactual propositions. In the quantum setting, however, we generalize this notion to counterfactual probabilities, which we refer to as \emph{supposabilities}  (\Su).  We use this similar-sounding neologism to avoid the potential misunderstanding that 
%these probabilities and expectations 
a counterfactual probability would be straightforwardly empirically observable. We note here that this idea (without the terminology) was inspired by the approach of Suresh \ea~\cite{suresh2023semantics}. 

In addition to these generalizations, our framework makes explicit the role of fixtures, mainly where Lewis’s account faces criticism. The fixtures form a broader class of possible assignments of classical variables, and the available evidence is used to update our knowledge of these assignments. The resulting distribution then determines how the counterfactual probabilities are weighted when calculating the supposability of the consequent.

With these generalisations in place, we now show how Lewis's four desiderata, introduced in Sec.~\ref{sec:2}, are realised in orthodox quantum theory.

%\ing{Rewrite the points similar to that in the thesis and paper one and then follow with that and then talk about generalization and qualitatively discuss about the supposabilities similar to suresh ea and say about updating the knwoledge abput the fixtures and the fixtures belong to broader calss of sets recalculate the counterfactual updates weighted the probabilities of the fixtures}
\begin{enumerate}[label=(\roman*)]
\item We choose the locus of change to be a measurement setting. This does not violate any law of quantum theory, since the theory permits free choices of measurement settings.

\item We chooose the fixtures to ensure maximal spatio-temporal match between the actual and counterfactual worlds. This requires us to keep fixed all classical variables that are not causally influenced by the counterfactually altered settings. This includes measurement settings and outcomes that are unknown to the counterfactual reasoner, or, more formally, that might not even be specified by the theory or included as ``evidence''~\cite{pearl2018book} in the counterfactual.

\item Not only does our framework not involve any large, widespread, or diverse violations of physical law (first desideratum), it likewise avoids any small or localized violations of law.

\item Moreover, we do not appeal to approximate similarity with respect to particular matters of fact.

\end{enumerate} 
This will become clear in the formalization of~\cref{The calculus}. But first we introduce Hybrid Structural Causal models (HSMs) which provide a unified framework incorporating both quantum and classical nodes. This framework is essential for extending the notion of quantum counterfactuals to general causal structures, as we do in this paper. 

\subsection{Introducing Hybrid Structural causal Models}\label{Sec:HSM}

Consider an experiment described by the set of all relevant events,  $\boldsymbol{\Omega} = \{(\Omega_\alpha,\alpha)\}_\alpha$. 
By an event we mean a classical variable (observable or hidden) at a particular points in space-time. Here $\alpha$ is a unique index for an event, which encodes the space-time location, and $\Omega_\alpha$ is the classical random variable describing it. (Not all variables need be random, but we allow it.) In Ref.~\cite{banerjee2025counterfactualquantummeasurements} we were specifically concerned with space-time locations, but in this paper we replace that with a causal structure described by a DAG, and so we will generally use $\Omega_\alpha$ as a shorthand for the pair $(\Omega_\alpha,\alpha)$ below. 

The set of events $\boldsymbol{\Omega}$ comprise four disjoint subsets: the 
measurement settings $\boldsymbol{Z}$, the matching outcomes $\boldsymbol{O}$ (so $|\boldsymbol{Z}|=|\boldsymbol{O}|$), the outputs $\boss{R}$ arising from classical processing of the outcomes, 
and the exogenous
``hidden'' variables $\boldsymbol{U}$, which can be thought of as classical noise. The variables $\boss{R}$ in classical registers are an addition to the set of events in Ref.~\cite{banerjee2025counterfactualquantummeasurements}; we introduce them because, as we will discuss in \cref{sec:4(e)}, we wish to consider some such variables as a consequent in a counterfactual. In terms of a causal model described by a DAG $G$, we can think of each of these as the variable labelling a classical edge as it exits a node (\ie, from the top). By this convention, we would have to add to $G$ a trivial node for each hidden variable $u_j$, whose only connection to the graph is the edge labelled $u_j$ that enters (from below) node $j$. However, unlike for Suresh \ea, the exogenous variables $\bm{U}$ do not play an important role in our approach, and in all of our examples $\bm{U}=\nullset$. We include them only to make our model as general as theirs, and to make this point of difference explicit in the comparison.

Naturally, for a quantum causal model framework, we also want edges on our DAG $G$ which can carry quantum information. We use labels $\boss{Q}$ for these edges, and again, for convenience, we can associate a given label $Q_j$ with the point where the edge  exits node $j$. One should think of $Q_j$ as representing a Hilbert space of dimension $d_j$, rather than a specific state, 
just as $O_j$, for example, labels a random variable with dim($O_j$) possible values, rather than a specific value, $o_j$. 
Given all of this, we define a labeled DAG (L-DAG) as a triple 
$\langle \bm{\Omega}, \bm{Q}, G \rangle$.
We also denote the sets of input and output edges for node $j$ as 
In$_G(j)$ and Out$_G(j)$ respectively, where the $G$ subscript indicates that these sets are defined by $G$. 

We now introduce what we call a Hybrid Structural causal Model (HSM). We use this idea, rather than the QSM of Ref.~\cite{suresh2023semantics}, to indicate the possibility (and, indeed, importance) of the communication and processing of classical information as well as quantum information, and also because we define things differently from in Ref.~\cite{suresh2023semantics}. A HSM is specified as the tuple 
$\langle \text{L-DAG}, \bm{\mathcal{S}}, \bm{\mathcal{M}}, \mathbf{p} \rangle$. Here $\bm{\mathcal{S}}$ is a function describing how the settings $\bm{Z}$ depend on other events in the set $\bm{\Omega}$. Specifically $Z_j =
\mathcal{S}_j(\text{In}_G([Z_j]))$, where $[Z_j]$ is the index for the node which outputs $Z_j$. This allows, for example, for adaptive measurements, where the setting depends on prior measurement results. Next, each $\mathcal{M}_j$ is a completely 
positive trace-preserving (CPTP) map from $\text{In}_{G}(j)$ to 
$\text{Out}_{G}(j)$. It describes what happens at node $j$, and satisfies
\begin{equation}
    (\mathcal{D}_{\alpha}\otimes \mathcal{I}_{{\rm Out}_{G}(j)\setminus\Omega_\alpha})\,\mathcal{M}_j = \mathcal{M}_j 
    \qquad \forall\, \alpha : \Omega_\alpha \in \text{Out}_{G}(j),
\end{equation}
where $\mathcal{D}_\alpha$ is the map that completely dephases the edge  $\alpha$ in the logical basis: $\mathcal{D}_\alpha \bullet = \sum_{i=1}^{\text{dim}(\Omega_\alpha)}\ket{i}\bra{i}\bullet\ket{i}\bra{i}$. 
The above construction allows us to use a single object (a quantum CPTP map) to describe both quantum and classical information processing at a node, ensuring that a classical output $\Omega_\alpha$ is classical in an operational sense~\cite{Watrous2018,Wilde2017}. Finally, the lack of knowledge about 
the exogenous classical noise variables is described by a set $\mathbf{p}$ of individual probability distributions as usual. 
%\ktl{Ok, I am confused by this part. Is the ${\cal D}_\alpha$ only acting on the classical register? Well I guess the CPTP map is also only acting on some ancilla in the node more generally. Sorry, I probably need to sit down and work through the L-DAG stuff in more detail to parse everything.}\hmw{Correct. See clarifying additions above (new red) in Eq.(7).}

\begin{figure}
    \centering
    \includegraphics[width=0.50\textwidth]{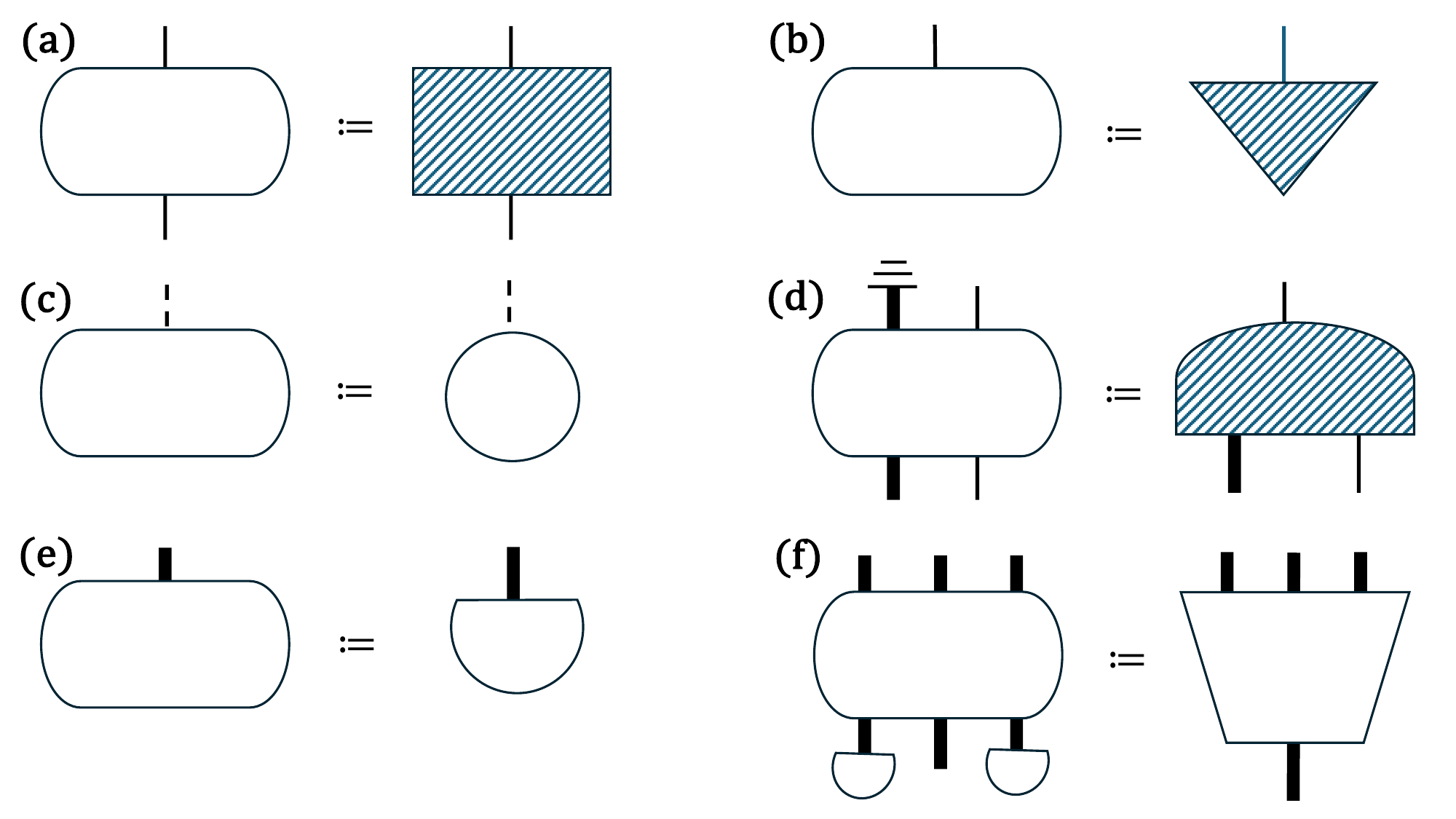}
    \caption{Notation for variables, causal influences, and information flow. (a) Classical register (hatched rectangle). (b) Classical measurement setting (hatched inverted triangle). (c) Classical hidden (uncontrolled exogenous) variable (rounded node). (d) Measurement device (hatched rectangle with a rounded top). (e) Quantum source (rounded cup). (f) General quantum-only processing block (trapezoid). Rounded rectangles denote generic nodes. The thick edges indicate
quantum information flow and thin edges indicate classical signals with the direction of flow being bottom to top.} \label{Terms}

\end{figure}

To make it easier to interpret the L-DAG in our HSM, we have introduced some graphical notation, used
throughout this work, \blk summarized in Fig.~\ref{Terms}. 
This distinguishes quantum, classical, and hybrid
information-processing components.

In all the diagrams, 
\emph{thick} black
edges denote quantum information flow, reflecting that a quantum system
carries strictly more information than any classical description, 
\emph{thin solid} edges represent classical information flow such as measurement outcomes, or control signals, and \emph{thin dashed} edges represent classical noise (uncontrolled exogenous variables). These edges connect nodes, and the direction of flow of information is always into the bottom of a node and out of its top. For the edges themselves, we try to keep their orientation to be generally upward in the direction of information flow, like in Fig.~\ref{Q-DAG}.
%[[A quantum causal DAG is an acylic graphical representation that encodes the causal connections between the random variables\ing{(quantum nodes)}, this is more general than the space time specifications and is not restricted to relativistic causation.]]\how{this may be useless now.}

Moving now to the nodes, rounded
rectangles denote generic nodes, without committing to
whether they are quantum or classical or hybrid; this distinction is conveyed
solely by the thickness of the edges entering and exiting. However, 
for ease of reading the graphs, we adopt the convention of hatching nodes which only have classical outputs (thin solid edges). Just as in the work of Suresh \ea, every node $j$ potentially also has a classical noise input $u_j$.
We also choose specific shapes for specific cases, as illustrated: (a) a classical register or processing element is represented
by a sharp-edged, hatched box with only classical edges incoming and outgoing; (b) a setting for a classical measurement or operation (a controlled exogenous variable), drawn as a hatched inverted triangle with only an outgoing classical signal; (c) a classical hidden (uncontrolled exogenous) variable: a rounded node with only outgoing dotted edges; (d) a measurement device (a hybrid
quantum--classical device, which accepts both classical and quantum inputs and outputs only a classical result), drawn as a rectangle with a rounded top; (e) a quantum source (an exogenous quantum variable), drawn as a rounded cup with a single thick quantum output; and (f) a general quantum-only processing block that aggregates quantum sources and a single quantum input to produce multiple quantum outputs, drawn as a trapezoid. 
%to indicate that it implements a general update map rather than representing a specific physical variable.
Together, these conventions provide a uniform
graphical language for describing causal structure and information flow in the cases following.

\subsection{Counterfactual measurement calculus \label{The calculus}}
%Here we review the counterfactual measurement calculus introduced in our previous work  and make the description more general by grounding the formalism with 

In our previous work~\cite{banerjee2025counterfactualquantummeasurements}, we assumed that there is some causal structure $\mathfrak{C}$ on the set $\boldsymbol{\Omega}$ such that, given a subset $\boss{S} \subset \boss{\Omega}$, there is a unique set of inclusive causal descendants $D_{\mathfrak{C}}(\boss{S}) \subseteq \boss{\Omega}$. In all of the examples we presented in that paper, we took $\mathfrak{C}$ to be determined by the space-time structure of the experiment, with $D_{\mathfrak{C}}(\boss{S})$ being the events in the union of the future light cones of the events in $\boss{S}$. But we noted that, more generally, one could take $\mathfrak{C}$ to be specified by a DAG, so that not all events in the future light cone need be in the causal future. That is the approach we will use in this paper, with the DAG $\DAG$ for the HSM introduced in Sec.~\ref{Sec:HSM}. %\how{Do we need to say anything more about how causal descendants are defined by a DAG, or is it obvious?}\ktl{Hmm, it is fairly obvious. Perusing through Pearl, even there, there is not an explicit definition for descendants.}

We assume that the experiment described by the DAG has associated probability distributions 
$\wp(\boldsymbol{z},\boldsymbol{o},\boldsymbol{r}|\boldsymbol{u},\boss{\mathcal{S}})$ and $\mathbf{p} = (\ldots,\,p_l(u_l), \ldots\,)$. %\hmw{I changed $\lambda$ to $u$ -- right?}\ing{I think so} 
These could be determined by the HSM introduced in Sec.~\ref{Sec:HSM}, but we also leave open that these probabilities could be determined purely empirically. We emphasize that, even if we don't use the HSM, we still use the associated causal framework, the DAG $\DAG$.   

Let the actual world be described by a measurement strategy $\boldsymbol{\mathcal{S}}$ specifying $Z_j =
\mathcal{S}_j(\text{In}_G([Z_j])$, as explained in Sec.~\ref{Sec:HSM}.  The evidence is defined by the value $\boss{e}$ for $\boldsymbol{E}$, which is the subset $\boldsymbol{E} \subseteq \boldsymbol{Z} \cup \boldsymbol{O}$. Following the convention of Suresh \ea\ (See Sec.~\ref{sec:Suresh}), 
we use primes for variables in the counterfactual world. 
Starting with the antecedent, this is defined by the value $\boss{a}'$ for $\boldsymbol{A}'$. The actual world 
counterpart is the subset of measurement settings $\boldsymbol{A} \subseteq \boldsymbol{Z}$, which we call the locus of change because it specifies the variables whose values are altered in the counterfactual scenario. \blk
From this scenario we can define the counterfactual measurement strategy $\boldsymbol{\mathcal{S}}'=\overline{\boldsymbol{\mathcal{S}}}\,\union\,\boldsymbol{a}'$, where $\overline{\boldsymbol{\mathcal{S}}}
\subset \boldsymbol{\mathcal{S}}$ is the subset of the actual strategy which is disjoint to $\boldsymbol{a}'$ so that $|\overline{\boldsymbol{\mathcal{S}}}| = |\boldsymbol{Z} \setminus
\boldsymbol{A}|$. Thus, for every $Z_j \in (\boldsymbol{Z} \setminus \boldsymbol{A})$ (\ie, not in the antecedent), the measurement setting is given by $Z_j = \overline{\mathcal{S}}_j(\text{In}_G([Z_j])$. 

Next, we define $\boldsymbol{D} = D_G(\boss{A}) \subset\boldsymbol{\Omega}$, the inclusive classical 
descendants of $\boss{A}$, relative to the causal structure described by the DAG $G$.  Similarly we define the counterfactual descendants $\boldsymbol{D}' = D_G(\boldsymbol{A}')$. This lets us define the fixtures, the set of events such that $\boldsymbol{F}\equiv\boldsymbol{F}'$, by $\boldsymbol{F}\equiv\boldsymbol{\Omega}\backslash\boldsymbol{D}$. Finally, there is some consequent $\boldsymbol{C}'\subseteq\boldsymbol{D}'$, 
for which we can consider any possible value, $\boldsymbol{C}'=\boldsymbol{c}'$.

The counterfactual question is therefore: given as background the probability distributions $\wp$ and $\mathbf{p}$, the strategy $\boldsymbol{\mathcal{S}}$, and the DAG
$G$, what is the supposability that $\boldsymbol{c}'$ would have pertained had $\boldsymbol{a}'$
pertained, given that $\boldsymbol{e}$ pertained? Following the argument in Sec.~\ref{sec:ReturnToLewis}, 
%\hmw{It really seems like there was supposed to be more qualitiative discussion in 3.1 (in particular about the role of unknown fixtures) than there is. Do you have a suitable discussion in your thesis? Our first paper must have had such a discussion.}\ing{I have now included a sentence similar to the discussion in the first paper, discuss more with respect to the comments from the examiner.} 
this is given by, 
\begin{align}
& \mathrm{Su}(\boldsymbol{C}' = \boldsymbol{c}' \mid \boldsymbol{A}' = \boldsymbol{a}' ||\boldsymbol{E} = \boldsymbol{e}) \coloneqq \nn \\
 &
    \sum_{\boldsymbol{f}}
     \Pr(\boldsymbol{C}' = \boldsymbol{c} \mid \boldsymbol{A}' = \boldsymbol{a}', \boldsymbol{F} = \boldsymbol{f})\,
      \Pr(\boldsymbol{F} = \boldsymbol{f} \mid \boldsymbol{E} = \boldsymbol{e}). \label{Supposability}
\end{align}
Importantly, the definition of supposability in~\erf{Supposability} distinguishes between two conceptually distinct types of conditioning. The conventional conditional bar ($|$) refers entirely to the counterfactual world: it expresses the probability of the counterfactual consequent $\boldsymbol{C}'=\boldsymbol{c}'$, given that the antecedent $\boldsymbol{A}'=\boldsymbol{a}'$ is supposed to hold. The double bar ($||$), by contrast, denotes conditioning on the evidence from the actual world. In the second line of \erf{Supposability}, the second factor reflects how the posterior distribution over the unknown fixtures $\boldsymbol{F}$ is updated using the available evidence. Since the fixtures are the only variables shared between the actual and counterfactual worlds, they provide the bridge through which actual-world evidence informs counterfactual supposability, via the first factor. As the values of the fixtures are unknown in general, the final supposability is obtained by marginalizing over all possible fixture values\footnote{We note that one could omit this marginalization and still obtain a distribution, over $\boldsymbol{c}$ and $\boldsymbol{f}$, with all the mathematical properties of a probability distribution. In this paper we stick to defining the Su distribution only for the consequent, \ie, that which is potentially different in the counterfactual world. But we note that the state in \cref{Eq:V-cFS} could naturally be regarded as a counterfactual expectation, in this extended sense, over a functional of fixtures $\boldsymbol{f}$ alone, which is not in the consequent.}. 
%\how{More discussion definitely needed, but how much of that will have been in Sec.~\ref{sec:ReturnToLewis} I don't know. Make sure that ``Su'' is thoroughly explained.}\ktl{I agree we can be more discursive here. It is fairly fast here and easy to get lost in the notation - although that seems like it is going to happen anyway, at least for me.}

\subsection{Example: Bell--CHSH experiment}

Having introduced our general counterfactual calculus, we now 
consider, as in Ref.~\cite{banerjee2025counterfactualquantummeasurements}, the simple example of a Bell--CHSH experiment. There are two major differences from the previous presentation. First, as we have (in this paper) already introduced the symbol $\boldsymbol{A}$ for the locus of change variable, we avoid potential confusion with the 
symbol for variables associated with one of the conventional parties, `Alice', by replacing her by `Tara'. We also replace `Bob' by `Vlad'. Second, we explicitly introduce an L-DAG $G$ and a HSM to describe the experiment, rather than relying on the space-time locations of various events to define the descendants in the counterfactual. 

\begin{figure}[H]
    \centering
    \includegraphics[width=0.5\textwidth]{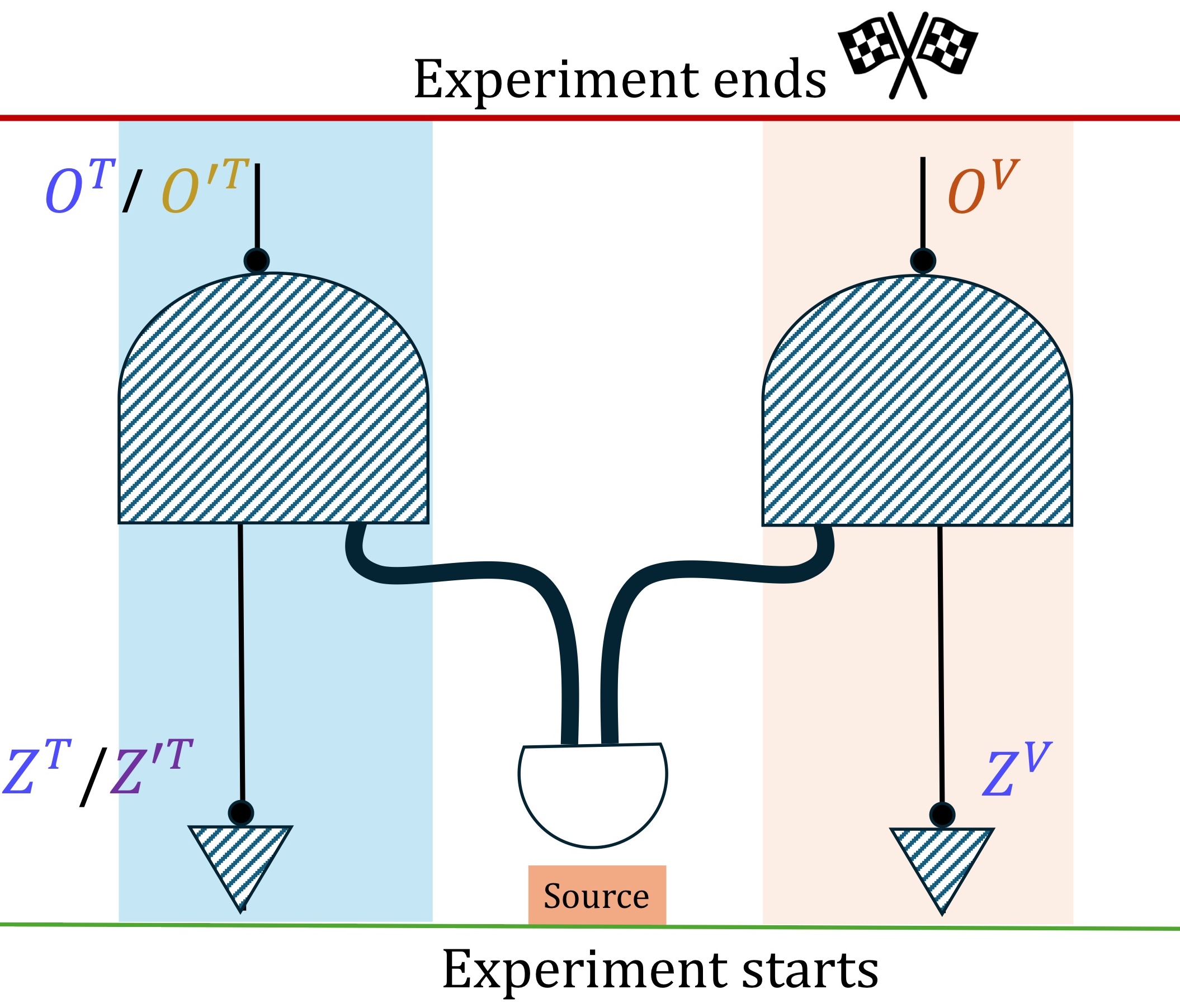}
    \caption{L-DAG representation of the causal structure underlying a CHSH scenario for counterfactual reasoning. On Tara's side the slash distinguishes the  actual/counterfactual scenario, with primed variables referring to the counterfactual scenario. The evidence variables are coded in blue, antecedent in purple, fixtures in brown and consequent in golden. % and differ conceptually from the outputs of exogenous variables. 
    Details of the graphical notation used here are provided in Fig.~\ref{Terms}. 
    \label{CHSH DAG}}
\end{figure}

Using these new observers, we re-represent the CHSH scenario in terms of the L-DAG illustrated in Fig.~\ref{CHSH DAG}. 
Counterfactual variables—representing outcomes or settings under alternative choices—are denoted using primes (e.g., $Z^{'\bullet}$, $O^{'\bullet}$) to distinguish them from the factual variables. This notation will be used consistently for the rest of the paper for setting and outcome variable.

The purpose of this representation is to emphasize how the DAG is used to identify the set $\boldsymbol{D} = D_G
(\boldsymbol{A})$, the complement of which denotes the fixtures $\boldsymbol{F}$. This identification is a key step in the general counterfactual calculus developed here. In the CHSH case, it is fairly straightforward to see that the only descendant of the antecedent, the setting of Tara $Z^T$, is just her own outcomes $T$.
In this scenario the most natural counterfactual question that would occur to Tara is as follows: ``Given the evidence $Z^T,O^T$ and $Z^V$ what is the supposability that, had I chosen $Z'^T$, I would have seen $O'^T$?'' We answered this question, in terms of supposability values, in our work~\cite{banerjee2025counterfactualquantummeasurements}, %.\blk 
%To summarize that we obtained in~\cite{banerjee2025counterfactualquantummeasurements} 
which we now translate into the notation of Tara and Vlad in place of Alice and Bob. The value of the variables we considered were: $Z^T=\updownarrow$, (\ie, measuring spin in the vertical direction), $O^T=\uparrow$ and $Z^V=$\rotatebox[origin=c]{135}{$\updownarrow$} for the actual scenario; and $Z'^T=\leftrightarrow$ and $O'^T=\rightarrow$ for the counterfactual scenario.  The counterfactual probability is evaluated by conditioning on the actual-world evidence and averaging over the unknown fixture variables $O^V$ shared between the actual and counterfactual worlds. Using the corresponding quantum probabilities yields a non-trivial supposability of $\frac34$. Considering instead $O'^T=\leftarrow$ naturally yields a supposability of $\frac14$. These differ from more naive approaches based either on unconditional recalculation or on simple collapse-based intuition, both of which instead predict $\frac12$. For the derivation of this result, we refer the reader to Ref.~\cite{banerjee2025counterfactualquantummeasurements}.

\subsection{Example: Continuous monitoring of an open quantum system \label{sec:System}} 
%\hmw{Motivation: continuous-in-time (or repeated) measurements on radiation from an OQS allows more interesting scenarios to be explored. Introduce Suspectation (explain as before). Reiterate example from previous paper. Just qualitative description of system and measurements. Give ST diagram, and a separate diagram showing DAG. Say what we calculated (suspectation) and why. No plot of it yet. Then compare this to CHSH: though more complicated, the party correspondence still holds.}

In an open quantum system that is continuously monitored through repeated measurements of its emitted radiation, a non-trivial calculation is needed to obtain informative statistical quantities using the counterfactual calculus. In continuous-monitoring scenarios, one is confronted with the cardinality problem. To compute the full supposability distribution of a time-dependent quantity such as a measurement record is usually challenging, and often infeasible, particularly for homodyne detection, where the resulting signal is nowhere differentiable. For this reason, we focus on a simpler and more tractable statistic: the {\em suspectation} ($\mathbb{S}$) of the time-dependent variable at each instant in time. As introduced in our previous work~\cite{banerjee2025counterfactualquantummeasurements}, the suspectation plays the role for expectation ($\mathbb{E}$) that supposability plays for probability.  That is, under a counterfactual proposition, the value one would have expected to observe at some particular time can be interpreted as what one suspect one would have observed at that time in that hypothetical scenario. 

\begin{figure}[H]
    \centering
    \includegraphics[width=0.6\textwidth]{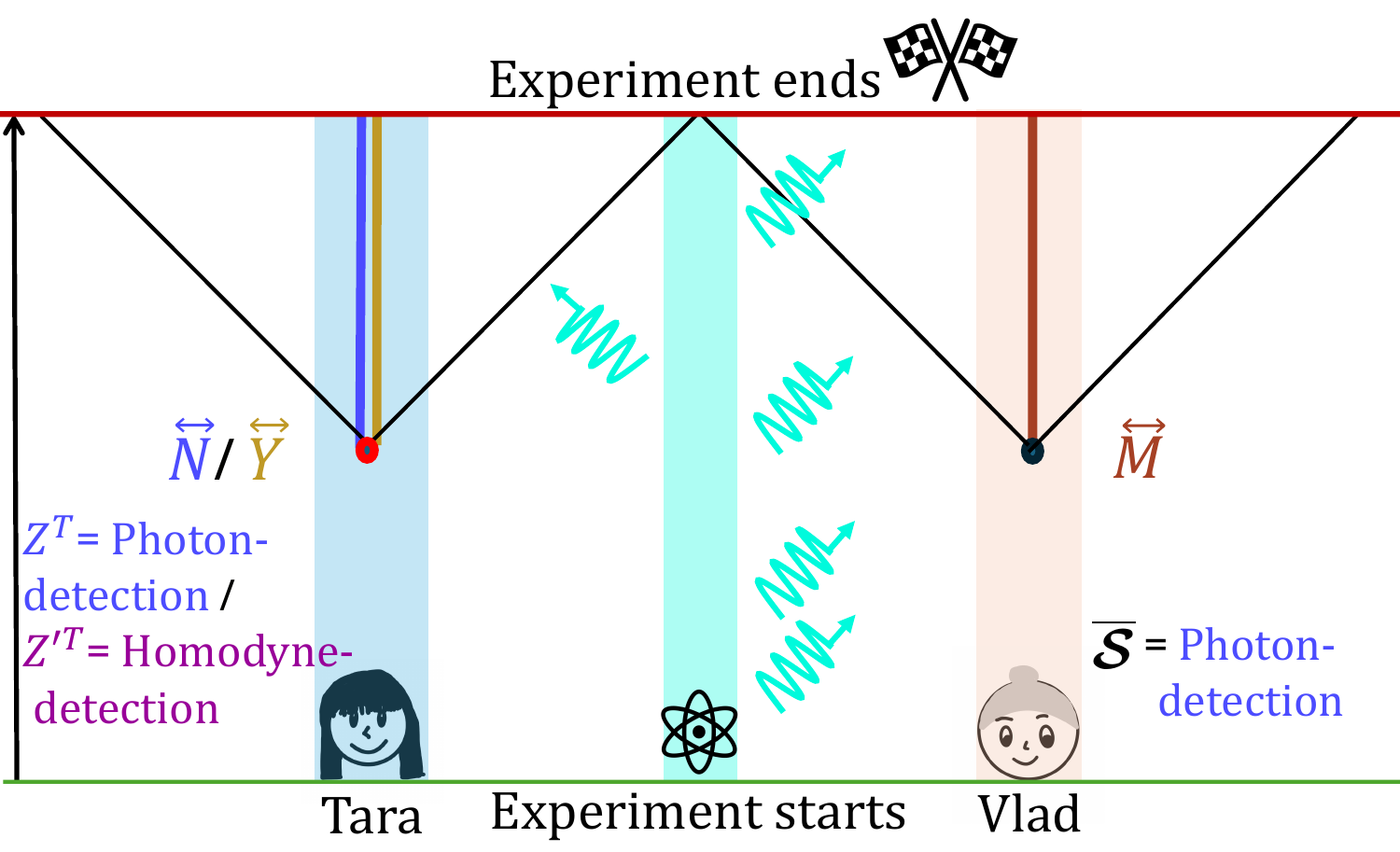}
    \caption{Space-time diagram 
        for the resonance fluorescence counterfactual. Tara's actual record of photo-detections $\both{O}^T=\overleftrightarrow{N}$ (blue, evidence) is space-like separated from Vlad's record of photo-detections $\both{O}^V=\overleftrightarrow{M}$ (brown, fixture). Tara's counterfactual measurement is homodyne, with record $\both{O}^{'T}=\overleftrightarrow{Y}$ (golden, consequent). The photons (cyan wiggles, not classical events but included to aid intuition) are emitted by the atom asymmetrically in accordance with the chosen parameters for the simulation: $\eta_T=0.2$, $\eta_V=0.8$, and $\Omega=2\gamma$, for an interval $[0,10\gamma\inv)$.
    \label{ST-Cont}}
\end{figure}

In our earlier work~\cite{banerjee2025counterfactualquantummeasurements}, we illustrated this using a resonantly driven two-level atom radiating, say via a leaky cavity, into two independent electromagnetic baths, each monitored continuously by two observers, here called Tara and Vlad. The space time diagram for this scenario is given in Fig.~\ref{ST-Cont}.

\begin{figure}[H]
    \centering
    \includegraphics[width=0.5\textwidth]{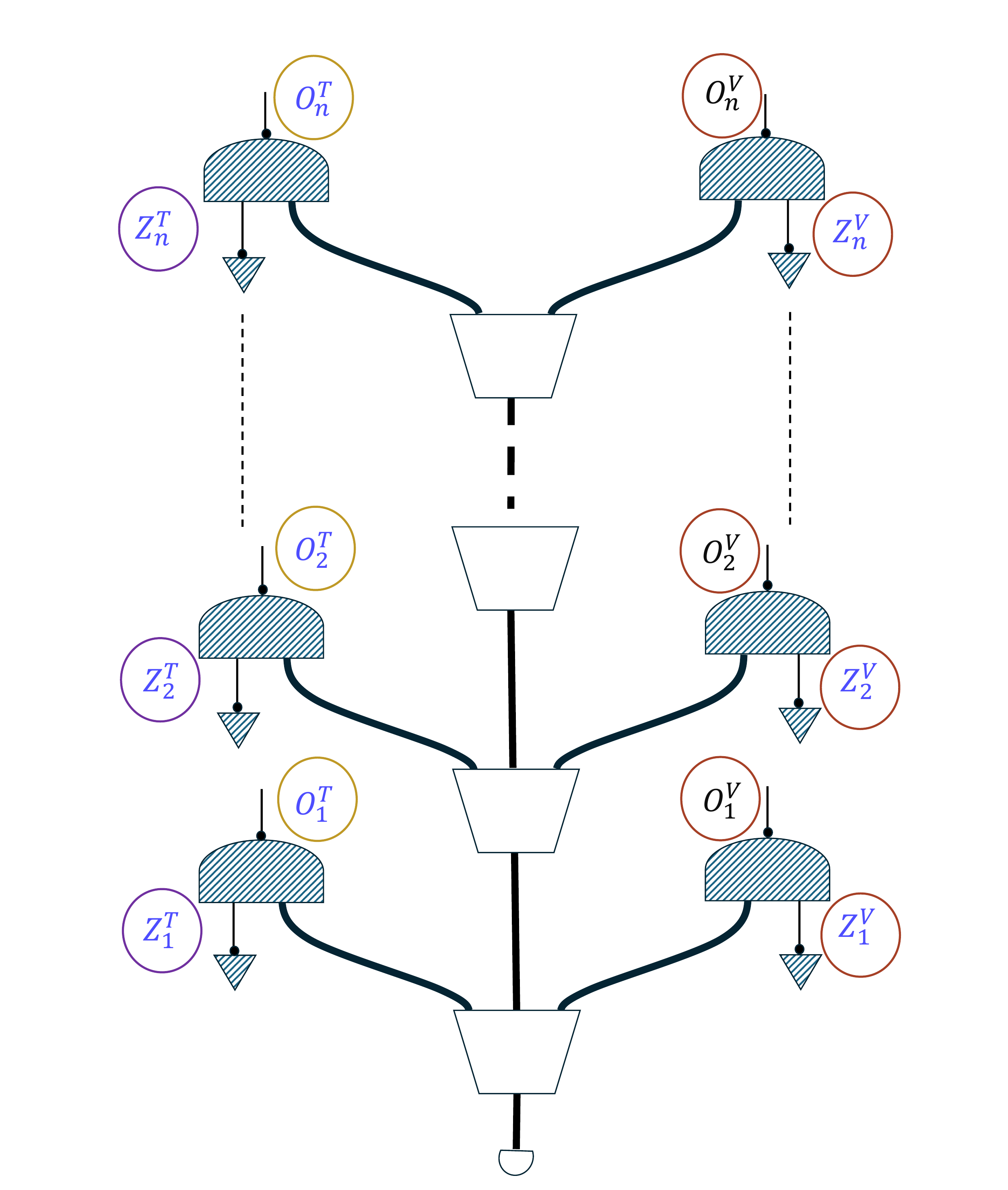}
    \caption{L-DAG representation of the causal structure for counterfactuals in continuous-in-time measurement scenario. The measurement process is discretized into fine-grained time steps $\{1,2,\ldots,n\}$ for representational convenience, with the source state being updated at each step according to the general quantum-only processing map. Blue variables represent the evidence $\bm{E}$. The ones in brown circles denote fixtures $\bm{F}$, whose values are held fixed under counterfactual evaluation. The purple circled variables represent the locus of change and turn into the antecedent $\bm{A}'$ for the counterfactual scenario. Golden circled variables similarly turn into the consequent $\bm{C}'$ whose values we compute the suspectation values of.  %computed in the counterfactual scenario due to the antecedent. 
    Thick dashed lines denote the continued causal influence mediated as quantum information by the update map, while thinner dashed lines serve only to indicate the presence of many intermediate discretized measurement steps. The graphical notations used here follows from Fig.~\ref{Terms}
    \label{Q-DAG1}}
\end{figure}

We assume that a fraction $\eta_T$ of the fluorescence is detected by Tara, while Vlad detects the rest, $\eta_V=1-\eta_T$. In the actual world, both Tara and Vlad monitor the fluorescence through photon detection, resulting in all-time records denoted by $\overleftrightarrow{N}$ and $\overleftrightarrow{M}$ respectively, in the notation of~\cite{PhysRevLett.115.180407}. The counterfactual problem we considered was, Tara asking what her suspected homodyne photo-current would be, at each point in time, had she been doing a homodyne detection instead of a photon-detection. Fig.~\ref{Q-DAG1} illustrates the, more complicated, L-DAG for this scenario. The evidence $\boldsymbol{E}$ that Tara has is her setting $(Z_t^T)$, her results $(O_t^T)$, and Vlad's setting $(Z_t^V)$, at each time point $t$, which in the L-DAG has been replaced with a discrete time index. We will use the same notation for these as for the all-time records introduced above. For instance, the locus of change $\bm{A}=\both{Z}^T$, Tara's measurement settings at all instances of time. In the time interval of the experiment, the only events in the union of the future light cones of $\both{Z}^T$ are Tara's outcomes $\both{O}^T$. Thus $\boldsymbol{D}$, the inclusive descendants of the locus of change, is the union of $\both{Z}^T$ and $\both{O}^T$. This is also derivable from the L-DAG in Fig.~\ref{Q-DAG1}.

Having defined $\boldsymbol{D}$, we define the fixtures $\boldsymbol{F}$ as its complement as usual. In this case, this comprises all of
Vlad's settings $\both{Z}^V$ and outcomes $\both{O}^V=\both{M}$. Finally,  
the consequent is Tara's counterfactual homodyne current record, $\boldsymbol{C}'\equiv\overleftrightarrow{Y}$. As discussed before, in Ref.~\cite{banerjee2025counterfactualquantummeasurements} we focussed on a simple statistic for it: the suspectation of $Y_t$ at each $t\in[0,t_f)$. %Here we have introduced suspectation  to be the counterfactual counterpart to expectation , just as supposability is to probability.
Specifically, we calculated the supectation value $\mathbb{S}[Y_t||\both{N}=\both{n}]$ conditioned on a typical actual record $\both{n}$ for Tara. 
For the details of the calculation, and the result, the reader may see Ref.~\cite{banerjee2025counterfactualquantummeasurements}, but these are also covered here, in Secs.~\ref{sec:QMT} and \ref{sec:BreakParty}. 

Although more complicated, this scenario maintains the party correspondence appearing in the CHSH scenario, because the essential causal structure remains the same. That is, in both scenarios, %the roles of the evidence, antecedent, and consequent carry over directly to the present scenario: 
the evidence is given by both parties' actual settings and the actual measurement record of Tara (here, her photon detection record) 
%the reasoning party (\ie, Tara’s photon detection record), 
the locus of change is in her measurement setting (here, replacing photon counting with homodyne detection), and the consequent is her counterfactual measurement outcome (here, a homodyne photocurrent). Likewise, the fixtures remain, as before, Vlad’s outcomes (here, a photon detection record). This simple correspondence is one of the things superseded by the new scenarios in this paper, as explained in the following Section.  

\section{Overview of new scenarios in this paper}  \label{sec:overnew}

In this section, we discuss three extensions of the counterfactual calculus: breaking the party correspondence by relaxing strict space-like separation (\cref{BPC}), extending suspectations to non-trivial summary statistics of the measurement records rather than just the measurement records themselves (\cref{sec:4(e)}), and combining both by introducing a third observer, Godot, whose conditioned quantum state for the system breaks the party correspondence while also being a non-trivial summary statistic (\cref{sec:4(f)}).

%\hmw{Check Outline section of intro and conclusion.}

\subsection{Suspectation values that break the party correspondence  \label{BPC}} 
%\ktl{up to here for S3 - will read S4 and come back. Honestly contemplating whether this part should be its own section. Currently the section is already fairly big (10 pages) and there is a lot to digest in it for a reader. And also, these are the parts original to this paper (excluding the HSM stuff).}\hmw{Also what I had concluded, Kiarn! Implemented now.}

%\hmw{more generic structure. Give ST diagram, and a separate diagram showing L-DAG, for the RF system, Tara and Vlad.}

\begin{figure}[H]
    \centering
    \includegraphics[width=0.65\textwidth]{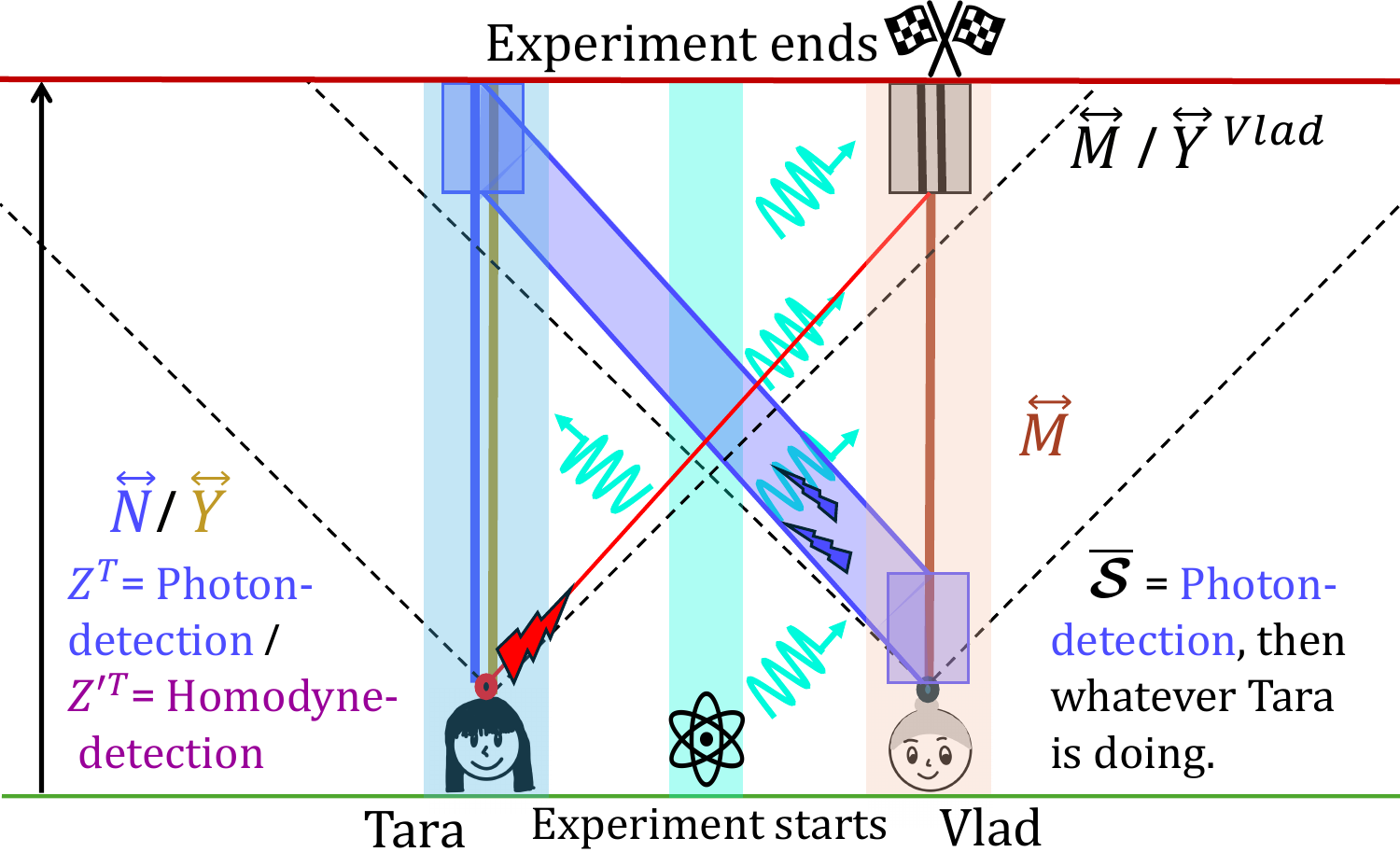}
    \caption{Space-time diagram of the counterfactual scenario with continuous monitoring by Tara and Vlad. The parties are space-like separated for most of the experiment and enter each other’s future light cones near the end. The red lightning bolt denotes a classical signal from Tara to Vlad which he uses to set his own settings as Tara's beyond that point. The blue lightning bolts are the classical signal from Vlad to Tara that indicates the portion of Vlad’s outcomes accessible to Tara (evidence $\boldsymbol{E}$), while the gray region denotes outcomes inaccessible to her (descendants $\boldsymbol{D}$). Primed variables refer to the counterfactual scenario. The slash distinguishes the actual from the counterfactual
scenario, with primed variables referring to the counterfactual
scenario. The black dashed lines are the light cones. The details of the colour coding and notation are similar to that in Fig.~\ref{ST-Cont}.
    \label{fig:Generic}}
\end{figure}

In this subsection, we move to a more generic situation—one in which the two observers are no longer completely isolated from one another. As time progresses, their actions and outcomes begin to fall within each other’s light cones, creating a more intertwined causal structure. To explore how counterfactual reasoning behaves in this setting, we walk through the same construction in this more general scenario Fig.~\ref{fig:Generic}. 
The blue shaded region in Fig.~\ref{fig:Generic} represents the subset of Vlad’s earlier outcomes that are later communicated to Tara and hence contribute to her evidence set $\boldsymbol{E}$. The gray shaded region corresponds to the final segment of Vlad’s measurement, during which his setting depends on Tara’s signal while the corresponding outcomes remain inaccessible to her; these outcomes now belong to the inclusive descendants $\boldsymbol{D}$ determined by the underlying causal structure. Consequently, not all of Vlad's outcomes can be held as fixed in this counterfactual scenario. This late time causal connection modifies the counterfactual  relevant to Tara’s reasoning.

\begin{figure}[H]
    \centering
    \includegraphics[width=0.5\textwidth]{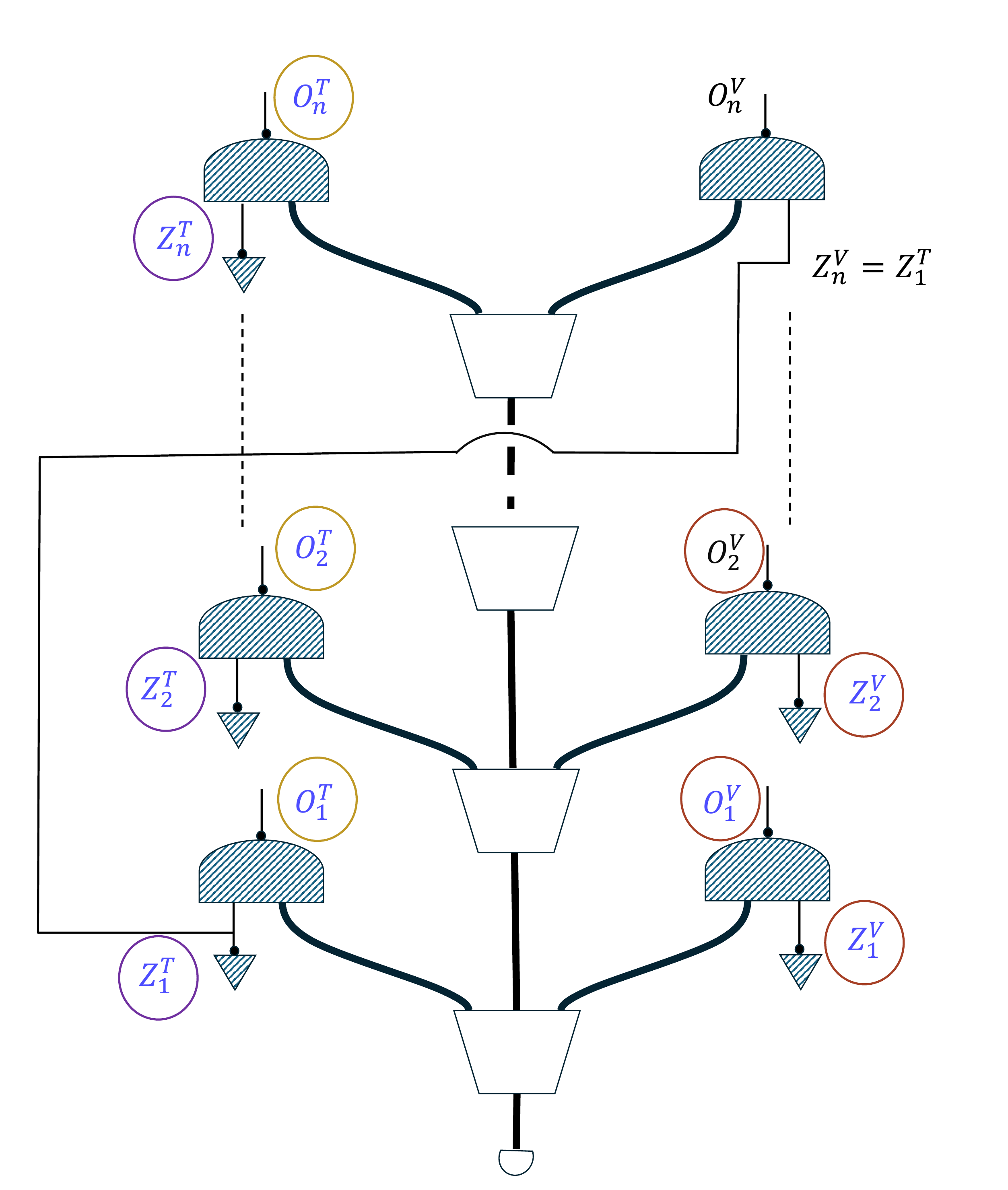}
    \caption{Labeled DAG representation of the causal structure for counterfactuals in continuous-in-time measurement and breaking the party correspondence scenario. The details of the colour coding are similar to that in Fig.~\ref{Q-DAG1}. The crucial difference is in a direct causal influence from Tara to a subset of Vlad's measurement setting and a subset of Vlad's outcomes that becomes accessible to Tara and therefore becomes a part of the evidence. This late time causal influence modifies the sets $\boldsymbol{E}$, $\boldsymbol{D}$ and $\boldsymbol{F}$, relative to the strict SLS. The graphical notations used here follows from Fig.~\ref{Terms}.}% \hmw{Text either blue (almost all) or black. Coloured circles in Purple, Gold, and Brown. We won't need primes.}
    \label{DAG-Gen}
\end{figure}

The corresponding L-DAG is shown in the Fig.~\ref{DAG-Gen}. In this scenario, Tara and Vlad perform continuous measurements on an open quantum system, in the same physical setting as described in the previous section. The key difference lies in the causal structure of the experiment. For the majority of the monitoring period, the two observers remain outside each other’s light cones. However, shortly before the experiment ends, they enter each other’s future light cones. This departure from strict SLS scenario complicates the evaluation of the counterfactual consequent, but at the same time, it illustrates the generality of the analysis and its practical relevance in actual experiments where there is often some degree of causal overlap involved between observers.

\subsection{Suspectation values for interesting summary statistics \label{sec:4(e)}}

%\hmw{Instead of suspectation of Y, suspectation of the filtered state $\rho_Y$. Be careful to note this is still technically estimating a classical quantity -- the CF Tara's estimate of the state. Relate to variables $\boss{R}$ introduced earlier. Include more DAGs}

In this subsection, we apply our counterfactual calculus in a new regime, where the elements that make up the set $\bm{R}$ introduced in~\cref{Sec:HSM} play an essential role. This set consists of classically processed outputs using information from the local readouts obtained at the output of the measurement nodes. This allows agents to compute conditioned quantum states.
Then we can consider suspectations of such conditioned states, for a counterfactual measurement setting, as an example of a (time-dependent) summary statistic of the counterfactual record. That is, it is a more interesting counterfactual suspectation than the suspected value of the measurement record itself. 

We start by discussing how Tara estimates the state $\rho^{{O}^{'T}}_t$ \footnote{The time subscript labels the time at which the relevant information is emitted by the system, rather than the time at which the state is reconstructed. Consequently, the state, measurement record, and measurement setting all carry the same time label.}, namely the state that Tara would have reconstructed  from her homodyne measurement data, had she been doing a homodyne measurement instead. 

\begin{figure}[H]
    \centering
    \includegraphics[width=0.5\textwidth]{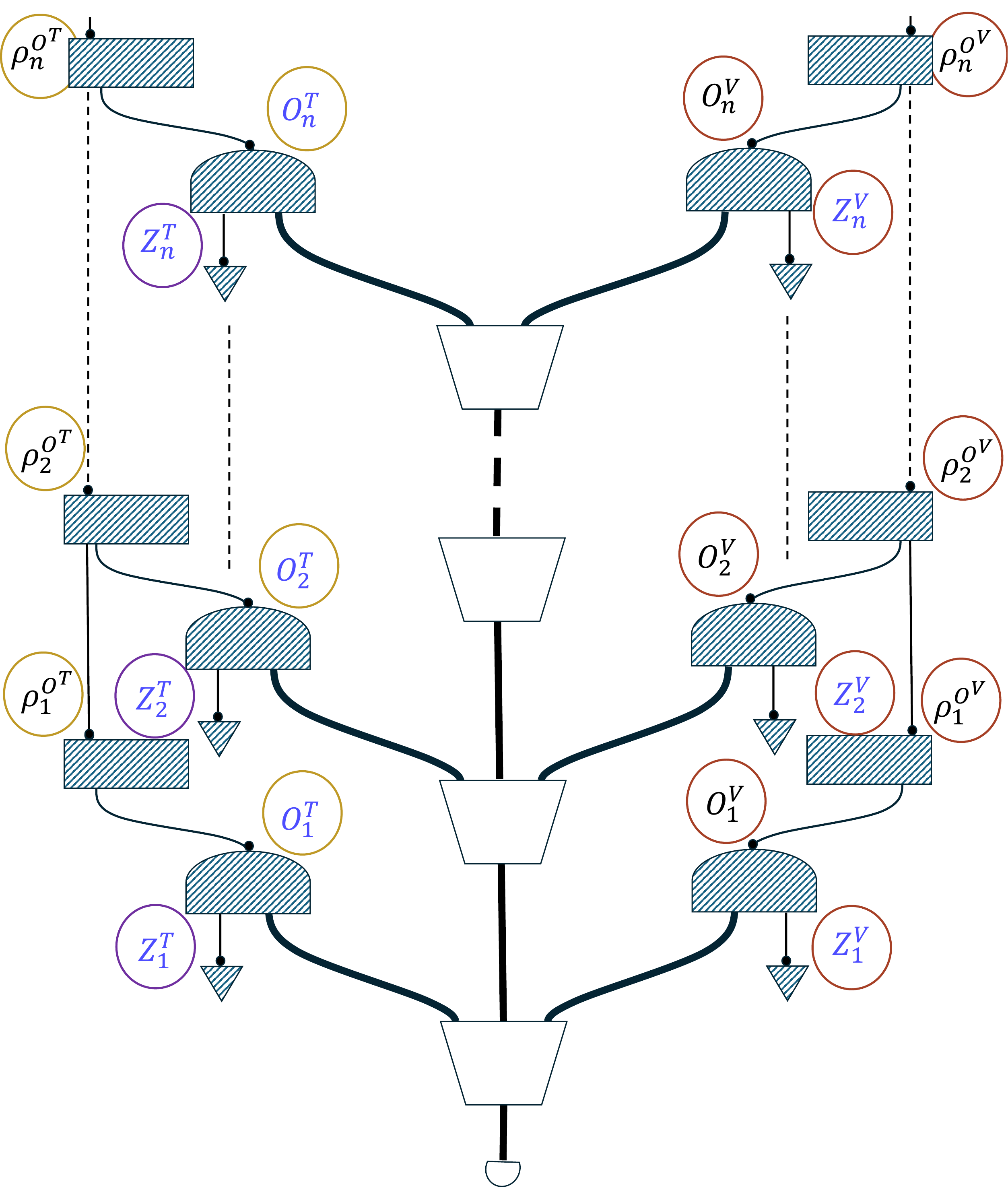}
    \caption{L-DAG representation of Tara’s reconstruction of the state needed for the consequent in a continuous-in-time measurement scenario. The details of the colour coding are similar to that in Fig.~\ref{Q-DAG1}. The new elements here are the golden circled states representing the reconstructed counterfactual state used to evaluate the consequent~\erf{TCF-S}. The graphical notations used here follows from Fig.~\ref{Terms}.}% \hmw{Similar to above, but this time there will be a lot more black text.}
    \label{F-QDAG}
\end{figure}

Fig.~\ref{F-QDAG} provides the corresponding L-DAG for this construction. For representational purposes, the continuous-time dynamics are approximated by discrete time steps. The diagram shows how the physical system undergoes continuous monitoring on both wings, corresponding to measurements performed by Tara and Vlad. 
While the full (Tara's and Vlad's combined) measurement record can generate a %fine-grained 
pure conditional state at every stage of the evolution, Tara does not have access to this complete description. Instead, she knows only her local readouts ($O_t^T$), as well as, we assume, the measurement settings for her ($Z_t^T$) and Vlad ($Z_t^V$) at each time point). These blue variables in the diagram, constitute the evidence set $\bm{E}$. The purple circled variables represent the locus of change $\bm{A}$ and corresponds to Tara’s measurement settings $Z^T$ that are modified in the counterfactual scenario to constitute the antecedent $\bm{A}'$. The fixtures $\bm{F}$ are given by the Vlad settings and outcomes (the brown circled variables), and the corresponding classically processed states, which remain fixed in the counterfactual scenario. The variables that form the consequent $\bm{C}$ (golden circled) include both her outcomes $(O^{T}_k)$ and the states $\rho^{O^{T}}_k$ that are evaluated in the counterfactual scenario. This state represents Tara's best counterfactual estimate of the system state at time $t$ that she constructs in the classical register nodes by the quantum filtering theory that will be described in~\cref{sec:QSF} using her counterfactual outcomes. This 
generates a mixed conditioned state, as, 
%conditioning both on the available evidence and on the  At 
at each time step $k$, Tara averages over the inaccessible information in Vlad’s measurement outcomes ($O^V_t$). 

\subsection{Suspectation values for interesting summary statistics that also break the party correspondence \label{sec:4(f)}}

\begin{figure}[H]
    \centering
    \includegraphics[width=0.5\textwidth]{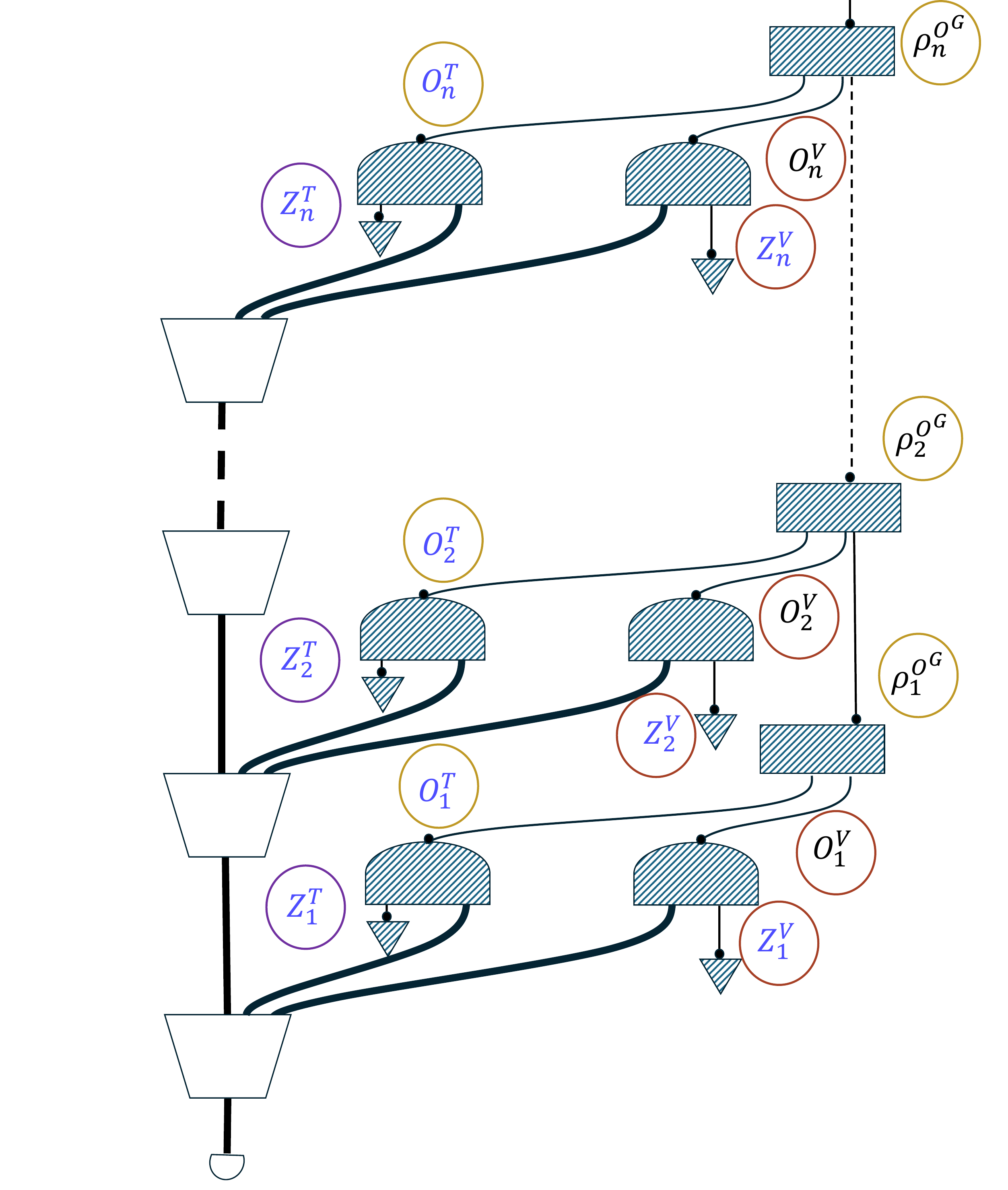}
    \caption{L-DAG representation of Godot’s reconstructed state obtained from the complete measurement record of both observers. In contrast to the preceding figures, the classical information generated by the measurement outcomes on Tara’s and Vlad’s wings is collected into a single classical register, reflecting Godot’s access to the full record at all times. The details of the colour coding are similar to that in Fig.~\ref{Q-DAG1}. The system evolution is otherwise identical to that shown previously. The graphical notation used here follows Fig.~\ref{Terms}.
    \label{Godot-DAG}}
\end{figure}

In the final subsection of this section, we introduce a third party, Godot, who has access to all of the measurement record generated by continuous monitoring from both the observers Tara and Vlad. Therefore at each point in time he has complete knowledge of the state of the quantum system which is Godot's fully conditioned state given by  $\rho^{{O}^{'G}}_t$. A schematic representation of this reconstruction, in which all measurement outcomes are collected into a single classical register, is shown in Fig.~\ref{Godot-DAG}.

Similarly to the situation in \cref{sec:4(e)}, Tara can now use the counterfactual calculus to calculate the suspectation for Godot's conditioned state. Since Godot breaks the party correspondence (his state is conditioned on fixtures but is also in the consequent), such a calculation illustrates both breaking the party correspondence as in \cref{BPC} and estimating more interesting summary statistics (conditioned state) as in \cref{sec:4(e)}.

Tara can estimate the counterfactual $\rho^{{O}^{'G}}_t$ both for
${O}^{'T}$, and when ${O}^{'T}=\nullset$. In the latter case, where Tara does no measurement (thus modifying the antecedent to a no-measurement case rather than a different measurement case), Godot's state in the counterfactual scenario is mathematically identical to Vlad's (actual but unknown) conditioned state. But, unlike Vlad's state, it is in the consequent and hence is a suitable object for calculating a suspectation value. In \cref{sec:states} we will relate this suspectation value to a more conventional type of estimate.

%\hmw{Say that now to do all this we need to review quantum trajectory theory, especially quantum (state) filtering and quantum state smoothing.}
In order to compute the various estimates (expectation values and states) discussed here and in the preceding sections, it is now necessary to review the formal framework of quantum trajectory theory. In particular, we require the notions of quantum (state) filtering and quantum state smoothing, which provide the technical tools for constructing conditional quantum states from measurement records. The purpose of the next section is to introduce the stochastic dynamical equations that govern the evolution of quantum states conditioned on continuous measurement data, and explain how to calculate the various state estimates. Readers familiar with this topic may still wish to survey this section to familiarise themselves with our notation, and in particular read \cref{CF_Smoothing} on an unexpected link between smoothing and our counterfactual calculus.

\section{Quantum state filtering and smoothing \label{sec:QMT}}
%\hmw{to self: All good}

\subsection{Open quantum systems}

In order to understand the general evolution of a quantum system undergoing continuous monitoring, it is necessary to consider its coupling to a quantum environment. Such a system is called an open quantum system, and the  
%since its dynamics is no longer governed by the deterministic Schr\"{o}dinger equation due to interactions with the environment. 
%For an open quantum system this 
coupling is typically continuous. In what follows, we will take the coupling to be one of ``quantum white noise''~\cite{LI20181}. Physically, this is often derived from two assumptions, conventionally called: (1) the Born approximation, valid when the system and environment coupling is weak; and (2) the Markov approximation, valid when the environment is much larger than the system, ensuring that at each moment the system effectively interacts with a different, independent, portion of the environment. 

 One must distinguish two types of evolution of open quantum systems, conditioned and unconditioned. When the information in the environment is ignored, 
 %state update is not conditioned on the information from the measurement, 
 the system evolution is deterministic and smooth, governed by a Lindblad master equation.  Working in the interaction picture with respect to the free Hamiltonian (the system and environment), the evolution of the system over an infinitesimal time step $\dd t$ it is given a completely positive trace-preserving (CPTP) superoperator or map~\cite{wiseman2009quantum}
\begin{equation}
    \rho_{t+\dd t}={\mathcal{E}}\rho_t\,, \label{me}
\end{equation} where one can start from the initial condition $\rho_{t_i}$. This equation is known as a master equation, and ${\cal E}\bullet = \bullet +{\cal L}\bullet\dd t$, with ${\cal L}$ being the Lindblad super-operator \cite{lindblad1976generators}. 
%Note that here we focus on the dynamics of the quantum state and hence used the Schr\"odinger picture approach to quantum dynamics instead of the Heisenberg picture which describes the dynamics of the observables. 
By contrast, when the information in the environment is not ignored, but rather gathered using detectors, the system evolution will be indeterministic. However, under the quantum white noise condition, the {\em average} evolution (averaging over the master equation) will still be given by \erf{me} because measuring the part of the environment that has already interacted with the system does not alter the parts of the environment that are yet to interact with it. Moreover, it is possible to monitor the environment continuously in time such that the evolution is still Markovian (like the master equation). This comes at the expense of the evolution becoming non-differentiable (quantum diffusion) or even discontinuous (quantum jumps)~\cite{wiseman2009quantum}. 

\begin{figure}[H]
    \centering
    \includegraphics[width=0.6\textwidth]{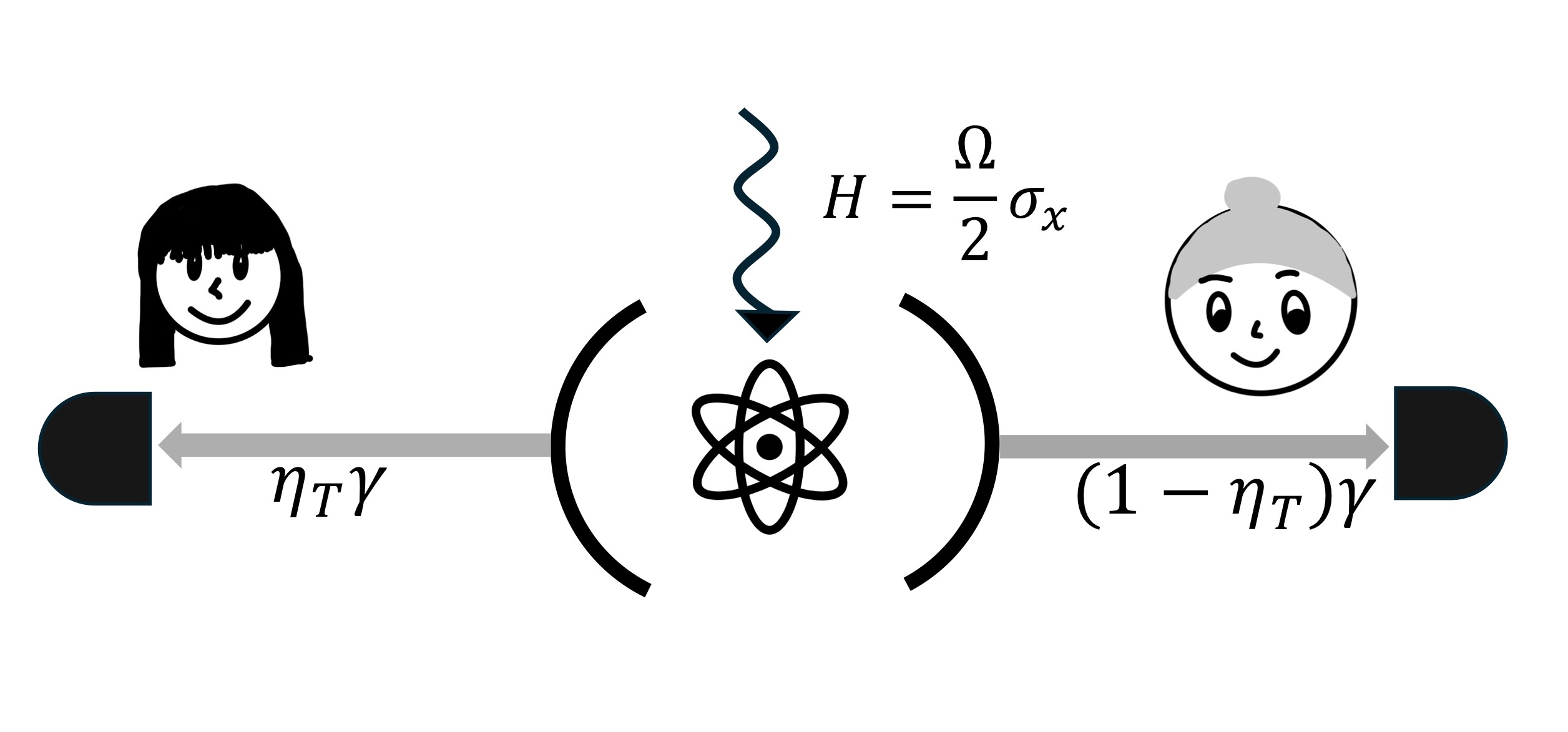}
    \caption[Open Quantum System]{ Example of an open quantum system consisting of a resonantly driven two-level atom radiating into two independent bath modes. The output channels are continuously monitored by two observers who share the total detection efficiency, $\Omega$ is the Rabi frequency of driving, and $\gamma$ is the total radiative damping rate, including emission into both output channels of the cavity (from both ends of the cavity). 
                               \label{Fig: OpenSystem}}
\end{figure}

The open quantum system we consider is a resonantly driven two-level atom radiating via a leaky cavity into two independent electromagnetic baths (another term for the environment). These separately undergo continuous-in-time measurements by two observers, Tara and Vlad. See Fig.~\ref{Fig: OpenSystem}. The atom is driven by an external classical field (laser) with a driving amplitude $\Omega$ about the $x$-axis of the Bloch sphere. The fluorescence of the atom is then measured with an efficiency $\eta_T$ by Tara and the remaining portion is measured by Vlad with an efficiency $\eta_V=1-\eta_T$, where $\gamma$ is the rate at which the excited atom decays into the ground state due to its interaction with the environment.

 Therefore the Lindblad super-operator ${\cal L}$ can be separated into three parts as
\begin{equation}
{\cal L}\rho = {\cal D}[\hat T]\rho + {\cal D}[\hat V]\rho - i[\hat H,\rho], \label{eq:me1}
\end{equation}
where ${\cal D}[\hat w]\rho = \hat w \rho \hat w^\dagger 
- \frac{1}{2}\big(\hat w^\dagger \hat w \rho + \rho \hat w^\dagger \hat w\big)$ denotes the standard Lindblad superoperator (unrelated to the ${\cal D}$ used for a decohering map in \cref{Sec:HSM}). The Lindblad operators 
\beq
\hat T=\sqrt{\gamma\eta_T}\,e^{-i\Phi}\hat{\sigma}_- \textrm{ and }\hat V=\sqrt{\gamma\eta_V}\,\hat{\sigma}_- \label{eq:LinOps}
\eeq 
correspond to the two decoherence channels associated with the independent electromagnetic baths monitored by Tara and Vlad, respectively, and account for the non-unitary evolution of the dynamics. The phase factor $e^{-i\Phi}$ in Tara's Lindblad operator is irrelevant here, but will play a role when we get to homodyne detection. 
The Hamiltonian in \cref{eq:me1} is $\hat{H}=\frac{\Omega}{2}\hat{\sigma}_x$, where we have set $\hbar=1$.

\subsection{Quantum state filtering \label{sec:QSF}}

In this section, we will discuss the causal conditioned evolution of an open quantum system where the state of the system is updated based on measurement outcomes. This evolution describes how the state $\rho_t$ depends on the continuous-in-time measurement record $O_t$, where each realization of the measurement record corresponds to a single trajectory traced by the quantum system. Under the quantum white noise condition, this trajectory can be thought of as arising from repeated measurements on the environment, on a time scale short compared to the interesting system evolution, but long compared to the bath correlation time (and hence without invalidating the master equation)~\cite{LI20181}.  This monitoring yields a record that is a stochastic function of time, $O_t$. The conditioned state $\rho^{\overleftarrow{{O}}}_t$  is a function of the past measurement record and is called the filtered quantum state. Here $\past{O}_t$ denotes the record $O_s$ for all $s$ prior to time $t$; a  detailed explanation of this notation, and other arrows we use, is given in~\ref{App:A1}. 
%\ktl{Hmmm, I think it would be better if we gave them when they appear and explain the conventions we have here.}\ing{I haven't shifted this yet, will wait for what Howard suggests}.
While 
$\rho^{\overleftarrow{{O}}}_t$ is a function of the past measurement record, its evolution is local in time, governed by the conditional forward map (recall from Sec.~\ref{Sec:HSM} that $Z_t$ is the time-dependent setting) 
\begin{equation} \label{foo}
    \tilde{\rho}^{\overleftarrow{{O}}}_{t+\dd t} = {\mathcal{F}}({O_t};Z_t)\tilde{\rho}^{\overleftarrow{{O}}}_{t},
\end{equation}
where we use a tilde to denote an unnormalized state. The normalized state $\rho^{\past{O}}_t$ is called the filtered quantum state.
% On average, the conditioned state reproduces the master equation evolution:
%     \begin{equation}
%         \mathbb{E}[\rho^{\overleftarrow{\bm{O}}}_t]=\rho_t=\text{exp}[\mathcal{L}(t-t_i)]\rho_{t_i}.
%     \end{equation} 
This framework constitutes the quantum analogue of classical filtering and is commonly referred to as quantum trajectory theory~\cite{wiseman2009quantum}.
The conditional forward map $\mathcal{F}(O_t; Z_t)$ represents the unraveling of the 
unconditioned channel $\mathcal{E}$ in~\erf{me}, i.e., 
$\mathcal{E}\bullet = \sum_{O_t} \mathcal{F}(O_t; Z_t)\bullet$. 
%= \sum_{O_t} \hat{M}_{O_t}\hat{U}\!\left(\sum_{l=0}^{L} \hat{K}_l \bullet \hat{K}_l^\dagger\right)\!
%\hat{U}^\dagger \hat{M}_{O_t}^\dagger$, which is the Kraus decomposition in terms of the
%measurement operators $\hat{M}_{O_t}$. The Kraus operators $\hat{K}_l$ account for any 
%decoherence arising from measurement inefficiency or unmonitored channels, while the 
%unitary propagator $\hat{U} = e^{-i\hat{H}\,\mathrm{d}t}$ captures the coherent evolution.

As discussed previously, in the scenario in Fig.~\ref{Fig: OpenSystem}, Tara and Vlad continuously monitor distinct environmental 
channels of a resonantly driven two-level atom. The CPTP map in Eq.~(\ref{me}) thus 
decomposes as
\begin{align}
    \mathcal{E}\bullet
    = \sum_{O^T_t, O^V_t} 
    \mathcal{T}(O^T_t; Z^T_t)\,\mathcal{U}\,\mathcal{V}(O^V_t; Z^V_t)\bullet\,,
\end{align}
where $\mathcal{T}(O^T_t; Z^T_t)$ and $\mathcal{V}(O^V_t; Z^V_t)$ are the superoperators relating to Tara and Vlad, respectively, and $\mathcal{U}(\bullet) = \hat{U}\bullet\hat{U}^\dagger$ where $\hat{U} = e^{-i\hat{H}\,\mathrm{d}t}$ with $\hat{H} = (\Omega/2)\hat{\sigma}_x$. Here the superoperators, 
 $\mathcal{T}$ and 
$\mathcal{V}$ are (like ${\cal U}$) purity-preserving CP maps. They are, for typical results $O^T_t$ and $O^V_t$ respectively, only infinitesimally different from the identity map. Thus the order of the maps does not matter, and $\mathcal{T}$ and $\mathcal{V}$ should be thought of as conditional forward maps, like $\mathcal{F}$, but for Tara's and Vlad's measurements only, and not including the Hamiltonian evolution captured by $\mathcal{U}$. 

There are, in general, two main types of measurements used for continuous monitoring~\cite{wiseman2009quantum}, often described as different unravellings of the master equation.  The first is a jump process, in which the quantum state undergoes sudden, 
discontinuous transitions at random detection times with a finite rate and finite 
measurement back-action; between jumps, the state evolves smoothly but 
non-unitarily. Photon detection from a radiating two-level atom is a canonical example: 
detecting a photon collapses the atom to a lower-energy state at the detection time. 
The second is a diffusive process, which yields a continuous state evolution 
arising from a sequence of weak measurements, each extracting only an infinitesimal 
amount of information in $\mathrm{d}t$ and causing correspondingly weak back-action. 
Homodyne detection is the standard example, where the emitted field is mixed with a 
strong local oscillator and measured by photodetectors, producing a continuous 
record with limited instantaneous information about the system.
%\ing{I have shortened the above description of the types of measurements compared to the one in the thesis, I can add more if seem insufficient.} 

Building on the above discussion, we now define three types of channels, based on the types of measurements that we consider,
\begin{enumerate}
\item Photon-detections: The channel with unit-efficiency photon-counting is:
\begin{equation}
    {\cal C}({\dd N^C_t};{\rm pc})\bullet = \bullet+\hat{C} \bullet \hat{C}\dg \dd N_t^C - \dd t\, \bar{\mathcal{H}}[\smallfrac{1}{2}\hat{C}\dg\hat{C}]\bullet , \label{eq:Pd-channel}
\end{equation}
where, following the notation of Refs.~\cite{wiseman1996quantum,wiseman2009quantum}, 
\begin{equation}
    \bar{\mathcal{H}}[\hat{w}] \bullet = \hat{w} \bullet  + \bullet \hat{w}^\dagger. \label{eq:linear H}
\end{equation}
\begin{comment}
represents the linearized version of the superoperator 
\begin{equation}
    \mathcal{H}[\hat{w}] \bullet = \hat{w} \bullet  + \bullet \hat{w}^\dagger  - \text{Tr}(\hat{w} \bullet  + \bullet \hat{w}^\dagger)\bullet. \label{eq:non-linear H}
\end{equation}
\end{comment}
The operator $\hat{C}$ is the relevant collapse operator--- $\hat T$ or $\hat V$ in \cref{eq:LinOps} for our two observers---and $C$ (in the superscript) is the matching letter ($T$ or $V$) indicating whether the channel for the photon detections is for Tara or Vlad. That is, in \cref{eq:Pd-channel}, the letter C appears in three different typographical forms; for a given observer, each occurrence should be consistently replaced by the correspondingly styled letter, \eg, ${\cal C} \to {\cal T}$, $C \to T$, and 
$\hat C \to \hat T$ for Tara. The stochastic increment in the number of photon detections, $\dd N_t^C$, equals either zero or one, which is captured by the equation 
\begin{align}
    (\dd N_t^C)^2=\dd N_t^C.
\end{align}
\begin{comment}
    Therefore $\dd N_t$ is either zero or one.
The collapse operators defining the maps for Tara (photon-detections) and Vlad (photon-detections) are given by
\begin{equation}
    \hat{T}=\sqrt{\gamma\eta_T}\hat{\sigma}_-\,\,\, \text{and}\,\,\, \hat{V}= \sqrt{ \gamma(1-\eta_{{\rm T}})}\,\hat{\sigma}_-, 
\end{equation}
Therefore, the channels for Tara and Vlad's photon detections are
\begin{equation}{\cal C}({\dd N^c_t};{\rm pc})\bullet=\begin{cases} {\cal T}({\dd N^T_t};{\rm pc})\bullet
     & \text{when}\,\,(\hat{C}=\hat{T})\\
{\cal V}({\dd N^V_t};{\rm pc})\bullet& \text{when}\,\,(\hat{C}=\hat{V})
\end{cases}, \label{eq:pc-map}
\end{equation}
respectively. 
\end{comment}

\item Homodyne detection: The channel corresponding to unit-efficiency homodyne detection is~\cite{wiseman2009quantum}: 
\begin{equation}
    {\cal C}({\dd Y_t^C};{{\rm hd}})\bullet  = \left[\bullet + \dd t {\cal D}[\hat C]\bullet + \dd Y_t^C\,\bar{\mathcal{H}}[-i\hat  C] \bullet\right]\times\frac{1}{\sqrt{2\pi\dd t}}{\rm exp}\left[-\frac{(\dd Y_t^C)^2}{2 \dd t}\right]\label{eq:hd-map}
\end{equation}
 where ${\cal D}$ is the Lindblad superoperator in~\erf{eq:me1} and $\dd Y_t$ is the stochastic increment in the integrated homodyne current, which is real-valued.
 % has been generated from a gaussian distribution with respect to the following statistics as that of an infinitesimal Wiener increment that obeys~\cite{wiseman2009quantum} 
 % \begin{align}
 %     \dd Y_t^2=\dd t & \quad \quad
 %     \mathbb{E}_{\text{ost}}\left[dY_t\right]=0. \label{Eq:Homo-statistics}
 % \end{align} 
 In the scenario we consider in this paper, only the observer Tara has a homodyne measurement. Thus we will drop the $C$ superscript on $Y$, and replace $\hat C$ by $\hat T$ as defined in \cref{eq:LinOps}. 
 %and the relevant collapse operator defining the map for homodyne detection is 
% \begin{equation}
%    \hat{T}=\sqrt{\gamma\eta_T}\hat{\sigma}_-e^{-i\Phi},
% \end{equation}
 Here, for the first time, the phase $\Phi$ in $\hat T$ becomes relevant: it is the local oscillator phase. For the rest of the paper we will consider a `$Y$' quadrature measurement, with $\Phi=\pi/2$, and this is what the 
 %which when substituted in~\erf{eq:hd-map} gives the appropriate 
 map ${\cal T}({\dd Y_t};{\rm hd})\bullet$ will indicate.
 
\item No detection: The channel corresponding to no measurement is the channel where the measurement outcomes are averaged over: 
\begin{align}
    {\cal C}({\dd \nullset};{{\rm nm}})\bullet = \bullet + \dd t {\cal D}[\hat C]\bullet&=\sum_{\dd N=0,1}{\cal C}({\dd N^C_t};{\rm pc})\bullet\\&=\int \dd Y_t \,\,{\cal C}({\dd Y_t};{{\rm hd}})\bullet \label{eq:nm-map}
\end{align}
\end{enumerate}

\subsection{The states to be counterfactually estimated} 
As well as being useful for generating an ensemble of possible measurement records, the filtering framework introduced above is also directly useful for  illustrating the power of our counterfactual calculus. 
The filtered states themselves, as functionals of records held by some party or parties, represent classical objects that can be estimated. There are three 
distinct filtered states considered 
in this work, each of which could be estimated under a variety of conditions, some corresponding to quantum state smoothing and others to counterfactual inference. We now present those three filtered states. 

\begin{enumerate}
    \item \textit{Tara's filtered state:} The state conditioned on Tara's complete past 
    record $\overleftarrow{O}^T$ is obtained by evolving the system state under the maps 
    ${\cal T}$ and ${\cal U}$, while marginalizing over Vlad's records $\overleftarrow{O}^V$. 
    This yields
    \begin{equation}
        \tilde{\rho}_{t+\dd t}^{\overleftarrow{O}^T} 
        = {\cal T}({O^T_t};{Z^T_t})\, {\cal U}\,
        {\cal V}({\dd \nullset};{{\rm nm}}) \tilde{\rho}_{t}^{\overleftarrow{O}^T}. \label{TFS}
    \end{equation}

    \item \textit{Vlad's filtered state:} The state conditioned on Vlad's complete past 
    record $\overleftarrow{O}^V$ is obtained by evolving the system state under the maps 
    ${\cal V}$ and ${\cal U}$, while marginalizing over Tara's records $\overleftarrow{O}^T$. 
    This yields
    \begin{equation}
        \tilde{\rho}_{t+\dd t}^{\overleftarrow{O}^V} = {\cal T}({\dd \nullset};{{\rm nm}})\, 
        {\cal U}\, {\cal V}({O^V_t};{Z^V_t}) 
        \tilde{\rho}_{t}^{\past{O}^V}. \label{VFS}
    \end{equation}

    \item \textit{Godot's filtered state:} Here we introduce a third observer, Godot, who 
    has access to the complete measurement records of both Tara and Vlad, yet performs 
    no independent measurement of his own. The state conditioned on the combined past 
    record $\past{\boldsymbol{{O}}}^G = (\overleftarrow{O}^V, \overleftarrow{O}^T)$ is obtained by evolving the 
    system state under all three maps ${\cal T}$, ${\cal V}$, and ${\cal U}$, giving
    \begin{equation}
        \tilde{\rho}_{t+\dd t}^{\past{\boldsymbol{{O}}}^G} = {\cal T}({O^T_t};{Z^T_t})\, 
        {\cal U}\, {\cal V}({O^V_t};{Z^V_t}) 
        \tilde{\rho}_{t}^{\past{\boldsymbol{{O}}}^G}. \label{GFS}
    \end{equation}
\end{enumerate}

\subsection{Quantum effect retrofiltering}

To implement the estimation techniques just dicsussed (smoothing and counterfactual inference), it is necessary to consider records subsequent to the time related to the state of interest. 
%can utilize past measurement records to predict the statistics of measurement outcomes at later times.  
This requires the quantum analog of retrofiltering, known as quantum effect retrofiltering~\cite{PhysRevLett.111.160401,PhysRevLett.115.180407}. Suggestive of the name, the quantum retrofiltered effect is an effect operator $\hat{E}_R(t):=\hat E_t^{\fut{O}}$. Here $\fut{\bullet}$ denotes the outcomes later to time $t$; a more detailed notation of the arrows is given in~\ref{App:A1}. That gives the probability density for the string of measurement outcomes that occurred at a later time, given a particular quantum state at the current time $t$, via 
\begin{equation}
    \wp(\fut{O}|\rho_t)=\text{Tr}[\hat E_t^{\fut{O}} \rho_t].
\end{equation}
Note that here the $\rho_t$ is not necessarily a solution of the unconditioned evolution; it could be conditioned on past records via \cref{foo}. 
The retrofiltered effect evolves backwards-in-time from a final uninformative effect $\hat E_{t_f}^{\fut{O}} = \hat{I}$, reflecting the absence of any measurement beyond the final time. Its backward evolution over an infinitesimal interval $\dd t$ is given by~\cite{8pc3-7pg5}
\begin{equation}
    \hat E_{t-\dd t}^{\fut{O}}={\mathcal{F}^\dagger }({O_t};Z_t)\hat E_t^{\fut{O}}, \label{retrofilmap}
\end{equation}
where the $\fut{O}$ in the left-hand side of the equation now refers to the outcomes in future of $t-\dd t$. In \cref{retrofilmap}, ${\mathcal{F}^\dagger }$ is the adjoint of the forward conditioning map in \cref{foo}. To explain what this means, a general expression for the forward map is 
\beq
\mathcal{F}(O_t) = \hat{M}_{O_t}\hat{U}\!\left(\sum_{l=1}^{L} \hat{K}_l \bullet \hat{K}_l^\dagger\right)\!
\hat{U}^\dagger \hat{M}_{O_t}^\dagger,
\eeq
where $\sum_{l=1}^L \hat K\dg_l \hat K = \hat I$,  and the corresponding adjoint map is 
\begin{equation}
    {\mathcal{F}^\dagger }({O_t};Z_t)\bullet=\sum_{l=0}^L\hat{K}_l\dg
\hat{U}\dg\hat{M}_{O_t}\dg\bullet\hat{M}_{O_t}\hat{U}\hat{K}_l\end{equation}

In this work, Tara acts as the primary observer and, as we will see later, she can utilize her whole measurement record (before and after $t$) to guess Vlad's outcomes before some time $t$. To do this, she needs to calculate her effect operator $E_t^{\fut{O}^T}$ which evolves according to

\begin{equation}
   E_{t-\dd t}^{\fut{O}^T}={\cal V}_{t-\dd t}\dg({\dd \nullset};{{\rm nm}})\,{\cal U}\dg\,{\cal T}_{t-\dd t}\dg({O^T_t};{Z^T_t}) E_{t}^{\fut{O}^T}, \label{eq:Taraeffect}
\end{equation}
where we have used the adjoint map of the measurement operation  ${\cal T}_{t+\dd t}\dg({O^T_t};{Z^T_t})$ for the record that is observed by Tara (being the primary observer in the counterfactuals) and an adjoint unconditioned map ${\cal V}_{t-\dd t}\dg({\dd \nullset};{{\rm nm}})$ corresponding to the records inaccessible to Tara, here arising from Vlad’s measurement channel. For reasons that will become  evident later, we will also need the effect operator for Vlad, the evolution of which is given by
\begin{equation}
    E_{t-\dd t}^{\fut{O}^V}={\cal V}_{t-\dd t}\dg({O^V_t};{Z^V_t})\,{\cal U}\dg\,{\cal T}_{t-\dd t}\dg({\dd \nullset};{{\rm nm}})E_{t}^{\fut{O}^V}, \label{eq:VRF}
\end{equation}
where now the adjoint unconditioned map is for Tara ${\cal T}_{t-\dd t}\dg({\dd \nullset};{{\rm nm}})$, corresponding to her results that are inaccessible to Vlad.

\subsection{Quantum state smoothing \label{Q-Smooth}}

%\hmw{Start with fact that Tara's filtered state is her optimal estimate of Godot's filtered state at time $t$, given her past record. ... leads naturally to why not use the whole record. Done generally, not with specific measurement types. Don't get into ostensible probability. But give expression with the unnormalized state. Give as a displayed equation a smoothed probability distribution i.e. for one record given another in the smoothing section {\em en route} to the QSS. Here define the SQS as Tara's best estimate of Godot's state (who knows ).} 

In our scenario, Godot is the party with the most information. Indeed, if the initial system state is pure then his filtered state will be a state of maximal information in quantum mechanics, \ie, a pure state. It is natural, therefore, to consider the best estimate that another party (Tara) can make of Godot's filtered state (\ref{GFS}), which we also refer to as the fully conditioned state. In this subsection we are not concerned with this in a counterfactual setting, but rather the actual setting where Tara knows her own record but not Vlad's. This will serve as a useful comparison with counterfactual calculations later. 

%The problem that Tara is trying to approach is to obtain the best estimate of Godot's filtered state (fully conditioned state). 

The above paragraph raises the question of what the `best estimate' should mean in a quantum context. Quantitatively it should be defined as the state that minimizes an expected cost function. A cost function in general is any function (functional) defined for the unknown variables (unknown records or unknown fully conditioned state) and their estimates, which is minimized if the estimate coincides with the true value. For quantum state estimation, the Trace-Square Deviation is a natural quantifier for the divergence between the estimate and the fully conditioned state~\cite{chantasri2021unifying}, although relative entropy serves just as 
well~\cite{PhysRevA.104.032213,PhysRevA.106.069903}. The reason these are to be preferred for defining the cost function is the following. Given her past records $\past{O}_t$ the optimal estimate Tara can make of Godot's filtered state, \erf{GFS}, at a time $t$, according to the above-mentioned cost functions, is the ensemble average of Godot's filtered state over Vlad's unknown records conditioned on $\past{O}_t^T$: 
\begin{equation}
     \rho^{\past{O}^T}_t = \mathbb{E}_{\overleftarrow{O}_t^V|{\past{O}_t^T}}  \left[ \rho_t^{\overleftarrow{O}^V,\overleftarrow{O}^T} \right].
\end{equation}
But it is easy to show~\cite{PhysRevLett.102.250403} that this is exactly the filtered state that Tara can calculate via~\erf{TFS}.

Now, intuitively, Tara can make a better estimate of Godot's state if she uses her whole record (past and future) $\both{O}^T$; see \ref{App:A1} for the details of the arrow notation. For the cost functions mentioned  above, it can again be written as the ensemble average of Godot's filtered state over Vlad's unknown records, but now conditioned on both past and future records of Tara, %$\both{O}^T$, 
\begin{align}
  \rho^{\both{O}^T}_t &= \mathbb{E}_{\overleftarrow{O}^V_t|{\overleftrightarrow{O}^T}}  \left[ \rho_t^{\overleftarrow{O}^V,\overleftarrow{O}^T} \right] \nonumber\\
  & \equiv \sum_{\overleftarrow{O}_t^V} P(\overleftarrow{O}_t^V | \overleftrightarrow{O}^T) \rho_t^{\overleftarrow{O}^V,\overleftarrow{O}^T},  \label{Smo-State}
\end{align}
where \( P(\overleftarrow{O}_t^V | \overleftrightarrow{O}^T) \) is the smoothed probability distribution for Vlad's record prior to time $t$, obtained by conditioning on Tara's whole record $\overleftrightarrow{O}^T$.
This method of estimation, in which one uses the whole record to estimate the quantum state, is called quantum state smoothing~\cite{PhysRevLett.115.180407}.
Note that the calculation of this smoothed state is non-trivial since it involves averaging over the set of records that are unknown. We will discuss the details of the calculation later in~\ref{App:TaraSmoothing}.  

\subsubsection{Aside: Smoothed state as a supposable state \label{CF_Smoothing}} 

%\hmw{And can discuss Valerio's new paper which says that quantum counterfactuals can be used to define the QSS! Unless we think this will confuse the flow, in which case just mention it and present it in an Appendix. Kiarn may have an opinion.} 
Since we have now introduced the major techniques used for the various estimations discussed so far, it is relevant to discuss a recent work by Liu \ea ~\cite{liu2026unifyingquantumsmoothingtheories} in which a direct connection is drawn between our theory of quantum counterfactuals and quantum state smoothing within orthodox quantum theory. Making this connection explicit requires a measurement framework that permits reconstruction of quantum states from observed statistics. Informationally complete positive operator valued measurements (IC-POVM) provide precisely such a framework. We therefore briefly recall the notion of an informationally complete POVM.

A POVM, ${\cal M}=\{\hat E_j\}$ ($\hat{E}_j\geq 0$ and $\sum_j \hat{E}_j =\hat{1}$) for a quantum system on a finite-dimensional Hilbert space is said to be informationally complete (IC)~\cite{GMDAriano_2004,article}, if its outcome statistics contain enough information to reconstruct any preparation (\ie, any density operator) on the Hilbert space. Such a POVM is said to constitute a frame, since it spans the operator space of the Hilbert space (the space of Hermitian operators). This means that any operator $\hat A$ can be written as 
\begin{equation}
    \hat A= \sum_j I_j(\hat A)\hat E_j\,, \label{Eq:IF}
\end{equation}
where $I_j(\hat{A})$ is the data processing function of the outcome $j$ which depends on the operator $\hat{A}$. 
In that case the expectation value of any operator $\hat{A}$ of the system in the state $\rho$ can be obtained as
\begin{equation}
    \langle \hat{A}\rangle=\Tr[\rho \hat{A}]=\sum_j I_j(\hat{A})\Tr[\rho\hat{E}_j]\,. 
\end{equation}

Corresponding to this operator frame $\{\hat E_j\}$, there exists a dual frame $\{\hat F_j\}$ of Hermitian (though not necessarily positive semi-definite) operators. This dual frame enables one to reconstruct any operator. For an arbitrary IC-POVM, the canonical dual frame with respect to the trace measure is the optimal dual frame for state reconstruction~\cite{article}.
 Accordingly, any arbitrary operator $\hat A$, admits an expansion in terms of the canonical dual frame as
\begin{equation}
    \hat A= \sum_j \hat F_j \Tr[\hat E_j\hat A]\,,\label{Eq:LM}
\end{equation}
where the coefficients are given by the outcome probabilities of the IC-POVM measurement. 
%(or expectation values) associated with the informationally complete measurement through the corresponding linear map.

Now consider the following counterfactual question ``What would have Tara seen %at time $t$ had someone 
had she performed an IC-POVM $\{\hat{E}_j\}$ at time $t$ on the system (the atom in our example), given her actual observation $\both{O}^T$?''. Our counterfactual calculus assigns supposabilities, for Tara's hypothetical measurement outcomes $j$, 
in the usual way: 
%by marginalizing over the probability distribution of the known and unknown fixtures. 
In this case, the fixtures include Tara's and Vlad's outcomes prior to $t$, $\past{O}_t^T$ and $\past{O}_t^V$ respectively. 
Thus, using ${\cal M}(t)$ to denote performing (counterfactually) the IC-POVM at time $t$, 
\begin{align}
    \Su(j|{\cal M}(t)||\both{O}^T) &=\sum_{\past{O}^V} P(j|{\cal M}(t),\past{O}_t^V,\past{O}^T_t)P(\past{O}_t^V|\both{O}^T) \nonumber\\
    &= \sum_{\past{O}^V} \frac{P(j,\past{O}_t^V,\past{O}^T_t|{\cal M}(t))}{P(\past{O}_t^V,\past{O}^T_t)}P(\past{O}_t^V|\both{O}^T) \nonumber \\
    &= \sum_{\past{O}^V} \Tr \left[{\hat{E}_j{\rho}_t^{\overleftarrow{O}^V,\overleftarrow{O}^T}}\right]P(\past{O}_t^V|\both{O}^T)\nonumber\\
    &= \Tr \left[ \hat{E}_j\left(\sum_{\past{O}^V}P(\past{O}_t^V|\both{O}^T){\rho}_t^{\overleftarrow{O}^V,\overleftarrow{O}^T}\right)\right], \label{Eq:CF=Sm}
\end{align}
% \red Going forward, we will take $t_p=0$, as we did implicitly in Sec.~\ref{sec:QSF}. \blk 
 % \ing{Change the tm and tp to simply t even in the formula}.
From~\erf{Eq:CF=Sm} one can ask if there is a state that generates these probabilities. It is easy to see that the term inside the bracket~\erf{Eq:CF=Sm} is exactly the smoothed state $\rho_t^{\both{O}^T}$.
Liu \ea~\cite{liu2026unifyingquantumsmoothingtheories} showed that, within the smoothing framework, these counterfactual probabilities (supposabilities) coincide exactly with the Born-rule probabilities evaluated on the smoothed state. 

We can now use the linear map in \erf{Eq:LM} together with the dual frame formalism to define the smoothed state by weighting the dual operators with the corresponding supposabilities, for the outcomes of an unperformed measurement $\{\hat{E}_j\}$ as
\begin{equation}
    \rho_t^{\both{O}^T} = \sum_j \hat F_j \,\Su(j|{\cal M}(t)||\both{O}^T)\,.
\end{equation}
In this way, supposability supplies the probabilistic content of the counterfactual antecedent, while the dual frame elevates this information to an operator-valued object, namely the density matrix. This explicit dual-frame construction of the smoothed state from counterfactual probabilities  is not discussed in Liu \ea~\cite{liu2026unifyingquantumsmoothingtheories}, but rather is original to this work. 
%This dual-frame representation is gives an elegant, explicit equation for the smoothed state in terms of supposabilities. 

\section{Suspectation value breaking the party correspondence \label{sec:BreakParty}}

%\hmw{Refer back to IV.D. Re-explain more rigorously (now with specific Hamiltonian and Lindblad operators) and what Tara and Vlad are doing, and so what their $Z_t^V$ looks like as a function of time. And what all the bold CF calculus variables are in terms of the notation of the preceding section.}  
In this section we apply the continuous measurement theory and counterfactual calculus introduced above to compute a suspectation value for Tara's (counterfactual) homodyne measurement record. This is the same calculation as we did in Ref.~\cite{banerjee2025counterfactualquantummeasurements}. The difference here is that we consider the more generic situation discussed in  \cref{BPC},  in which the two observers are no longer in strict causal isolation from each other. 
The modified causal structure changes the set of variables that enter into the counterfactual calculation. In particular, part of Vlad’s outcomes from the beginning of the experiment becomes causally accessible to Tara due to this modification. 

Let us denote by $\tau_V$ the duration of Vlad's measurements, before which the outcomes become causally accessible to Tara before she poses her counterfactual question (at the end of one run of the experiment).
Additionally, an equal-duration (for simplicity) segment from the end of Vlad’s experiment---comprising both settings and outcomes---lies within the future light cone of Tara’s counterfactual setting (the antecedent).  Consequently, Vlad’s timeline can be divided into three disjoint intervals, $[t_i, t_i^+)$, $[t_i^+,t_f^-)$, and $[t_f^-, t_f)$, for the initial, middle, and final segments respectively, 
where $t_i^+ = t_i+\tau_V$ and $t_f^-=t_f-\tau_V$.
 This partition naturally defines the corresponding disjoint subsets of Vlad’s settings,
\begin{equation}
     \both{Z}^V \supset\,
    \begin{cases}
        \past{Z}_{t_i^+}^V & \text{in the initial segment}\,\,\,[t_i, t_i^+)\\
        \fut{Z}_{t_i^+}^V \cap \past{Z}_{t_f^-}^V & \text{in the middle segment}\,\,\,[ t_i^+,t_f^-)\\
        \fut{Z}_{t_f^-}^V\blk & \text{in the final segment}\,\,\,[t_f^-, t_f),
    \end{cases}
\end{equation}
and similarly for the corresponding outcomes,
\begin{equation}
    \both{O}^V \supset\,
    \begin{cases}
        \past{O}_{t_i^+}^V \blk & \text{in the initial segment}\,\,\,[t_i, t_i^+)\\
         \fut{O}_{t_i^+}^V \cap \past{O}_{t_f^-}^V & \text{in the middle segment}\,\,\,[ t_i^+,t_f^-)\\
         \fut{O}_{t_f^-}^V\blk & \text{in the final segment}\,\,\,[t_f^-, t_f).
    \end{cases}
\end{equation}

The crucial observation is that the modified causal structure restricts which variables can be included in the counterfactual element sets. Specifically, the set of available evidence is modified to
\begin{equation}
    \bm{E}= (\both{O}^T \union \both{Z}^T \union \past{O}_{t_i^+}^V\union \both{Z}^V ), %\!\setminus Z_{[t_i,t_f-t_V)}^V) ,_{t_f^-}
\end{equation} 
where $\past{O}_{t_i^+}^V$ is the subset of Vlad's initial portion of the outcomes that is now known to Tara. %Note that $Z^T$ is a single element set and hence not bolded.
The inclusive descendants of the locus of change, $\both{Z}^T$, become, 
\begin{equation}
    \bm{D}=(\both{Z}^T\union \both{O}^T\union \fut{O}^V_{t_f^-}\union \fut{Z}^V_{t_f^-}),
\end{equation}  Consequently, the fixtures set is given by 
\begin{equation}
    \bm{F}=( \past{Z}_{t_f^-}^V \union\; \past{O}_{t_f^-}^V) .
\end{equation} 
Now we are ready to update Tara’s counterfactual question in the generic scenario.

%\hmw{Delete the rest of this sentence. Tara, in the real world, does know what Vlad does for the whole time of the measurement, if she knows the HSM  in Fig.~7.7, which is our usual assumption.  It is just that the last part is not relevant any more for fixtures. So I think Figure 7.7 should have the the blue circle still for Vlad's last measurement. Are you keeping track of these changes which will be necessary for the paper too? Copy my comments and paste them in a file.} 
We now consider the same physical problem as in \cite{banerjee2025counterfactualquantummeasurements} with the modification as specified in \cref{BPC}. Tara and Vlad perform continuous measurements on a shared open quantum system using photon detection (\cref{sec:System}). Tara is interested in the following counterfactual question:``Given that Vlad and I are both using photon counters, and I detected the photon click record $\both{O}^T=\overleftrightarrow{N}$ in interval $[t_i,t_f)$, and I know Vlad's outcomes $\past{O}^V_{t_i^+}$ in the sub-interval $[t_i,t_i^+)$, what is the suspectation value for  $\both{O}^{'T}\!=\overleftrightarrow{Y}$ in the interval $[t_i,t_f)$, had I used a Y-homodyne detector instead, also causing Vlad to alter his monitoring beyond $t_f^-$ ?'' %Note that the interval $[t_i,t_i^+)$ is the initial segment of Vlad's outcomes that is now causally accessible to Tara.

In this scenario, Vlad’s measurement settings take the time-dependent form
\begin{equation}Z^V_t=\begin{cases}
  \text{Photo-detection}   & t\in [t_i,t_f^-)\\
Z^T& t\in [t_f^-,t_f),
\end{cases}
\end{equation} 
where we use $Z^T$ because, in the above scenario, $Z^T_t$ does not depend on $t$ (though it does depend on whether we are considering the actual or counterfactual world). 
 Note that while this seems trivial in the actual scenario where both Tara and Vlad perform photon detections, it becomes non-trivial in the counterfactual scenario where Tara's setting changes to homodyne detections.

In the actual world, both Tara and Vlad monitor the fluorescence through photon counting, yielding records 
$\both{O}^T=\overleftrightarrow{N}$ and 
$\both{O}^V=\overleftrightarrow{M}$ respectively. 
Conditioned on her past record $\past{N}_t$, Tara calculates her filtered state using the map \erf{TFS}. For the specific model of \cref{sec:System}, the conditional evolution of the (unnormalized) state takes the explicit form~\cite{wiseman2009quantum}
\begin{equation}
\tilde{\rho}^{\past{N}}_{t+\dd t}={\cal T}({\dd N^T_t};{\rm pc})\,{\cal U}\,
{\cal V}({\dd \nullset};{{\rm nm}})\tilde{\rho}^{\past{N}}_{t}=
\begin{cases}
     \gamma\eta_T \dd t \mathcal{J}[\hat{\sigma}_-]\tilde{\rho}^{\past{N}}_t  \text{ if her detector clicks},\\ \{1+(\mathcal{L}-\gamma\eta_T\mathcal{J}[\hat{\sigma}_-])\dd t\} \tilde{\rho}^{\past{N}}_t  \text{ otherwise},
\end{cases}     \label{jumptraj}
\end{equation}
where $\tilde{\rho}$ denotes the unnormalized conditional state, whose norm gives the probability for the observed record.

As illustrated in Fig.~\ref{fig:Generic}, Tara considers the counterfactual scenario in which she instead performs $Y$-homodyne detection, producing a continuous record $\both{O}^{'T}=\both{Y}$. 
The analogous 
conditioning equation to \erf{jumptraj} can be obtained via the map ${\cal T}({\dd Y_t};{\rm hd})\,{\cal U}\,
{\cal V}({\dd \nullset};{{\rm nm}})\tilde{\rho}_t^{\overleftarrow{Y}}$, and the infinitesimal change due to that is~\cite{wiseman2009quantum}
\begin{equation}
     \dd \tilde{\rho}_t^{\overleftarrow{Y}}=\dd t\mathcal{L}\tilde{\rho}^{\overleftarrow{Y}}_t+\sqrt{\eta_T\gamma}\;Y_t \dd t\,\overline{\cal H}[i\hat{\sigma}_{-}]\tilde{\rho}^{\overleftarrow{Y}}_t, \label{diffusion}
\end{equation}
where $\overline{\cal H}$ is the superoperator in~\erf{eq:linear H}.
Normalizing at each time step yields the standard stochastic master equation
\begin{equation}
    \dd \rho^{\past{Y}}_t
    =
    \dd t\,\mathcal{L}\rho^{\past{Y}}_t
    +
    \dd W_t\,\mathcal H[i\hat{\sigma}_{-}]
    \rho^{\past{Y}}_t ,
\label{NormYequation}
\end{equation}
where $\mathcal H[\hat{w}]\bullet= \hat{w} \bullet  + \bullet \hat{w}^\dagger-\Tr(\hat{w} \bullet  + \bullet \hat{w}^\dagger)\bullet$ is the nonlinear (traceless) version of $\overline{\cal H}$ and $\dd W_t$ is Wiener noise, which is related to the photocurrent (with a convenient choice of normalisation) by
\begin{equation}
    Y_t=\text{Tr}[\rho^{\overleftarrow{Y}}_t\hat{\sigma}_y]+\frac{1}{\sqrt{\eta_T\gamma}}\frac{\dd W_t}{\dd t}. \label{homodynecurrent}
\end{equation}

%\hmw{Then explain how to calculacte the suspectation value for the homodyne current. i.e. generalised weak value. Briefly talk about how we generated the ensemble ${\mathfrak E}$ in previous paper, and if possible refer back to a smoothed probability distribution for one record given another in the smoothing section.} 
In the scenario of this section, the consequent is just Tara's counterfactual homodyne current record, $\boldsymbol{C}'\equiv\overleftrightarrow{Y}$. As in our previous paper~\cite{banerjee2025counterfactualquantummeasurements}, rather than analyzing the full supposability distribution, we focus on a simple statistic for it: the {\em suspectation} of $Y_t$ at each $t\in[t_i,t_f)$.
Using the notation similar to \erf{Supposability}, the suspectation value is given by
\begin{align}
&\mathbb{S}[Y_t|\boldsymbol{A}'=\boldsymbol{a}'||\both{N}=\both{n}] \
\coloneqq \int \dd y \, 
y \,  \dbd{y}\Su[Y_t \leq  
%\Su[Y_t \geq
y|\boldsymbol{A}'=\boldsymbol{a}'||\both{N}=\both{n}] \label{defsuspect}
\end{align}
where $\boldsymbol{a}'=$ Y-homodyne as before. 
By \erf{Supposability}, this equals
\begin{align}
\sum_{\both{m}} \mathbb{E}[Y_t|\boldsymbol{A}'=\boldsymbol{a}',\both{M}=\both{m}]
 \Pr(\both{M}=\both{m}|\both{N}=\both{n}), \label{twotermsinsuspect}
\end{align} 
which makes explicit the decomposition into a conditional expectation value and a smoothed probability distribution. The former
can be viewed as a weak-value-like estimate of the homodyne signal conditioned on both prior and posterior measurement information. Unlike conventional weak measurements~\cite{PhysRevLett.60.1351}, the atom--field interaction responsible for the homodyne signal is dissipative, resulting in additional measurement back-action. Consequently, the conditioned mean current is determined by a generalized weak value~\cite{chantasri2021unifying} involving the atomic lowering operator $-i\s{-}$ rather than solely the measured field quadrature, as~\cite{banerjee2025counterfactualquantummeasurements}
\beq 
\mathbb{E}[\text{Y}_t|\overleftarrow{M}=\past{m},\overrightarrow{M}=\fut{m}] = {\rm Re}\frac{\Tr[\hat E^{\overrightarrow{m}}_t (i\hat{\sigma}_{-})\rho^{\overleftarrow{m}}_t ]}{\Tr[\hat E^{\overrightarrow{m}}_t \rho^{\overleftarrow{m}}_t]} . \label{eq:WV}
\eeq
As for the distribution $\Pr(\both{M}=\both{m}|\both{N}=\both{n})$, this is precisely the smoothed probability for a single record of Vlad, conditioned on both the past and future record of Tara, as introduced in \cref{Q-Smooth}.

The averaging in~\erf{twotermsinsuspect} is over 
an ensemble of $\both{m}$ with the appropriate probability (Pr).
In our previous work \cite{banerjee2025counterfactualquantummeasurements}, this ensemble, denoted by ${\mathfrak E}$, was generated numerically by simulating large collections of compatible joint trajectories and retaining those consistent with the observed record of Tara. The present calculation follows the same logic and the suspectation value therefore emerges as a time-resolved generalized weak value~\cite{PhysRevA.65.032111,chantasri2021unifying}, combining forward (filtered) and backward (retrofiltered) state information in the sense of quantum state smoothing~\cite{PhysRevLett.115.180407,chantasri2021unifying}.
We can numerically evaluate \erf{eq:WV} (See~\ref{App:A} for details) using 
the probability-normalized conditional state evolution introduced above, by adopting techniques from quantum state smoothing~\cite{PhysRevLett.115.180407,chantasri2021unifying} and generalized weak value theory~\cite{PhysRevA.65.032111,chantasri2021unifying}. We consider the case where Tara's actual record is a single click in her detector, at time $t_T=6.25 \gamma\inv$. We notate this by $\both{N}=\delta_{t,6.25}$, where we set $\gamma=1$ in all such subscripts. This is a typical record for the parameters we choose, $\eta_T=0.2$, and $\Omega=2\gamma\inv$, for an interval $[0,10\gamma\inv)$. The time $t_i^+=1.2\gamma\inv$ denotes the portion of the experiment for which Vlad’s measurement outcomes are causally accessible to Tara, owing to the more general spacetime configuration illustrated in Fig.~\ref{fig:Generic}. For simplicity we further assume that in the interval $[t_i=0,t_i^+=1.2\gamma\inv)$, Vlad does not detect any photons in his detector such that $\past{M}_{t_i^+}=0$. Under these assumptions, \erf{defsuspect} can be rewritten as
\begin{equation}
    \mathbb{S}[Y_t|Y\text{-homodyne}||\both{N}=\delta_{t,6.25},\past{M}_{1.2}=0]\label{Eq:GenericSus}
\end{equation}

%\hmw{Give result. On same plot give the previous result where V and T were fully SLS}. 

\begin{figure}[H]
    \centering
    \includegraphics[width=0.50\textwidth]{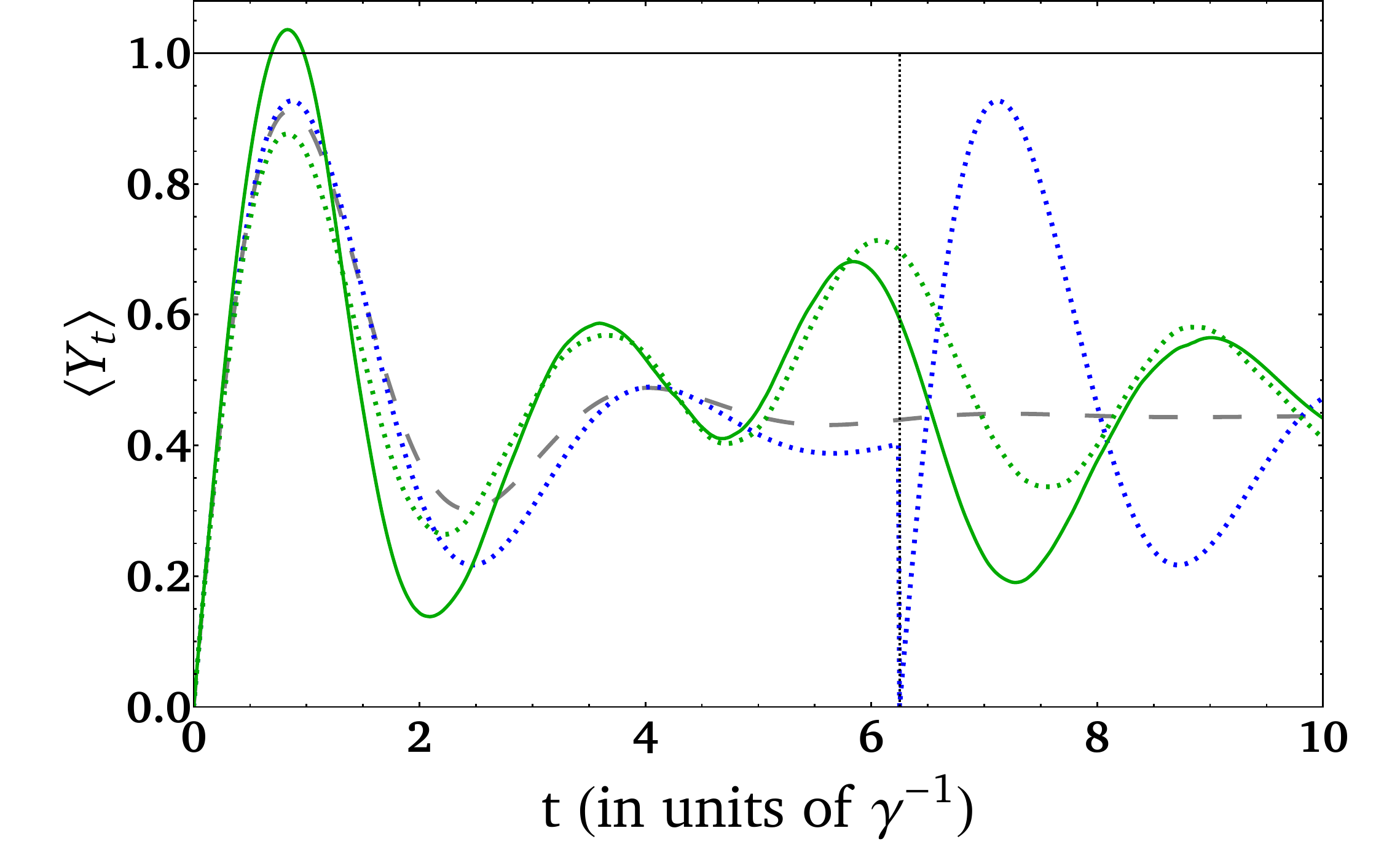}
    \caption{Dynamics of various estimates by Tara. The expectation of $\s{y}$ is shown for the unconditioned atomic state (gray dashed line) and Tara's filtered state conditioned on her actual record (a photo-detection at $t_T=6.25\gamma\inv$) (blue dotted line). 
    The gray dashed line is also the unconditioned expectation ${\mathbb E}[Y_t]$ of Tara's Y-homodyne current \erf{homodynecurrent}. The green curves are the suspectations ${\mathbb S}[Y_t]$ for \erf{homodynecurrent}, conditioned on her actual record $\overleftrightarrow{n}$. The green (dotted) curve is for the case of strict SLS, as in Fig.~\ref{ST-Cont} while the green solid curve is the suspectation as in \erf{Eq:GenericSus} illustrated in Fig~\ref{fig:Generic}, for the more generic case. 
    }
    \label{fig:Res-Generic}
\end{figure} 
%\hmw{Discuss.}\ktl{Up to here}

The results are shown in Fig.~\ref{fig:Res-Generic}; details of calculation of the results can be found in~\ref{App:A}. The gray curve shows the unconditional evolution of $\Tr[\rho_t \s{y}]$, where $\rho_t$ obeys the Lindblad master equation \erf{me} with the atom initially in the ground state. Via \erf{homodynecurrent}, this quantity is also the mean homodyne photocurrent that would be observed, at any given time, if homodyne detection were performed. The blue curve corresponds to the same quantity evaluated for Tara’s filtered state $\rho_t^{\past{n}}$, which evolves according to the photon-counting quantum trajectory \erf{jumptraj} for the specific record containing a single click at time $t_T = 6.25\gamma^{-1}$. At this time the atomic state undergoes a quantum jump back to the ground state, leading to $\Tr[\rho_t\hat{\sigma}_y]=0$. No homodyne measurement is actually carried out in this scenario.
The green curves represent the suspectation of Tara's counterfactual homodyne current conditioned on the observed (actual) photon count record. The dotted green line shows Tara’s suspectation in the strict SLS setting of Ref.~\cite{banerjee2025counterfactualquantummeasurements}, while the solid green line shows Tara’s suspectation in the generic setting we are considering here, where causal influence between observers breaks the party correspondence for the relevant counterfactual elements. 

In both the cases, the suspected value exhibits a pronounced peak in the vicinity of the actual jump time, with oscillatory “echoes” at the Rabi frequency $\Omega$, appearing before and after this event. This behavior reflects the strong temporal correlations imposed by the photon-count record on the set of compatible counterfactual homodyne trajectories.
In the generic setting, an additional feature is observed: at early times, the suspected homodyne current can transiently exceed the eigenvalue range associated with $\hat{\sigma}_y$. This is a characteristic signature of weak-value-like behavior arising from conditioning on both pre- and post-selection. In this case the pre-selection on the ensemble %$\mathfrak{E}$ \how{I don't think this has been introduced yet.} 
of Vlad's records 
is stronger because we retain only those trajectories for which no photon detections are recorded by Vlad in the interval $[0,1.2\gamma^{-1})$. %The resulting suspectation is therefore not constrained to lie within the operator spectrum, even though it remains fully consistent with the underlying quantum dynamics. 
While there is no reason this characteristic feature of weak values could not have been seen in the strict SLS scenario, for the parameters chosen it appears only in the generic scenario. The peak around $t_T$ in the generic scenario appears earlier than in the strict SLS case. This shift may arise from the absence of Vlad’s late-time fixed jumps, which would otherwise constrain the weak value; as a result, greater weight is effectively assigned to Vlad’s earlier fixed jumps.

\section{Suspectation values for interesting summary statistics} \label{sec:issi}

%\hmw{Discuss which earlier subsection we are returning to. (4.2) Remind reader about ``Interesting Summary Statistics'', and Godot. Say that although we are going to concentrate on estimating Tara's CF filtered state, for the convenience of presentation we will also discuss Godot's state here.}\ing{Done}
We now return to the idea introduced in \cref{sec:4(e)}, where we first considered suspectations of `interesting summary statistics' that go beyond the bare measurement record itself, and where Godot was introduced as a third party with access to the complete measurement record of both observers. In this section we develop this idea further, focusing our attention primarily on Tara's counterfactually filtered state, which constitutes the central case of interest for the remainder of this work. However, for convenience of presentation, and because the construction is closely related, we will also discuss Godot's fully conditioned state here rather than deferring it to a separate treatment.

\subsection{Various counterfactual  estimates \label{sec:states}}
%\hmw{This paragraph does not seem to belong here. It seems to belong in Sec. 4. Indeed, it refers to something that will ``become more explicit'' in a Figure already presented in Sec. 4. Either move there, or in Sec. 8 or rewrite. Or some combination of those!} \ing{Moved to section 8}
% Recall that in the counterfactual calculus introduced in~\cref{The calculus}, the consequent was defined as a set of elements belonging to the counterfactual descendants $\bm{D'}$. However, Vlad's filtered state does not belong to this set, whereas Godot's state does. This is because Godot's state is constructed using readouts from Tara's measurement outcomes, in addition to Vlad's outcomes, which are causal descendants of Tara's setting choice --- the locus of change for the counterfactual. This will become more explicit in the L-DAG representation in Fig.~\ref{Godot-DAG}. Note, however, that Vlad's filtered state coincides with Godot's filtered state when there is no information from Tara, \eg, when her counterfactual setting corresponds to no measurement. 

To characterise the different suspected states available to Tara, we first introduce a unified expression within the counterfactual setting calculus and then specify its explicit forms for the cases of interest. We denote suspected counterfactual estimates by $\varrho^{S}(\cdot||\cdot)$, distinguishing them from standard quantum states $\rho$, where the conditioning denotes that defined in~\cref{The calculus}, 
\begin{align} \label{generalsuspectedCFfilteredrho}
\varrho^S_t({Z'}^T||\overleftrightarrow{O}^T) 
&= \mathbb{S}\left[\rho^S_t(\overleftarrow{O}'{}^T,\overleftarrow{O}^{V})|{Z'}^T||\overleftrightarrow{O}^T \right].
\end{align}
Here $S=T$ or $G$, corresponding respectively to Tara's or Godot's filtered states estimated by Tara via the counterfactual calculus. These states depend on $\overleftarrow{O}'{}^T$ and, for $S=G$, also on $\overleftarrow{O}^{V}$. While Vlad's results are fixed (and hence not in the consequent), Godot's calculated state is in the consequent and can therefore be assigned a suspected value within the formalism of~\cref{The calculus}. We suppress the dependence on the settings, ${Z}^T$ and ${Z}^V$, part of the evidence which remain unchanged. Note also that, as above, we suppress any temporal indications in the notation for the settings because they are time-independent in the scenario we are considering. Finally, we replace statistical events such as $\bm{E}=\bm{e}$ with the random variables $\bm{E}$ themselves for brevity.

The three relevant cases, presented together for ease of comparison, are:
\begin{enumerate}
    \item Tara's suspected estimate of her own counterfactual filtered state:
    \begin{equation}
        \varrho_t^T({Z'}^T||\overleftrightarrow{O}^T)\!=\!\!\sum_{\overleftarrow{O'}^T,\overleftrightarrow{O}^V}\!\!\Pr(\overleftarrow{O'}^T|\overleftrightarrow{O}^V) \Pr(\overleftrightarrow{O}^V|\overleftrightarrow{O}^T)\rho_t^{\overleftarrow{O'}^T}. 
        \label{TCF-S}
    \end{equation}

    \item Tara's suspected estimate of Vlad's filtered state: in the special case where the antecedent corresponds to no measurement on Tara's side, Vlad's filtered state coincides with Godot's filtered state,
    \begin{equation}
        {\varrho}^G_t({Z'}^T\!\!=\nullset\,||\overleftrightarrow{O}^T)=
        \sum_{\overleftarrow{O}^V}\Pr(\overleftarrow{O}^V|\overleftrightarrow{O}^T)
        \rho_t^{\overleftarrow{O}^V}.
        \label{VS}
    \end{equation} 

    \item Tara's suspected estimate of Godot's counterfactual filtered state:
    \begin{equation}
        {\varrho}^G_t({Z'}^T||\overleftrightarrow{O}^T)=
        \sum_{\overleftarrow{O'}^T,\overleftrightarrow{O}^V} 
        \Pr(\overleftarrow{O'}^T|\overleftrightarrow{O}^V)
        \Pr(\overleftrightarrow{O}^V|\overleftrightarrow{O}^T)
        \rho_t^{\overleftarrow{O}^V,\overleftarrow{O'}^T}.   
        \label{GS}
    \end{equation}
\end{enumerate}

In this section we are concerned only with case (i). We will return to cases (ii) and (iii) in \cref{sec:3party}.

% In the next section we evaluate case 1, the suspectation of Tara's counterfactual filtered state. The remaining two cases, involving Godot's filtered state, are treated afterwards. Although conceptually more involved, those latter cases are computationally simpler than case 1.

\subsection{Computing case (i)\label{sec:QTestimates}} 

%In~\cref{sec:4(e)} and~\cref{sec:states}, we discussed the evaluation of .
In this subsection we show explicitly how the counterfactual calculus is used to calculate Tara's suspected estimate of her counterfactual filtered state [\erf{TCF-S}] under the same physical setting as in \cref{sec:System}.
To formalize the problem, we define Tara's accessible variables as the evidence
\begin{equation}
    \bm{E}= (\both{O}^T;{Z^T};{Z^V}),
\end{equation}
with known (to Tara) values
\begin{equation}
    \bm{e} = (\both{n};{\rm pd};{\rm pd}),
\end{equation}
where $\both{O}^T=\both{n}$ is Tara's actual photon-detection record, and ${Z^T},{Z^V}$ are the actual measurement settings for Tara and Vlad respectively. The antecedent corresponds to Tara instead performing homodyne detection, ${Z'}^T={\rm hd}$. Vlad's outcomes $\both{m}$ act as fixtures, while Tara's counterfactual homodyne outcomes $\both{O'}^T$ are denoted by $\past{Y}$.

The quantity of interest is Tara's suspectation of the counterfactual filtered state $\rho^{\past{O}^{'T}}_t$ of the system at times $t$ given by (from~\erf{generalsuspectedCFfilteredrho} and \erf{TCF-S})
\begin{align}
   \varrho_t^T({\rm hd}||\both{n})
   = \mathbb{S}\left[\rho^F_t(\past{Y},\emptyset)\Big|{\past{Z}^T}'={\rm hd}\Big|\Big|\bm{E}=\bm{e}\right]\
   = \sum_{\past{Y},\both{m}} 
   \wp(\past{Y}|\both{m})
   \wp(\both{m}|\both{n})
   \rho_t^{\past{Y}} .
   \label{eq:TCFS}
\end{align}
Here $\rho_t^F(\bullet)$ denotes the filtered state conditioned on the past records appearing in its argument. The average over the unknown fixtures, $\sum_{\both{m}} \wp(\both{m}|\both{n})\bullet\,$, 
corresponds to averaging over the ensemble $\mathfrak{E}$ introduced in~\cref{sec:BreakParty}~\cite{banerjee2025counterfactualquantummeasurements}. For each sampled Vlad record $\both{m}$, an ensemble of homodyne trajectories $\past{Y}$ is generated using an appropriate statistics, allowing evaluation of both $\rho_t^{\past{Y}}$ and $\wp(\past{Y}|\both{m})$ (see~\ref{App:B} for details).
As evidence, we again consider a typical Tara record $\both{O}^T=\both{n}$ containing a single click at time $t_T=6.25\gamma^{-1}$. The results are presented through the Bloch vector components,
\begin{equation}
    \langle\hat{\sigma}_k\rangle
    =
    \Tr[\varrho_t^T({\rm hd}||\both{n})\hat{\sigma}_k].
\end{equation}

Note that in the plots we use $\an{\sigma_y}$ and $\an{\sigma_z}$ as the two Bloch-vector components shown on the $y$-axis, which we denote $\langle Y_t\rangle$ and $\langle Z_t\rangle$ respectively.
%\begin{align}
%\langle Y_t\rangle &\equiv \an{\sigma_y}, \\
%\langle Z_t\rangle &\equiv \an{\sigma_z}. \label{eq:YZdef}
%\end{align}
The third component, $X_t \equiv \an{\sigma_x}$, remains identically zero throughout the evolution, since the atomic dynamics is confined to the $Y$--$Z$ plane of the Bloch sphere.

\subsection{Results and analysis}
\begin{figure}[H]
    \centering
    \begin{subfigure}{0.4958\textwidth}
        \centering
        \includegraphics[width=\linewidth]{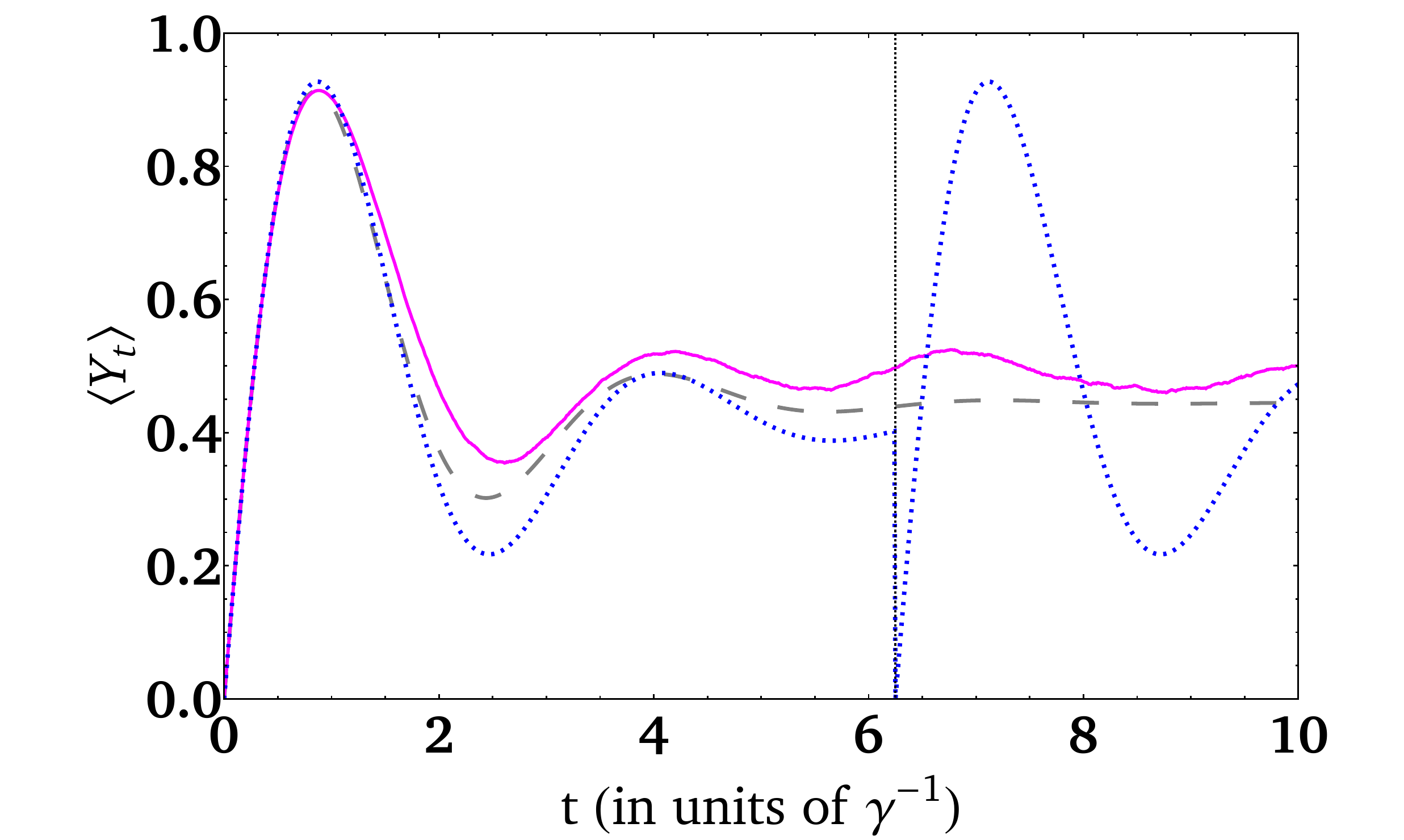}
        \caption{}
        \label{fig:ACF-Y}
    \end{subfigure}
    \hfill
    \begin{subfigure}{0.4958\textwidth}
        \centering
        \includegraphics[width=\linewidth]{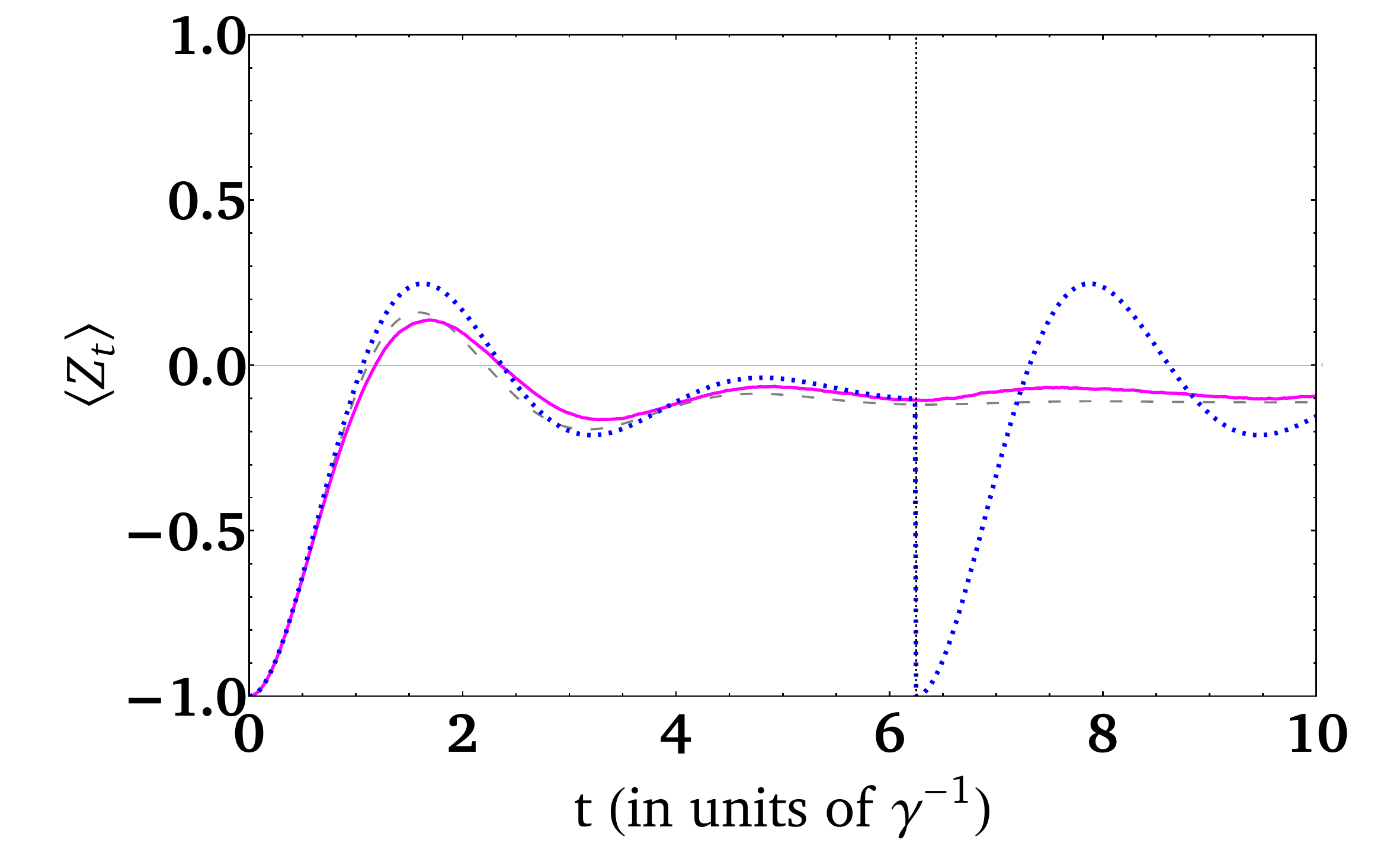}
        \caption{}
        \label{fig:ACF-Z}
    \end{subfigure}
    \caption[Results for Tara's CF-filtered state]{The suspectations for Tara's counterfactual filtered state (magenta solid) compared with Tara's actual filtered state (blue dotted) curve. The unconditioned evolution (gray dashed) has been plotted for an overall comparison with the conditioned curves.}
\end{figure}

The results are shown in Fig.~\ref{fig:ACF-Y} and Fig.~\ref{fig:ACF-Z}. For the scenarios considered here, the expectation value of $\hat{\sigma}_x$ is always zero. 
The suspected counterfactual filtered state $\varrho_t^T({\rm hd}||\overleftrightarrow{n})$ (magenta) follows the master equation solution (gray dashed) more closely than the actual filtered trajectory (blue dotted), $\rho_t^{\overleftrightarrow{n}}$. At the detection time, the sharp jump in the actual trajectory is replaced in the counterfactual case by a much weaker feature in $\langle \hat{\sigma}_z\rangle$, as seen more clearly in Fig.~\ref{fig:ACF-Z}. This is consistent with the discussion in~\cite{banerjee2025counterfactualquantummeasurements}, where the counterfactual conditioned atomic state was shown to remain closer to the ground state. Following the detection event, the counterfactual trajectory exhibits damped oscillations absent in the unconditioned evolution. These arise from indirect conditioning via Vlad's records on the actual detection events, which provides additional information to the estimate. The comparatively weak features of the counterfactual filtered state are likely related to Tara's low detection efficiency ($\eta_T = 0.2$), which limits the information available from her measurement record alone. The deviation from the unconditioned state will be maximal at some intermediate value of $\eta_T<1$, but we did not determine that value\footnote{If Tara's efficiency were perfect, Vlad's records would provide no additional conditioning information in the counterfactual scenario. Conversely, for $\eta_T=0$, Tara would have no evidence from which to infer Vlad's fixed records. In both limits, the suspected estimate reduces to the master equation solution. Consequently, the largest deviation from the unconditioned evolution will be at an intermediate efficiency; its precise value will depend on the particular counterfactual question considered.}.

\begin{figure}[H]
    \centering
    \includegraphics[width=0.50\textwidth]{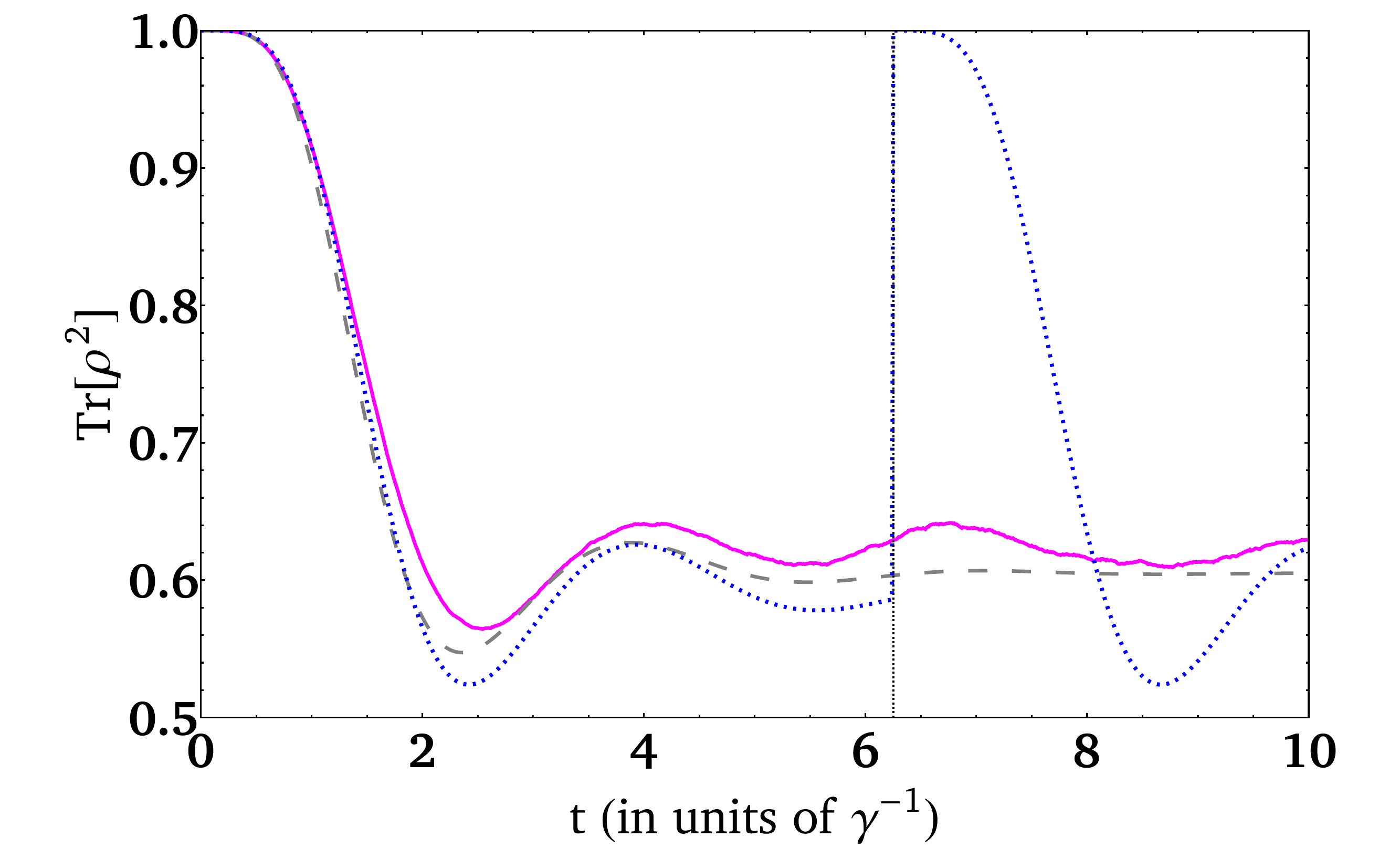}
    \caption{The purity of the Tara's suspected counterfactual filtered state (magenta solid), her actual filtered state (blue dotted) and the unconditioned state (gray dashed). The level of purity for the counterfactual filtered state remains higher than the actual filtered state at all times except for a period of about $2\gamma^{-1}$ after the jump time $t_T=6.25\gamma^{-1}$ (at which there is a black dotted vertical line).}
    \label{fig:ACF-Purity}
\end{figure}

Fig.~\ref{fig:ACF-Purity} shows the purity of the suspected counterfactual filtered state $\varrho_t^T({\rm hd}||\both{n})$ as a function of time. Interestingly, the purity is higher than that of Tara's actual filtered state $\rho_t^{\past{n}}$ for all of the time, except immediately after the quantum jump at $t_T=6.25\gamma^{-1}$, where the atom resets to the ground state and the purity of the actual filtered state briefly returns to one. Shortly afterwards, the purity again falls below that of the suspected state. The higher purity of the counterfactual state can be understood from the additional conditioning on Tara's future record implicit in the counterfactual construction. Although Tara does not have access to Vlad's records, each sampled homodyne trajectory is conditioned on a particular underlying emission history for Vlad, corresponding to a record in the ensemble $\mathfrak{E}$. Averaging over these trajectories conditioned on Tara's whole actual record reintroduces mixing, but not to the same extent as in Tara's actual filtered state which is inferred solely from her measurement record up to the time of interest. It is important to note, however, that the time-averaged purity of the actual filtered state remains larger due to the unit purity immediately following the jump.

\section{Suspectation values for interesting summary statistics that also break the party correspondence \label{sec:3party}}

Recall that, 
%in the counterfactual calculus introduced in~\cref{The calculus}, the consequent was defined as a set of elements belonging to the counterfactual descendants $\bm{D'}$.Following this, 
in \cref{sec:4(f)}, we introduced the third party, Godot who has access to all the measurement record generated by continuous monitoring of the same physical setting as in \cref{sec:System}, by both Tara and Vlad. In \cref{sec:states}, we introduced Tara's counterfactual estimate of Godot's filtered state, but we did not calculate it. In this section we turn to that calculation. 

Here calculating the suspected state is not only a case of a suspectation of a summary statistic of the records; it also breaks 
 the party correspondence 
 (see  \cref{sec:BreakParty}). 
 Recall from ~\cref{The calculus} that consequent was defined as a set of elements belonging to the counterfactual descendants $\bm{D'}$. 
Godot's filtered state is element of $\bm{D'}$ because Godot's state is constructed using readouts from Tara's measurement outcomes (See Fig.~\ref{Godot-DAG}). But, it is also constructed from Vlad's outcomes, and so Tara is not simply estimating a function of her own counterfactual results.

%which are causal descendants of Tara's setting choice --- the locus of change for the counterfactual (evident from Fig.~\ref{Godot-DAG}). 

As foreshadowed in \cref{sec:states}, 
we consider the suspectation Tara could make about Godot's filtered state, under two different counterfactuals. The first, \erf{VS}, is when Tara's counterfactual measurement is the null measurement. The second, \erf{GS}, is when Tara's counterfactual measurement is Y-homodyne, as usual. The evidence $\bm{E}$, locus of change $\bm{A}$, fixtures $\bm{F}$, and descendants $\bm{D}$ are the same as in the previous section. 

As noted in \cref{sec:states}, in the former case Vlad's filtered state coincides with Godot's filtered state when there is no information from Tara, \eg, when her counterfactual setting corresponds to no measurement. 
Also, recall that Vlad's measurement record is a fixture in the counterfactual~\cref{sec:System}, \ie, it is identical in the actual and counterfactual scenarios. Thus Tara's estimate also has an interpretation in the actual world: it is her {\em smoothed} estimate of Vlad's filtered state, since it uses both her past and future measurement records, $\overleftarrow{O}^T$ and $\overrightarrow{O}^T$~\cref{Q-Smooth}
\blk

We begin with the suspectation of Godot's filtered state in the case where Tara performs no measurement in the counterfactual scenario. The corresponding estimate, \erf{VS}, is evaluated by
\beq
\begin{split}
    {\varrho}^G_t({Z'}^T\!\!=\nullset\,||\both{n})
    &= \mathbb{S}\left[\rho^{G}_t(\past{M},\nullset)\Big|{\past{Z}^T}'\!=\nullset \Big|\Big|\bm{E}=\bm{e}\right]  
    \\
    & = \mathbb{S}\left[\rho^{F}_t(\past{M})\Big|\Big|\bm{E}=\bm{e}\right] \\
    &= \sum_{\past{m}}\wp(\past{m}|\both{n})\rho_t^{\past{m}}\,.
    \label{Eq:V-cFS}
\end{split}
\eeq
The average is again evaluated over the ensemble $\mathfrak{E}$, while the states $\rho_t^{\past{m}}$ are obtained from the filtered evolution conditioned on Vlad's record $\past{m}$. As stated above, ${\varrho}^G_t({Z'}^T\!\!=\nullset\,||\both{n})$ is a type of smoothed estimate involving only actual-world quantities. 
It is thus natural to compare it with Tara's actual smoothed state~\cref{Q-Smooth},
\begin{equation}
    \mathbb{E}\left[\rho^{F}_t(\past{M},\past{N})\Big|\bm{E}=\bm{e}\right]
    =
    \sum_{\past{m}}
    \wp(\past{m}|\both{n})
    \rho_t^{\past{m},\past{n}} .
    \label{eq:TSM}
\end{equation}
This is calculated using the same ensemble average over $\mathfrak{E}$, with the conditioned states $\rho_t^{\past{m},\past{n}}$ obtained from the corresponding forward conditioned evolution.

Next we consider the suspectation value of Godot's counterfactual filtered state when  the counterfactual antecedent is $Z^{'T} = {\rm hd}$. Here the state~\erf{GS} is evaluated as follows: 
\beq
\begin{split}
{\varrho}^G_t({\rm hd}||\both{n}) & =  \mathbb{S}\left[\rho_t^F(\past{M},\past{Y}{})\Big|{\past{Z}^T}{}'={\rm hd}\Big|\Big|\bm{E}=\bm{e}\right] \\
     &=\sum_{\past{Y}{},\both{m}} \wp(\past{Y}|\both{m})\wp(\both{m}|\both{n})\rho_t^{\past{m},\past{Y}} \\ 
     &=\sum_{\both{m}} \wp(\both{m}|\both{n}) \sum_{\past{Y}} \wp(\past{Y}|\both{m})\rho_t^{\past{m},\past{Y}}.  \label{eq:GS2}
\end{split}
\eeq
To compute this suspectation, Tara must average over two sets of outcomes: (i) Vlad's fixed photon-detection records $\both{O}^V=\both{M}$, and (ii) the counterfactual homodyne records $\past{O}^{'T}=\past{Y}$. Using her actual record $\past{n}$, corresponding to a single detection event at $t_T=6.25\gamma^{-1}$, Tara first generates the ensemble $\mathfrak{E}$ of hypothetical Vlad records $\both{M}$ as introduced in~\cref{sec:BreakParty}. For each sampled record, an ensemble of counterfactual homodyne trajectories $\past{Y}$ is then generated to evaluate the remaining conditional states and probabilities appearing in~\erf{eq:GS2}, $\sum_{\past{Y}} \wp(\past{Y}|\both{m})\rho_t^{\past{m},\past{Y}}$. 

It is worth noting that ${\varrho}^G_t({\rm hd}||\both{n})$ in \cref{eq:GS2} can also be given a real-world interpretation in terms of smoothed states. It is Tara's best estimate, using her whole record $\both{n}$ and only her record, of the smoothed state Vlad would calculate if he {\em wrongly} assumed that her detection method was homodyne rather than photon counting. The phenomenon of quantum state smoothing when one party makes such a wrong assumption was explored in Ref.~\cite{Chantasri_2025}.

Details of the numerical evaluation of all the states (including Tara's actual smoothed state) are given in~\ref{App:TaraSmoothing} to \ref{App:D}.

\subsection{Results and analysis}

The estimates in this section show significantly larger deviations from the unconditioned evolution than those considered previously, possibly reflecting the stronger conditioning provided by Vlad's higher-efficiency measurement record. Before Tara's detection event, both suspected estimates closely track Tara's actual smoothed state~\erf{eq:TSM}, while after the jump they deviate from it due to the absence of a corresponding jump in the counterfactual worlds.

\begin{figure}[H]
    \centering
    \begin{subfigure}{0.485\textwidth}
        \centering
        \includegraphics[width=\linewidth]{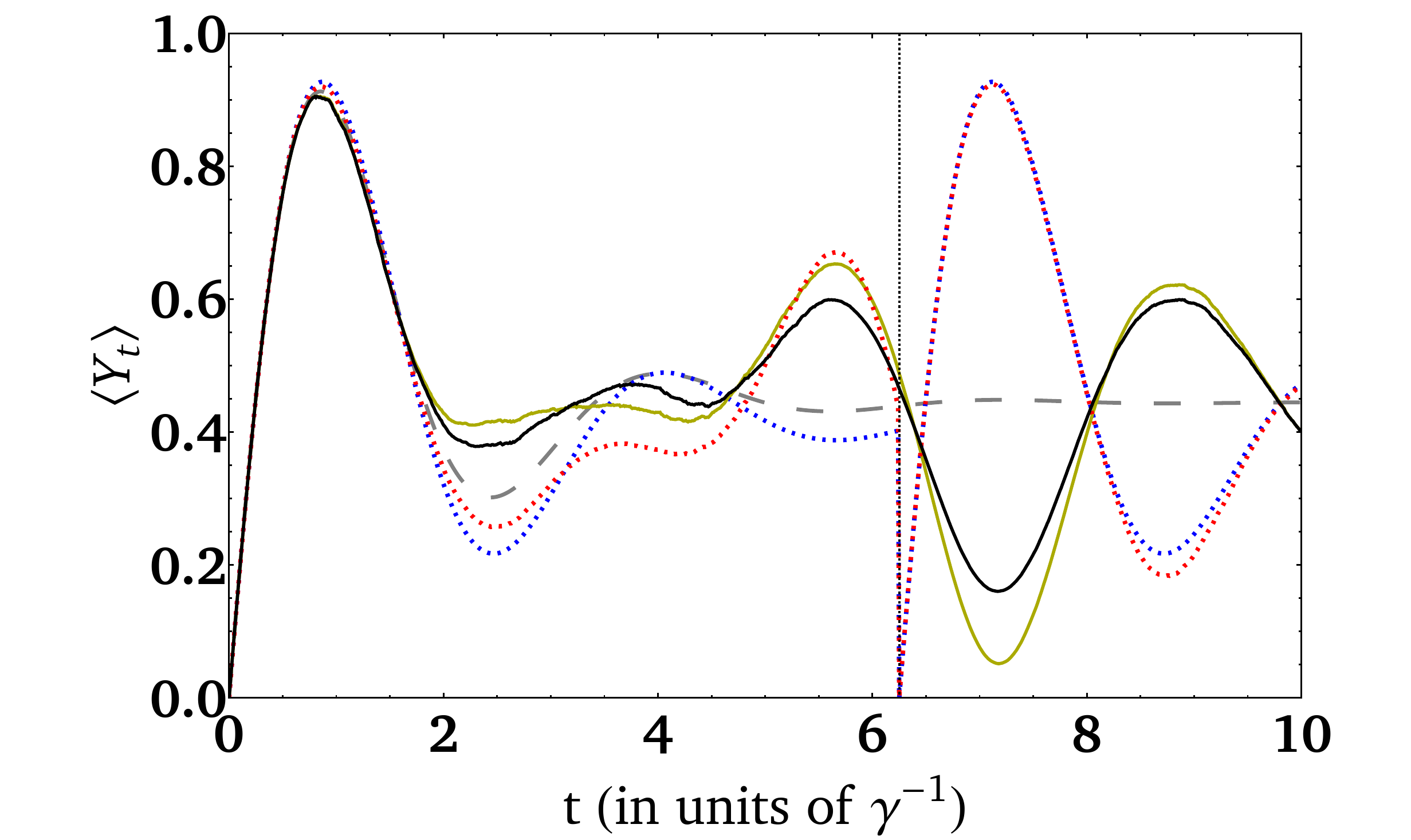}
        \caption{}
        \label{fig:GCF-Y}
    \end{subfigure}
    \hfill
    \begin{subfigure}{0.485\textwidth}
        \centering
        \includegraphics[width=\linewidth]{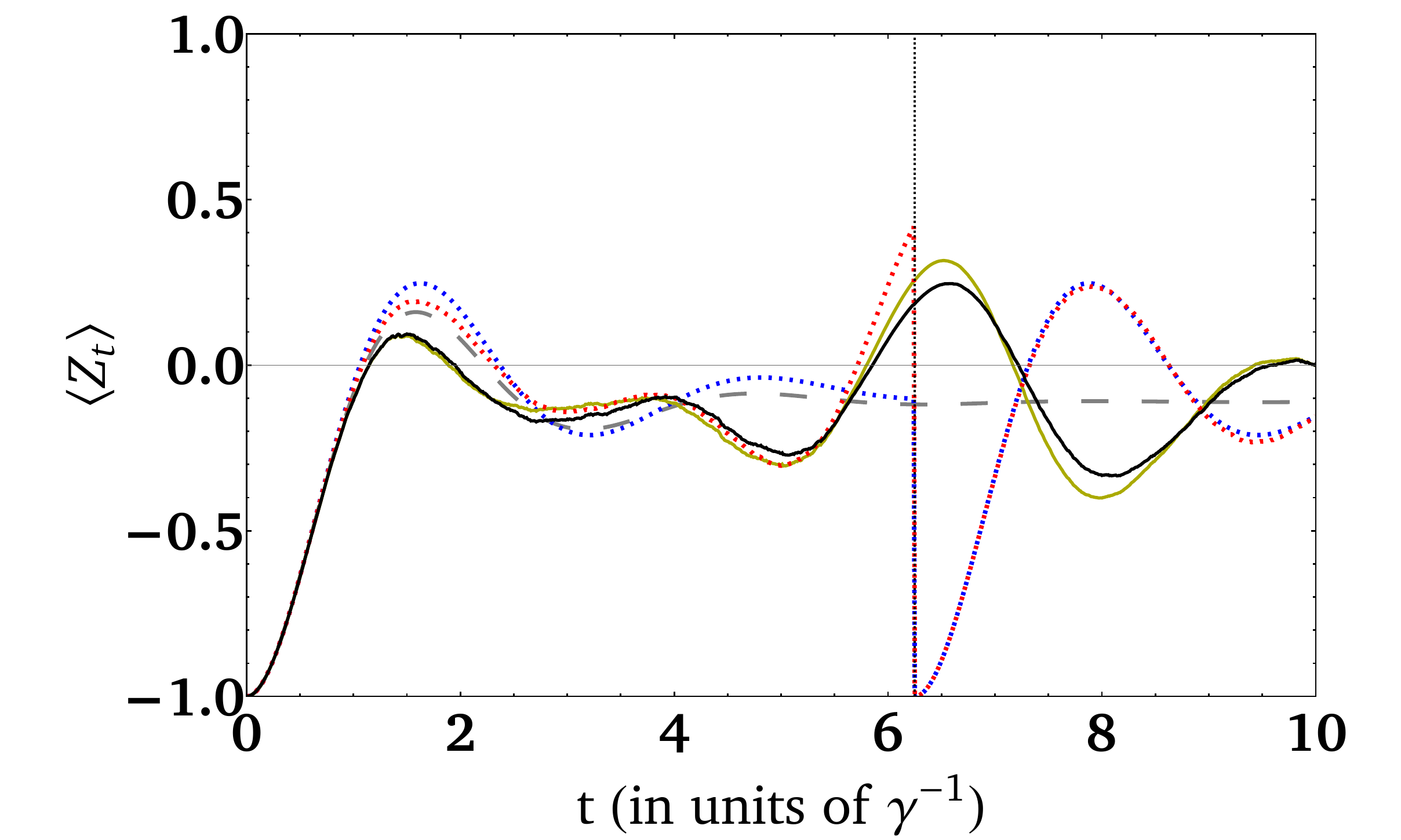}
        \caption{}
        \label{fig:GCF-Z}
    \end{subfigure}
    \caption[Results for Tara's suspected estimate of CF-state]{Tara's suspected estimate of Godot's counterfactual filtered state (black solid) and Vlad's filtered state (olive green) compared with Tara's actual smoothed state (red dotted) curve and Tara's actual filtered state (blue dotted) curve. The unconditioned evolution (gray dashed) has been plotted for an overall comparison with the conditioned curves.}
\end{figure}

We begin by comparing Tara's actual filtered and smoothed states. As seen in Figs~\ref{fig:GCF-Y} and \ref{fig:GCF-Z}, in the lead-up to the jump, the smoothed state (red dotted) is noticeably purer and biased towards the excited state relative to the filtered state (blue dotted), since it is additionally conditioned on future information indicating an impending detection event. After the jump, however, the two states nearly coincide because the future record contains little additional information. 

Now consider the suspected estimates, also shown in Fig.~\ref{fig:GCF-Y} and Fig.~\ref{fig:GCF-Z}. Following Tara's detection, the atom collapses to the ground state, suppressing nearby emission events in Vlad's detection record. Vlad's detections are therefore most likely approximately half a Rabi cycle before and after Tara's jump, where the excitation probability is maximal. Consequently, the averages over Vlad-conditioned trajectories produce dips in the suspected excitation of the states around these regions, while these smoothly peak near Tara's actual jump time where Vlad's jump probability is minimal.

This explanation is clarified by writing Tara's suspectation of Godot's counterfactual filtered state in the case $Z^{'T} = {\rm hd}$ as 
\begin{align}
    \sum_{\past{Y},\both{m}} 
    \wp(\both{m}|\both{n})
    \wp(\past{Y}|\both{m})
    \rho_t^{\past{m},\past{Y}}
    =
    \sum_{\both{m}}
    \wp(\both{m}|\both{n})
    \rho_t^{\both{m}({\rm CF})},
    \label{CF-sm}
\end{align}
where $\rho_t^{\both{m}({\rm CF})}$ denotes Vlad's hypothetical smoothed estimate in the counterfactual world (equivalently, his wrongly-assumed~\cite{Chantasri_2025} smoothed state in the actual world). The states in~\erf{eq:GS2} and~\erf{Eq:V-cFS} exhibit similar behaviour because both are averaged over the same ensemble $\mathfrak{E}$ and because, since Tara's efficiency is relatively low ($\eta_T=0.2$), the counterfactual smoothed $\rho_t^{\both{m}({\rm CF})}$ and filtered $\rho_t^{\past{m}}$ states  do not differ substantially. Both suspected estimates remain approximately in phase with Tara's actual smoothed state before the jump, but gradually drift out of phase afterwards because the continuous Rabi evolution in the counterfactual worlds is not interrupted by Tara's actual detection event.

\begin{figure}[H]
    \centering
    \includegraphics[width=0.50\textwidth]{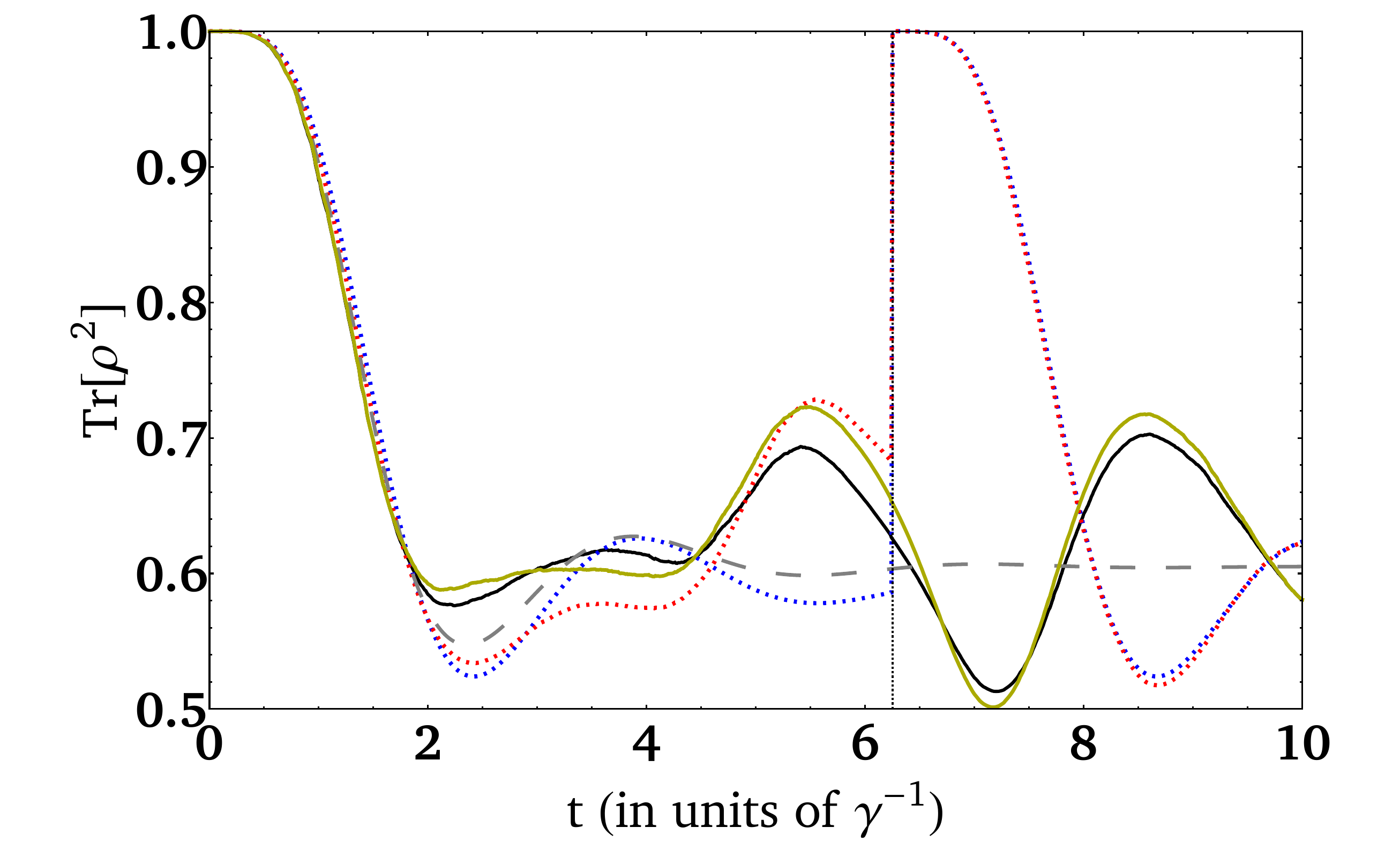}
    \caption[Purity of the CF states]{Purity $\mathrm{Tr}[\rho^2]$ of the counterfactual states. % obtained after averaging over Vlad’s hypothetical records. 
The black solid curve corresponds to the state constructed Tara's suspected estimate of Godot's counterfactual filtered state, while the olive solid curve corresponds to Tara's suspected estimate of Vlad’s filtered states. The states for comparison are Tara's actual smoothed (red dotted) and filtered (blue dotted) states. The gray dashed curve shows the unconditional (master equation) evolution, and the vertical dotted line marks Tara’s detection time at $t_T = 6.25\,\gamma^{-1}$.} 
%Despite individual smoothed trajectories being more pure, their ensemble-averaged state exhibits a lower purity.}
    \label{Purity after averaging}
\end{figure}

Fig.~\ref{Purity after averaging} compares the purities of the different state estimates. The red solid curve corresponds to Tara's actual smoothed state~\erf{eq:TSM}, the black curve to Tara's estimate of Godot's counterfactual filtered state, and the olive-green curve to Tara's estimate of Vlad's filtered state. Surprisingly, the estimate of Godot's filtered state does not, most of the time, exhibit a higher purity than that of Vlad's filtered state, despite the state to be estimated being conditioned on more information. We conclude this section by attempting to explain this counter-intuitive result.

Consider a state obtained by mixing over trajectories, 
\begin{equation}
\rho = \sum_{m} \wp(m|{n}) \, \rho^{m},
\end{equation}
where, to simplify notation, we have omitted the arrows since they do not affect the following analysis. Its purity is given by
\begin{equation}
\mathrm{Tr}[\rho^2] 
= \sum_{m} \wp(m|n)^2 \, \mathrm{Tr}\!\left[(\rho^{m})^2\right] 
+ \sum_{m \neq m'} \wp(m|n) \wp(m'|n) \, \mathrm{Tr}\!\left[\rho^{m} \rho^{m'}\right].
\end{equation}
The first term (diagonal contribution) depends on the individual purities of the states, while the second term (cross contribution) depends on their mutual overlaps.

For Vlad's counterfactual smoothed states, each $\rho_t^{\both{m}({\rm CF})}$ in~\erf{CF-sm} is conditioned on more information and is therefore more pure, so the diagonal term is larger. However, these states are also more distinguishable (as compared to the filtered states $\rho^{\past{m}}$ in~\erf{Eq:V-cFS}) from one another, implying smaller overlaps,
\begin{equation}
\mathrm{Tr}\!\left[\rho_t^{\both{m}({\rm CF})} \rho_t^{\both{m}'({\rm CF})}\right] \;\text{is smaller for } m \neq m',
\end{equation}
and hence a reduced cross-term contribution. In contrast, filtered states $\rho^{\past{m}}$ in~\erf{Eq:V-cFS} are less pure individually, but are more similar to each other, leading to larger overlaps and hence larger cross terms. Consequently, when averaging is performed at the level of density matrices, the reduced cross-term contribution for the smoothed states seems to outweigh their higher individual purities, resulting in a lower overall purity compared to the corresponding mixture of filtered states. This interpretation is consistent with the observed behaviour in Fig.~\ref{Purity after averaging}. 
%, although a deeper explanation for the reduced correlations between smoothed states remains an open question.

\begin{figure}[H]
    \centering
    \includegraphics[width=0.50\textwidth]{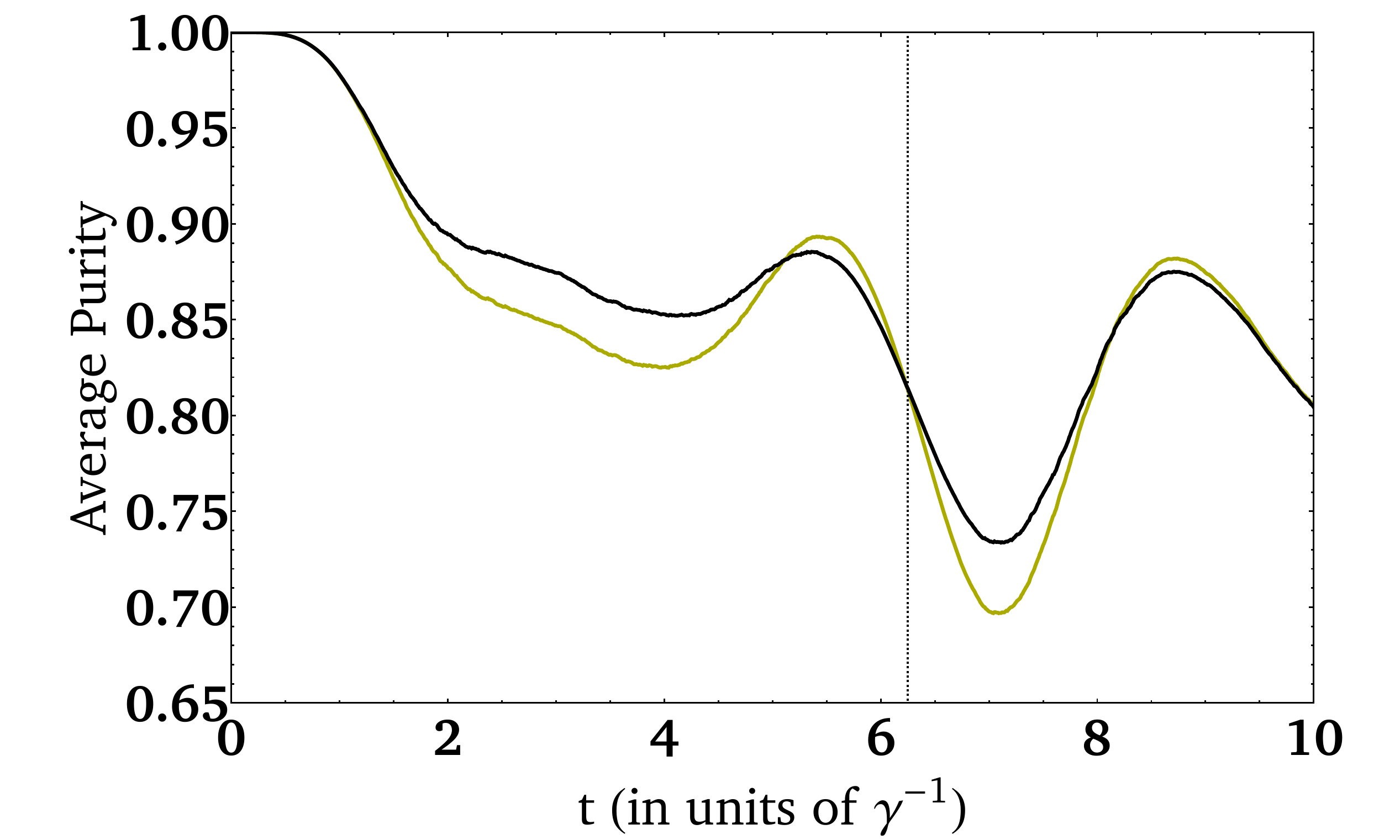}
    \caption[Average Purity of the CF states]{
Ensemble-averaged purity, $\mathbb{E}_{\mathfrak{E}}[\mathrm{Tr}(\rho^2)]$, computed by first evaluating the purity of individual trajectories and then averaging over Vlad’s hypothetical records. 
The black solid curve corresponds to suspected estimate of Godot's counterfactual states~\erf{eq:GS2}, while the olive solid curve corresponds to suspected estimate of Vlad's filtered states~\erf{Eq:V-cFS}. The vertical dotted line indicates Tara’s detection time at $t_T = 6.25\,\gamma^{-1}$. 
As expected, the smoothed trajectories are generally more pure than the filtered ones due to their additional conditioning on future information.} \label{fig:Avg Purity}
\end{figure}

To confirm that the purity of Vlad's smoothed states generally exceeds that of his filtered states, we calculated the ensemble average of the purities calculated at the level of individual trajectories. As we see in Fig.~\ref{fig:Avg Purity}, for the great majority of the time we have
\begin{equation}
    \sum_{\both{m}} \wp(\both{m}|\both{n}) \mathrm{Tr}[(\rho_t^{\both{m}({\rm CF})})^2]
>
\sum_{\past{m}} \wp(\past{m}|\both{n}) \mathrm{Tr}[(\rho^{\past{m}})^2].
\end{equation}

\section{Conclusion\label{sec:conc}}

\subsection{Summary}

In this work we have generalized, in three directions, the calculus of quantum counterfactuals that we introduced in Ref.~\cite{banerjee2025counterfactualquantummeasurements},
addressing each of the limitations of the examples presented there.

First, we have shown that the calculus does not rely on the space-time arrangement of events. By introducing Hybrid Structural causal Models (HSMs), together with the labeled directed acyclic graphs (L-DAGs) that represent them, we grounded the identification of the counterfactual element sets---evidence, antecedent, fixtures, and consequent---in an arbitrary acyclic causal structure. The HSM treats quantum and
classical information flow within a single object (a CPTP map at each node, constrained to dephase its classical outputs), and thereby naturally accommodates classical communication between parties and classical processing of measurement
records. This distinguishes it from the QSM of Suresh \ea~\cite{suresh2023semantics}, with which we made a detailed comparison, and the graphical conventions we introduced
allowed us to give explicit causal diagrams both for the scenarios of our earlier work and for the new ones treated here.

Second, we broke the party correspondence that characterized our previous examples, in which the evidence, antecedent, and consequent all pertained to one party while the fixtures pertained to the other. Allowing the observers to enter each other's future light cones near the end of the experiment modifies all of the counterfactual element sets: part of Vlad's record becomes evidence for Tara, while his late-time settings and outcomes become descendants of her antecedent and are therefore excluded from the fixtures. Evaluating the suspectation of Tara's counterfactual homodyne current in this generic scenario, we found that conditioning on the causally accessible portion of Vlad's record acts as a post-selection on the compatible ensemble, producing weak-value-like features---most strikingly, a transient excursion of the suspected current beyond the eigenvalue range of the measured quadrature. Since some degree of causal contact between observers is the norm rather than the exception in real experiments, this generalization considerably widens the practical applicability of the formalism.

Third, we moved beyond the suspectation of the instantaneous measurement record to more informative summary statistics: the suspectations of filtered quantum states.
Because a filtered state is a functional of a classical record---a state of knowledge---it is a legitimate classical object within our paradigm, computed at the classical register nodes $\boss{R}$ of the HSM. We gave a unified expression for the suspected state estimates accessible to Tara, and evaluated three cases: her own counterfactual filtered state, Godot's counterfactual filtered state, and, as a special case of the latter in which her counterfactual antecedent is the null measurement, Vlad's filtered state. 
Our numerical results revealed nontrivial behaviour of these suspected states: the suspected counterfactual filtered state tracks the unconditioned evolution more closely than the actual filtered state, while exhibiting conditioning-induced features such as damped oscillations and dips correlated with Vlad's most probable detection times. The comparison of purities, in particular the fact that the mixture of more-pure smoothed states can be less pure than the corresponding mixture of filtered states, illustrates how the interplay between individual-trajectory purity and inter-trajectory distinguishability shapes ensemble-level estimates.

Tara's estimates of these various counterfactual filtered states have close relations with actual-world smoothed estimates, either conventional~\cite{PhysRevLett.115.180407}, or where one party makes a wrong assumption about the type of measurement the other party is using~\cite{Chantasri_2025}. This is in addition to the explicit connection between our counterfactual calculus and quantum state smoothing identified in Ref.~\cite{liu2026unifyingquantumsmoothingtheories}. 
We also strengthened that latter connection by deriving an explicit dual-frame representation of the smoothed state, in which the supposabilities for an unperformed informationally complete measurement directly supply the parameters defining the smoothed quantum state. 

\subsection{Future directions}

Several natural extensions of the counterfactual calculus developed here suggest themselves. First, adaptive measurement scenarios, in which earlier outcomes are fed forward to determine subsequent measurement settings, introduce a dynamical causal structure; analysing counterfactuals in this context would clarify how the formalism accommodates feedback and time-dependent interventions. Second, while we have evaluated suspectation 
values for counterfactual filtered states, the smoothed state generally provides a more accurate estimate of the underlying quantum state, and extending the analysis to counterfactual smoothed states is a natural next step. Third, the continuous-monitoring calculations would benefit from being revisited in a system with a $Z$-type coupling ($\propto \hat{\sigma}_z$) in place of $\hat{\sigma}_-$. In the atomic case, a detection event projects the system to the ground state, anti-correlating the pre- and post-jump states; combined with the dynamics repopulating the excited state, this competition 
complicates interpretation. A $\hat{\sigma}_z$ measurement, by contrast, preserves its eigenstates and may yield more transparent dynamics.

More ambitious open questions concern the scope of the formalism itself. One direction is to relax the restriction that counterfactual antecedents correspond solely to alternative measurement settings, by also allowing a subset of counterfactual outcomes. Another would be to extend the formalism to modifications of dynamics, initial states, or environmental interactions. This would broaden the class of counterfactual worlds under consideration, bringing the framework closer in spirit to counterfactuals as originally 
conceived in Lewisian analysis. 
One could further consider counterfactual analogues of other trajectory-level functionals of the record---for instance, thermodynamic quantities such as work and heat along
quantum trajectories \cite{alonso2016thermodynamics,elouard2019work,naghiloo2020heat}---for which the suspectation could quantify the energetic imprint of a measurement that was never performed. Consideration of thermal noise may require giving a nontrivial role to %Finally, while our 
the exogenous variables $\boss{U}$, 
which were non-existent in our examples. This would connect our formalism more closely to the broader program of quantum causal inference. 
Finally, exploring the counterfactual calculus within other interpretations of quantum theory could provide new insights, or expose deficiencies in those interpretations.

\ack
We thank Eric Cavalcanti, Ryan Savage and Emily Adlam for discussions.
This work was supported by the ARC Centre of Excellence project number CE170100012  and an Australian Government RTP Scholarship. K.T.L. acknowledges the Plan France 2030 through the project NISQ2LSQ (Grant ANR-22-PETQ-0006) and the project OQuLus (Grant ANR-22-PETQ-0013).

%\bibliography{References.bib}

%Estimate Tara's filtered state.
\appendix

\section{Notation for measurement records and conditioned states \label{App:A1}}
Estimation theory aims to describe the system by conditioning upon measurements. The system takes \textit{different quantum trajectories} or paths, depending on the measurement records $O_t$ obtained in each infinitesimal time $[t+\dd t)$. 
Here we reintroduce the following notation for the records:
\begin{align}
    \past{{O}}_t=\{{O}_s:s\in[t_i,t)\}\,, &\\
    \overrightarrow{{O}}_t=\{{O}_s:s\in[t,t_f)\}\,, &\\
    \both{{O}}=\{{O}_s:s\in[t_i,t_f)\}\,.
\end{align}
Similarly for the conditioned quantum state we use the notation
\begin{equation}
    \rho_t^{\past{O}},
\end{equation}
which denotes the state conditioned on all measurement records prior to time $t$. Since the time label already appears in the state subscript, we suppress the explicit time subscript on the record where no ambiguity arises.

\section{Evaluating ~\erf{Eq:GenericSus} 
\label{App:A}}

Recall the map that we introduced in Eq.~\eref{eq:Pd-channel}. Here we rewrite the map in the following way
\begin{equation}
    {\cal C}({\dd N^C_t};{\rm pc})\bullet = \frac{\hat{C} \bullet \hat{C}\dg}{\lambda} \dd N_t^C +(1-\dd N^C_t)\times\left[\bullet - \frac{\dd t}{2}\, \bar{\mathcal{H}}[\hat{C}\dg\hat{C}-\lambda]\bullet\right] , \label{Newjumptraj}
\end{equation}
where $\lambda$ corresponds to an appropriate ostensible probability such that 
\begin{equation}
    \mathbb{E}_{\text{ost}}\left[\dd N_t^V\right]=\lambda \dd t.
\end{equation} 
% Note that for Tara there is no defined ostensible distribution, this is because a typical record for Tara's photon detection  is always chosen throughout the thesis which will be explained in the next chapter. Note also that Vlad's operator $\hat V$ here is different from that mentioned below \erf{6.2}, precisely because we have introduced an ostensible distribution for him. 
% \ing{Until here shhift in the appendix}
Note that for Tara the map is~\erf{jumptraj} and there is no defined ostensible distribution, this is because a typical record for Tara's photon detection  is always chosen. The above defined map is used for our simulations.

The evaluation of the suspectation value~\erf{Eq:GenericSus} follows exactly the same methodology as evaluating the suspected homodyne photo current in strict SLS scenario~\cite{banerjee2025counterfactualquantummeasurements}. As noted before in~\erf{eq:WV} we rewrite here the first factor in~\erf{twotermsinsuspect} as the pre and post selected mean homodyne current at time $t$
\begin{equation}
    \mathbb{E}[\text{Y}_t|\overleftarrow{M}=\past{m},\overrightarrow{M}=\fut{m}] = {\rm Re}\frac{\Tr[\hat E^{\overrightarrow{m}}_t (i\hat{\sigma}_{-})\rho^{\overleftarrow{m}}_t ]}{\Tr[\hat E^{\overrightarrow{m}}_t \rho^{\overleftarrow{m}}_t]}. \label{Eq:A1}
\end{equation} 
We can use the new map~\erf{Newjumptraj} (${\cal C}({\dd N^C_t};{\rm pc})={\cal V}({\dd N^V_t};{\rm pc})$) to update Vlad's filtered state $\rho^{\overleftarrow{m}}_t$, for each record $\both{m}$ in the ensemble, starting from a ground state as 
\begin{equation}{\label{VFS-evolution}}
\begin{split}
    \tilde{\rho}^{\past{M}}_{t+\dd t} &= {\cal T}({\dd \nullset};{{\rm nm}})\, {\cal U}\,{\cal V}({\dd N^V_t};{\rm pc})\tilde{\rho}^{\past{M}}_t\\
    &=
\begin{cases}
\cu{1 + \dd t\ro{ {\cal L} + \lambda - \eta_{\rm V}\gamma{\cal J}[\sigma_{-}]}}\tilde{\rho}^{\past{M}}_t  & \text{if no click,}\\
  (\eta_{\rm V}\gamma/\lambda){\cal J}[\sigma_{-}] \tilde{\rho}^{\past{M}}_t & \text{if click.}
\end{cases}  
\end{split} 
\end{equation} 
with $\tilde{\rho}_{t_i}=\rho_{t_i}$. 
Then starting from $\hat{E}_{t_f}=\hat I$, the effect $\hat E^{\overrightarrow{m}}_t$ can be evolved via the {\em adjoint} of the maps in \erf{VFS-evolution} [as in~\erf{eq:VRF}] 
\begin{equation}
\begin{split}
\hat E^{\fut{M}}_{t-\dd t} & = {\cal V}\dg({\dd N^V_t};{\rm pc})\,{\cal U}\dg{\cal T}\dg({\dd \nullset};{{\rm nm}})\hat E^{\fut{M}}_{t}\\
& = \begin{cases}
\cu{1 + \dd t\ro{ {\cal L}\dg + \lambda - \eta_{\rm V}\gamma{\cal J}[\sigma_{-}]}}E^{\fut{M}}_{t}  & \text{if no click,}\\
  (\eta_{\rm V}\gamma/\lambda){\cal J}[\sigma_{-}] E^{\fut{M}}_{t} & \text{if click.}
\end{cases} 
\end{split}
\end{equation}

The first difference in the generic scenario is that the average over $\both{m}$ in~\erf{twotermsinsuspect} is performed over a restricted sub-ensemble $\bar{\mathfrak{E}}\subset\mathfrak{E}$ (\cref{sec:BreakParty},\cite{banerjee2025counterfactualquantummeasurements}), consisting of the set of all the measurement records where Vlad does not detect any photons in the interval $[0,t_i^+)$ (with $t_i^+=1.2\gamma\inv$). Explicitly this corresponds to all the records $\past{M}_{1.2}=0$.
A second and conceptually distinct modification arises from Tara's lack of knowledge of what measurement Vlad is performing in the final portion of his interval $[t_f^-,t_f)$. Therefore, in this interval, Tara can no longer condition her inference on Vlad's measurement outcomes and the only solution that she can use to compute the suspectation is the uncondition evolution of the state (the master equation solution). Consequently, the conditional expectation~\erf{Eq:A1} can be replaced by a piecewise expression
\begin{equation}\mathbb{E}[\text{Y}_t|\overleftarrow{M}=\past{m},\overrightarrow{M}=\fut{m}] =\begin{cases}
  {\rm Re}\frac{\Tr[\hat E^{\overrightarrow{m}}_t (i\hat{\sigma}_{-})\rho^{\overleftarrow{m}}_t ]}{\Tr[\hat E^{\overrightarrow{m}}_t \rho^{\overleftarrow{m}}_t]}&[t_i,t_f^-)\\
\Tr[\hat \sigma_y \rho_t]& [t_f^-,t_f).
\end{cases}
\end{equation}
The second factor in~\erf{twotermsinsuspect} is the probability of the equiprobable ensemble  of Vlad's photon-detection records ${\mathfrak{E}}=\{\both{M}|\both{n}\}$, that  Tara generates  using her actual record $\past{n}$ \cite{banerjee2025counterfactualquantummeasurements}. As a result, any average involving the conditional probability $\wp(\both{m}|\both{n})$ can be replaced by
\begin{equation}
    \sum_{\both m} f(\both{m})\times \blk \text{Pr}(\both{M}=\both{m}|\both{N}=\both{n})  =  \lim_{|{ \mathfrak{E}}|\to\infty} \sum_{\both m \in \mathfrak{E}} f(\both{m})\times \frac{1}{|{ \mathfrak{E}}|}\,, \label{eq:Equ-Average}
\end{equation}
where $f(\both{m})$ can be any expression depending on $\both{m}$. 

When the analysis is restricted to a sub-ensemble (e.g. $\bar{\mathfrak{E}}$), the same replacement holds with $\mathfrak{E}$ replaced by the corresponding sub-ensemble.

\section{Tara's estimate of her counterfactual filtered state \label{App:B}}

We begin with the information available to Tara, namely her evidence, 
\begin{equation}
    \boss{e} = (\both{n};{\rm pc};{\rm pc}),
\end{equation}
where pc denotes photon counting. We consider the typical record consistently used throughout our calculations where Tara, in her run of the experiment, records a single click in her detector at time $t_T=6.25\gamma\inv$ and the total time interval is discretized into 1000 points with an infinitesimal interval $\dd t=0.01\gamma^{-1}$. 

Next we rewrite the state in~\erf{eq:TCFS} as
\begin{equation}
   \sum_{\past{Y},\both{m}}\!\! \wp(\both{m}|\both{n})\wp(\past{Y}{}|\both{m})\rho_t^{\past{Y}{}}. \label{eq:A2}
\end{equation}
We describe how each term in this expression is evaluated numerically.
The first factor, $\wp(\both{m}|\both{n})$, is obtained from the equiprobable ensemble $\mathfrak{E}$ introduced in~\cref{sec:BreakParty} and \cite{banerjee2025counterfactualquantummeasurements}.

To compute the second factor in~\erf{eq:A2}, we independently generate an ensemble of homodyne measurement records
\(\past{Y}=\{\dd Y_t\}\). Each record is a sequence of homodyne current increments, which for numerical simulations are sampled from the usual Gaussian ostensible distribution,
\begin{equation}
    \wp_{\rm ost}(\past{Y})
    =
    \frac{1}{\sqrt{2\pi \dd t}}
    \exp\!\left(
    -\frac{\dd Y_t^2}{2\dd t}
    \right). \label{Ost-homo}
\end{equation}

These records are used to evolve the unnormalized conditional state according to the update equation~\erf{GFS}. For each pair
\((\past{m},\past{Y})\),
\begin{equation}
    \tilde{\rho}^{\past{m},\past{Y}}_{t+\dd t}
    =
    \mathcal{T}(Y_t;{\rm hd})\,
    \mathcal{U}\,
    \mathcal{V}(m_t;{\rm pc})\,
    \tilde{\rho}^{\past{m},\past{Y}}_{t},
\end{equation}
yielding \(\tilde{\rho}_t^{\past{m},\past{Y}}\) at time \(t\).

In parallel, for each photon-detection record \(\both{m}\), the corresponding effect matrix
\(\hat{E}_t^{\fut{m}}\) is obtained by backward evolution,
\begin{equation}
    \hat{E}_{t-\dd t}^{\fut{m}}
    =
    \mathcal{V}^{\dagger}_{t-\dd t}(m_t;{\rm pc})\,
    \mathcal{U}^{\dagger}\,
    \mathcal{T}^{\dagger}(\dd \nullset;{\rm nm})\,
    \hat{E}_{t}^{\fut{m}}. \label{Vladeffect}
\end{equation}

The conditional distribution of homodyne records given the photon-detection record \(\both{m}\) is then sampled using the weights
\begin{equation}
    \wp(\past{Y}|\both{m})
    =
    \frac{
    \Tr\!\left[
        \hat{E}_t^{\fut{m}}
        \tilde{\rho}_t^{\past{m},\past{Y}}
    \right]
    \wp_{\rm ost}(\past{Y})
    }
    {
    \mathbb{E}_{\past{Y}\sim{\rm ost}}
    \!\left[
        \Tr\!\left(
            \hat{E}_t^{\fut{m}}
            \tilde{\rho}_t^{\past{m},\past{Y}}
        \right)
    \right]
    },
\end{equation}
where \(\mathbb{E}_{\past{Y}\sim{\rm ost}}\) denotes an average over the ostensible homodyne ensemble.

Using the same ensemble of homodyne records, the normalized filtered states
\(\rho_t^{\past{Y}}\) are obtained by evolving the stochastic master equation~\erf{NormYequation}. The expression in~\erf{eq:A2} is then evaluated numerically as the nested average
\begin{align}
    \sum_{\past{Y},\both{m}}
    \wp(\both{m}|\both{n})
    \wp(\past{Y}|\both{m})
    \rho_t^{\past{Y}}
    =
    \mathbb{E}_{\past{M}|\both{N}}
    \left[
    \frac{
    \mathbb{E}_{\past{Y}\sim{\rm ost}}
    \!\left[
        \Tr\!\left(
            \hat{E}_t^{\fut{m}}
            \tilde{\rho}_t^{\past{m},\past{Y}}
        \right)
        \rho_t^{\past{Y}}
    \right]
    }
    {
    \mathbb{E}_{\past{Y}\sim{\rm ost}}
    \!\left[
        \Tr\!\left(
            \hat{E}_t^{\fut{m}}
            \tilde{\rho}_t^{\past{m},\past{Y}}
        \right)
    \right]
    }
    \right].
\end{align}

\section{Tara's actual smoothed estimate of the state \label{App:TaraSmoothing}}

We rewrite Tara's smoothed estimate~\erf{eq:TSM} of the fully conditioned state as
% \begin{equation}
%     \sum_{\past{m}}
%     \wp(\past{m}|\both{n})
%     \rho_t^{\past{m},\past{n}},
% \end{equation}

% where the smoothed probability density can be written down as 
% \begin{align}
%     \frac{\sum_{k} \text{Tr} \left[ \tilde{\rho}^{(k)}_{\overleftarrow{m},\overleftarrow{n}}(t)\hat{E}_{\overrightarrow{n}} (t)\right]\frac{\tilde{\rho}^{(k)}_{\overleftarrow{m},\overleftarrow{n}}(t)}{\text{Tr}\left[\tilde{\rho}^{(k)}_{\overleftarrow{m},\overleftarrow{n}}(t)\right]}}{\text{Tr}\left[\sum_{k} \text{Tr} \left[ \tilde{\rho}^{(k)}_{\overleftarrow{m},\overleftarrow{n}}(t)\hat{E}_{\overrightarrow{n}} (t)\right]\frac{\tilde{\rho}^{(k)}_{\overleftarrow{m},\overleftarrow{n}}(t)}{\text{Tr}\left[\tilde{\rho}^{(k)}_{\overleftarrow{m},\overleftarrow{n}}(t)\right]}\right]}
% \end{align}
\begin{align}
    \wp(\overleftarrow{m} | \overleftrightarrow{n}) \rho_t^{\overleftarrow{m},\overleftarrow{n}}
    &= \frac{P(\overleftarrow{m} , \past{n},\fut{n})}{P(\both{n})}
    \rho_t^{\overleftarrow{m},\overleftarrow{n}}
    \label{Sm:1}\\
    & \propto P(\overleftarrow{m} , \past{n},\fut{n})
    \rho_t^{\overleftarrow{m},\overleftarrow{n}}
    \nonumber 
    \propto P(\fut{n}|\past{m} , \past{n})
    P(\past{m} , \past{n})
    \rho_t^{\overleftarrow{m},\overleftarrow{n}}
    \nonumber\\
    & = \Tr \left[\hat{E}_t^{\fut{n}}\rho_t^{\overleftarrow{m},\overleftarrow{n}}\right]
    \wp_{\rm ost}(\past{m} , \past{n})
    \tilde{\rho}_t^{\overleftarrow{m},\overleftarrow{n}}
    \nonumber \\
    & = \Tr \left[\hat{E}_t^{\fut{n}}\tilde{\rho}_t^{\overleftarrow{m},\overleftarrow{n}}\right]
    \wp_{\rm ost}(\past{m} , \past{n})
    {\rho}_t^{\overleftarrow{m},\overleftarrow{n}}
    \label{Sm:2}\\
    & \propto \Tr \left[\hat{E}_t^{\fut{n}}\tilde{\rho}_t^{\overleftarrow{m},\overleftarrow{n}}\right]
    \wp_{\rm ost}(\past{m})
    {\rho}_t^{\overleftarrow{m},\overleftarrow{n}},
    \label{Sm-prob}
\end{align}
where, in~\erf{Sm:1}, we used introduced the proportionality by using the fact that Tara's record is fixed and therefore $P(\both{n})$ is constant. Similarly, the proportionality in the last line, \erf{Sm-prob}, was using the fact that Tara's ostensible distribution $\wp_{\rm ost}(\past{n})$ is fixed.
The choice of the ostensible distribution $ \wp_{\rm ost}(\past{m})$ for Vlad's jumps are
\begin{equation}
    \wp_{\text{ost}}(\dd N_t^V:=1)=\lambda(t) \dd t \;\; \wp_{\text{ost}}(\dd N_t^V:=0)=1-\lambda(t) \dd t\,.
\end{equation}
To reduce statistical errors, it is advantageous to choose an ostensible distribution that closely approximates the true distribution, thereby maximizing the effective ensemble size~\cite{chantasri2021unifying}. In the present case, a natural choice, given Tara's observed record, is an ostensible rate equal to the count rate predicted at each instant by Tara's filtered state $\rho^{\past{n}}_t$ (\erf{jumptraj}),
\begin{equation}
\lambda(t)=\gamma(1-\eta_{{\rm T}})\text{Tr}[\hat{\sigma}_{+}\hat{\sigma}_{-}\rho^{\past{n}}_t].
\label{ost}
\end{equation}
We evaluate Tara's effect matrix as piecewise solutions in the two intervals defined by the detection time $t_T$ and using the evolution map~\erf{eq:Taraeffect}
which can be explicitly written down as 
\begin{equation}
    -\dd \hat{E}_t^{\fut{n}} = + i\dd t \left[ \frac{\Omega}{2} \hat{\sigma}_x , \hat{E}_t^{\fut{n}} \right] + \, \dd t \, \mathcal{D}^\dagger \left[ \hat{V} \right] \hat{E}_t^{\fut{n}} - \, \dd t \, \frac{1}{2} \left( \hat{T}^\dagger \hat{T} \hat{E}_t^{\fut{n}} + \hat{E}_t^{\fut{n}} \hat{T}^\dagger \hat{T} \right),
\end{equation}
where $\hat{T}=\sqrt{\gamma\eta_T}\hat{\sigma}_-\,\,\, \text{and}\,\,\, \hat{V}= \sqrt{ \gamma(1-\eta_{{\rm T}})/\lambda}\,\hat{\sigma}_-,$ with $\hat{V}$ redefined using the appropriate ostensible distribution for Vlad's jumps. The initial conditions for the two intervals are 
\begin{equation}
    \begin{cases}
        \hat{E}^{\fut{n}}_{t_T}=\gamma \delta t |e\rangle \langle e| & \text{in the interval $[0,t_T)$}\\
        \hat{E}^{\fut{n}}_{t_f}=\hat{I} & \text{in the interval $[t_T,t_f)$}.\\
    \end{cases}
\end{equation}
Similarly, we evaluate the true state ${\rho}_t^{\overleftarrow{M},\overleftarrow{n}}$ for each record $m$ in the ensemble $\mathfrak{E}$ using the piecewise solution solved in the two intervals defined by $t_T$ using
\begin{equation}
    \tilde{\rho}^{\past{M},\past{n}}_{t+\dd t} = 
\begin{cases}
\cu{1 + \dd t\ro{ {\cal L} + \lambda - \gamma\eta_{\rm V}{\cal J}[\hat{\sigma}_{-}]-\gamma\eta_{\rm T}{\cal J}[\hat{\sigma}_{-}]}}\tilde{\rho}^{\past{M},\past{n}}_t  & \text{if no jumps\,,}\\
  (\gamma\eta_{\rm V}/\lambda){\cal J}[\hat{\sigma}_{-}] \tilde{\rho}^{\past{M},\past{n}}_t & \text{if Vlad records a jump\,,}\\
\text{arbitrary constant} \times {\cal J}[\hat{\sigma}_{-}] \tilde{\rho}_t^{\past{M},\past{n}}  & \text{when Tara records a jump at} \; (t=t_T)\,,
  \end{cases} \label{TS}
\end{equation}
with $\tilde{\rho}_{t_i}=\rho_{t_i}$, which is chosen to be the ground state. 

\section{Tara's estimate of Godot's filtered state \label{App:C}}

Following the similar logic discussed in~\erf{eq:Equ-Average} of ~\ref{App:A} we write down the estimate~\erf{eq:GS2} as,
\begin{align}
    \sum_{\past{Y}{},\both{m}} \wp(\both{m}|\both{n})\wp(\past{Y}{}|\both{m})\rho_t^{\past{O}^G}\nonumber\\
    =\lim_{|{ \mathfrak{E}}|\to\infty} \sum_{\past{Y}{},\,\both m \in \mathfrak{E}} \frac{1}{|{ \mathfrak{E}}|}\times\wp(\past{Y}{}|\both{m})\rho_t^{\past{O}^G}\,, \label{eq:B1}
\end{align}
where $\rho_t^{\past{O}^G}=\rho_t^{\past{m},\past{Y}{}}$, is Godot's conditioned state and the sum over $\both{m}$ corresponds to an average over the equiprobable ensemble $\mathfrak{E}$ and will not be repeated here. The remaining terms $\sum_{\past{Y}} \wp(\past{Y}|\both{m})\rho_t^{\past{m},\past{Y}}$ represents an average over the ensemble of homodyne records $\past{Y}$ conditioned on a fixed record $\both{m}$. We could have evaluated this approximately using the ensemble generated from the ostensible distribution of $\past{Y}$ homodyne records as in \erf{Ost-homo}. However, we can obtain an exact solution for this average by  exploiting the fact that the conditional state $\rho_t^{\past{m},\past{Y}}$ can be parametrized by a single variable, namely the Bloch angle $\theta$, such that
\begin{equation}
    \rho_t^{\past{m},\past{Y}} = \hat{S}(\theta) = \frac{1}{2}\left(\hat{I} + \sin{\theta}\,\hat{\sigma}_y + \cos{\theta}\,\hat{\sigma}_z\right).
\end{equation}
The conditional probability $\wp(\theta|\both{m})$ for Godot's counterfactual state $\hat{S}(\theta)$, given Vlad's record $\both{m}$, can then be expressed as %a function of $\theta$,
\begin{equation}\label{P(Y|m)}
    \wp(\theta|\both{m}) %\wp(\past{Y}|\both{m})
    =\frac{\tilde{\wp}_{\past{m}}%,\past{Y}
    (\theta)\Tr[\hat{E}_t^{\fut{m}}\hat{S}(\theta)]}{\int \dd\theta\,\tilde{\wp}_{\past{m}}%,\past{Y}
    (\theta)\Tr[\hat{E}_t^{\fut{m}}\hat{S}(\theta)]}.
\end{equation}
% where $\tilde{\wp}_{\past{m}}(\theta)$ is obtained as the solution to the partial differential equation~\erf{PDE} (see \cref{App:D2} for details of the parametrization and derivation of the partial differential equation). The operator $\hat{E}_t^{\fut{m}}$ is Vlad’s effect operator, defined in~\erf{eq:VRF} for the corresponding record $\both{m}$.
% To determine the conditional state $\rho_t^{\past{m},\past{Y}{}}$, we we adopt an alternative approach to solving the evolution equation~\erf{eq:TS-SME}. Since the atomic state is driven about the $x$-axis, its dynamics are confined to the $y$--$z$ plane of the Bloch sphere. This allows us to reparametrize the (unnormalized) state by the Bloch angle $\theta$, such that 
% \begin{equation}
%     \tilde{\rho}_t^{\past{M},\past{Y}{}} =(\Lambda_t/2)(\hat{I}+\sin{\theta}\,\hat{\sigma}_y+\cos{\theta}\,\hat{\sigma}_z)\,, \label{ParaTrueState}
% \end{equation} 
% where, $\Lambda_t\equiv \text{Tr}[\tilde{\rho}_t^{\past{M},\past{Y}{}}]$ takes care of the state's norm. 
% As shown in Section~5.2 of Ref.~\cite{chantasri2021unifying}, one can use this parametrized form of the unnormalized state in \erf{eq:TS-SME} to obtain an unnormalized probability density function (PDF) $\tilde{\wp}_{\past{M},\past{Y}{}}(\theta)$ for $\theta$ given the past records, which gives the normalized PDF simply by 
% \begin{equation}
%     \wp_{\past{M},\past{Y}{}}(\theta)=\frac{\tilde{\wp}_{\past{M},\past{Y}{}}(\theta)}{\int \dd\theta\tilde{\wp}_{\past{M},\past{Y}{}}(\theta)}. \label{NPDF}
% \end{equation}
The probability density function $\tilde{\wp}_{\past{m}}(\theta)$ is periodic and satisfies the boundary condition $\wp_{\past{m}}(\theta=0)=\wp_{\past{m}}(\theta=2\pi)\,\forall \,t$ and its initial condition is a $\delta$-function at $\theta=\pi$ reflecting the ground state. 
Specifically, one can obtain $\tilde{\wp}_{\past{m}}(\theta)$ by solving the partial differential equation (PDE)
\beq
\begin{split}
    \partial_t\tilde{\wp}_{\past{m}}(\theta)=&
- \partial_\theta\!\big[ A_\theta(\theta)\, \tilde{\wp}_{\past{m}}(\theta) \big] 
- \frac{\gamma \eta_{\rm V}}{2}\,(1 + \cos\theta)\, \tilde{\wp}_{\past{m}}(\theta) \\
&+ \frac{1}{2}\,\partial_\theta^2\!\Big[ B_\theta(\theta)^2\, \tilde{\wp}_{\past{m}}(\theta) \Big]
+ \partial_\theta\!\Big[ B_\theta(\theta)\,\sqrt{\gamma \eta_{\rm T}}\,\sin\theta \, \tilde{\wp}_{\past{m}}(\theta) \Big], \label{PDE}
\end{split}
\eeq
where
\begin{align*}
    A_\theta(\theta)&= 
-\Bigg(
\Omega \;-\; \frac{\gamma  \eta_{\rm V}}{2}\,\sin\theta \;+\; 
\frac{\gamma \eta_{\rm T}}{2}\,\cos\theta \,\sin\theta
\Bigg),\\
\ 
B_\theta(\theta)&= 
- \sqrt{\gamma \eta_{\rm T}}\,\big(1 + \cos\theta\big).
\end{align*}
This can be solved numerically (using Mathematica's `NDSolve' package) with the $\delta$-function initial condition approximated by a suitably narrow-width Gaussian function $\wp_{t=0}(\theta)=\frac{1}{\sqrt{2\pi\sigma^{2}}} \exp\,[-\frac{(\theta - \mu)^{2}}{2\sigma^{2}} ]$. For the initial ground state, we have chosen $\theta_0=\pi$ and $\sigma=0.01$, which gives the peak to be $\wp_{t=0}(\pi)\approx39.89$ instead of infinity. The effect matrix $\hat{E}_t^{\fut{m}}$ can be obtained from~\erf{Vladeffect}. 

This allows the sum over homodyne records $\past{Y}$ to be replaced by an integral over the parameter $\theta$,
\begin{equation}
\sum_{\past{Y}} \wp(\past{Y}|\both{m})\rho_t^{\past{m},\past{Y}}
=\frac{ \int \dd \theta \tilde{\wp}_{\past{m}}(\theta)\Tr[\hat{E}_t^{\fut{m}}\hat{S}(\theta)]\hat{S}(\theta)}{\int \dd\theta\,\tilde{\wp}_{\past{m}}(\theta)\Tr[\hat{E}_t^{\fut{m}}\hat{S}(\theta)]}
\, . \label{parametrized 2nd factor}
\end{equation}

\section{Tara's estimate of Vlad's filtered state \label{App:D}}

Tara's counterfactual estimate of Vlad's filtered state~\erf{Eq:V-cFS}, is equivalently her counterfactual estimate of Godot's filtered state in the scenario where Tara is not performing any measurement
\begin{equation}
    {\varrho}^G_t({Z'}^T\!\!=\nullset\,||\both{n})=
        %\varrho_t^{'\past{O}^V_t,T:\varnothing}
        \sum_{\past{m}}\wp(\past{m}|\both{n})
        \rho_t^{\past{m}} \label{eq:D1}
\end{equation}
The conditioned state $\rho_t^{\past{m}}$ can be obtained directly by evolving the initial ground state according to the update equation~\erf{VFS} 
\begin{equation}{\label{S10}}
    \tilde{\rho}^{\past{m}}_{t+\dd t} = 
\begin{cases}
\cu{1 + \dd t\ro{ {\cal L} + \lambda - \eta_{\rm B}\gamma{\cal J}[\sigma_{-}]}}\tilde{\rho}^{\past{M}}_t  & \text{if no click,}\\
  (\eta_{\rm B}\gamma/\lambda){\cal J}[\sigma_{-}] \tilde{\rho}^{\past{m}}_t & \text{if click.}
   
  \end{cases}
\end{equation}
with the initial condition $\tilde{\rho}_0=\rho_{t_i}$, and where $\lambda$ is the appropriate ostensible probability~\cite{banerjee2025counterfactualquantummeasurements}. 

Thus, for each measurement record $\past{m}$ in the equiprobable ensemble $\mathfrak{E}$, one can solve the above evolution to get the conditioned state $\rho_t^{\past{m}}$. The final counterfactual state is then obtained by averaging these conditioned states with respect to probability as in~\erf{eq:D1}.

\section*{References}

\bibliographystyle{iopart-num}
\bibliography{References}
\end{document}